%% file: estimating-assessing-ode-models.tex
\documentclass[journal = jacsat, manuscript=article]{achemso}
\usepackage{achemso}
\usepackage[utf8]{inputenc}
\usepackage{natbib}
\usepackage{graphicx}
\usepackage{hyperref}
\usepackage{xcolor}
\usepackage{geometry}
\usepackage{tikz,bm,amsmath, amsfonts,amssymb}

\usepackage[T1]{fontenc}
\usepackage{color}
\usepackage{fancyvrb}
\usepackage{upquote}

\DefineVerbatimEnvironment{Highlighting}{Verbatim}{commandchars=\\\{\}}
\usepackage{framed}
\definecolor{shadecolor}{RGB}{248,248,248}
\newenvironment{Shaded}{\begin{snugshade}}{\end{snugshade}}

\newcommand{\AttributeTok}[1]{\textcolor[rgb]{0.77,0.63,0.00}{#1}}

\newcommand{\CommentTok}[1]{\textcolor[rgb]{0.56,0.35,0.01}{\textit{#1}}}

\newcommand{\ConstantTok}[1]{\textcolor[rgb]{0.00,0.00,0.00}{#1}}
\newcommand{\ControlFlowTok}[1]{\textcolor[rgb]{0.13,0.29,0.53}{\textbf{#1}}}

\newcommand{\DecValTok}[1]{\textcolor[rgb]{0.00,0.00,0.81}{#1}}

\newcommand{\FloatTok}[1]{\textcolor[rgb]{0.00,0.00,0.81}{#1}}
\newcommand{\FunctionTok}[1]{\textcolor[rgb]{0.00,0.00,0.00}{#1}}

\newcommand{\NormalTok}[1]{#1}

\newcommand{\OtherTok}[1]{\textcolor[rgb]{0.56,0.35,0.01}{#1}}

\newcommand{\SpecialCharTok}[1]{\textcolor[rgb]{0.00,0.00,0.00}{#1}}

\newcommand{\StringTok}[1]{\textcolor[rgb]{0.31,0.60,0.02}{#1}}

\usepackage{setspace}

\title{Estimating and Assessing Differential Equation Models with Time-Course Data}

\author{Samuel W.K. Wong}
\affiliation{Department of Statistics and Actuarial Science, University of Waterloo, Waterloo ON, Canada}
\altaffiliation{Contributed equally to this work}
\author{Shihao Yang}
\affiliation{H. Milton Stewart School of Industrial and Systems Engineering, Georgia Institute of Technology, Atlanta GA, USA}
\altaffiliation{Contributed equally to this work}
\author{S. C. Kou}
\affiliation{Department of Statistics, Harvard University, Cambridge MA, USA}
\email{kou@stat.harvard.edu}

\begin{document}

\maketitle
\singlespacing

	\begin{abstract}
		Ordinary differential equation (ODE) models are widely used to describe chemical or biological processes. This article considers the estimation and assessment of such models on the basis of time-course data. Due to experimental limitations, time-course data are often noisy and some components of the system may not be observed. Furthermore, the computational demands of numerical integration have hindered the widespread adoption of time-course analysis using ODEs. To address these challenges, we explore the efficacy of the recently developed MAGI (MAnifold-constrained Gaussian process Inference) method for ODE inference.  First, via a range of examples we show that MAGI is capable of inferring the parameters and system trajectories, including unobserved components, with appropriate uncertainty quantification. Second, we illustrate how MAGI can be used to assess and select different ODE models with time-course data based on MAGI's efficient computation of model predictions. Overall, we believe MAGI is a useful method for the analysis of time-course data in the context of ODE models, which bypasses the need for any numerical integration.
	\end{abstract}

	\section{Introduction}
	
	It is fair to say that the advances in single-molecule and single-cell experiments have profoundly enhanced our understanding of biological processes \citep{xie1998optical,xie1999single,moerner2002dozen,sunney2002single,chung2012single,zong2012genome,hou2013genome,chen2017single,wu2022highly}. These advances, in particular, enable researchers to follow a system of interest over the course of the biological/chemical process, generating time-course data \citep{duggleby1986progress,radmacher1994direct,xue1995differences,brehm2004single,palmier2007rapid,locke2009using}. These rich sources of data have propelled scientific investigation: in chemistry, time-course analysis can reveal the effects of products and intermediates on a reaction, such would not be possible with rate measurements only \citep{duggleby2001quantitative}; in biology, time-courses have enabled the study of dynamic changes in gene expression \citep{van2020trajectory} and the determinants of single-cell outcomes \citep{spiller2010measurement}, to list just a few examples.
	
	This article considers the analysis of time-course data. As the dynamics of chemical or biological processes are often modeled by ordinary differential equations (ODEs), this article focuses on analyzing time-course data in the context of ODE models. Two broad questions are considered: (i) given an ODE model, how to infer the unknown parameter values as well as the unobserved components of the system; (ii) how to assess whether a specific ODE model adequately describes the underlying dynamics, or which one of the competing models best describes the dynamics given the available time-course data. These questions could be challenging to answer, since (a) the time-course data generated in the experiments are often quite noisy due to experimental uncertainties and/or measurement error, (b) it is often the case that due to various experimental limitations, not all the system components are observed in the experiments (i.e., some components are entirely unobserved during the course of the experiment), and (c) models associated with time-course data tend to be complex to analyze.
	
	This article explores a method to infer ODE models from time-course data, which was called MAGI (MAnifold-constrained Gaussian process Inference) \cite{yang2021inference}. The method employs two key ingredients: (i) a Gaussian process (GP) Bayesian prior on the trajectories (either observed or unobserved) of the ODE system, and (ii) placing a manifold constraint on the GP that satisfies the ODEs, which enable it to completely bypass the need for any numerical integration. The next section reviews the MAGI method.
	
\begin{figure}[ht!]
		\includegraphics[width=\textwidth]{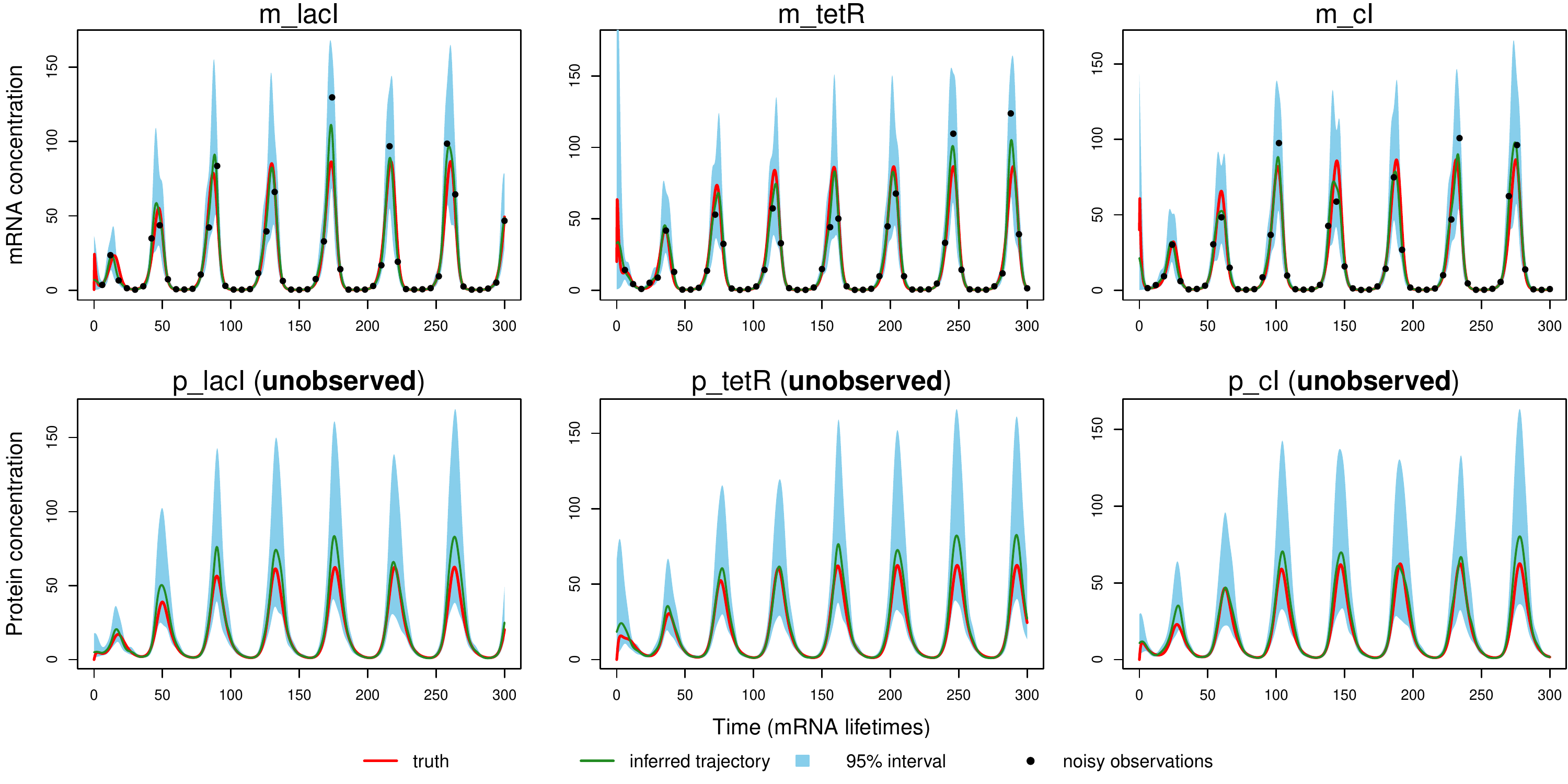}
		\caption{Inferred trajectories for a sample dataset from the repressilator gene regulation model. The black points are the noisy measurements. In this example, the protein concentrations are never observed. The red curves are the true trajectories. The inferred trajectories are shown by the green curves, with the blue shaded areas representing 95\% intervals. Both mRNA and protein concentrations are normalized as in ref \citenum{elowitz2000synthetic}.}
		\label{fig:rep-infer-single-dataset}
	\end{figure}

\begin{figure}[ht!]
		\centering
		\includegraphics[width=\textwidth]{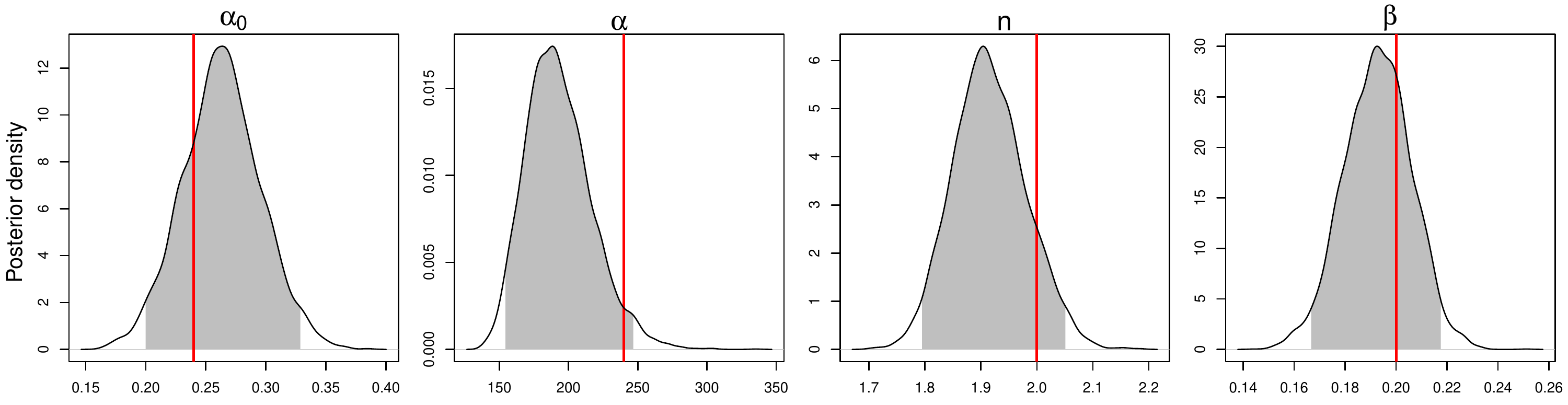}
		\caption{Bayesian posterior probability densities of system parameters for a sample dataset from the repressilator gene regulation model obtained by MAGI. The prior distributions of the parameters were the Lebesgue measure over the positive real numbers. The red vertical lines show the true parameter values used in the simulation. The shaded area represents the 95\% interval estimate of each parameter.}
		\label{fig:rep-param-single-dataset}
	\end{figure}

	To illustrate the MAGI method on the inference of ODE models from time-course data, we consider the model for a repressilator gene regulation network proposed by \citet{elowitz2000synthetic}, which represented one of the first successful attempts to engineer a \textit{de novo} synthetic network that could exhibit stable oscillatory behavior. The network was built as a three-gene loop of the successive repressors $lacI$, $tetR$, and $cI$ in \textit{E.~coli}. A time-course experiment then measured the expression of single cells (through multiple cell-division cycles) and confirmed the occurrence of stable oscillations. The work motivated subsequent studies on stochastic gene expression in single cells, e.g., ref \citenum{elowitz2002stochastic}.  An ODE model provides a simple and useful description for the dynamics of mRNA and protein levels corresponding to the three genes in the network, via six coupled differential equations \citep{elowitz2000synthetic}:
	
	\begin{minipage}{.2\textwidth}
		\begin{tikzpicture}
		\path 
		(0,0) node (cI) {$cI$}
		(0.75*1.25,1.25) node (tetR) {$tetR$}
		(1.5*1.25,0) node (lacI) {$lacI$};
		\draw [|-,thick](cI) -- (tetR);
		\draw [|-,thick](tetR) -- (lacI);
		\draw [|-,thick](lacI) -- (cI);
		\end{tikzpicture}
	\end{minipage}
	\begin{minipage}{.5\textwidth}
		\begin{eqnarray}\label{eq:represillator}
		\begin{split}
		\frac{dm_i}{dt} &= -m_i + \frac{\alpha}{1 + p_j^n} + \alpha_0 \\ 
		\frac{dp_i}{dt} &= -\beta (p_i - m_i)
		\end{split}
		\end{eqnarray}
		\medskip
	\end{minipage}
	
	\noindent where $i = (lacI, tetR, cI)$, $j = (cI, lacI, tetR)$, $m_i$ is the (scaled) mRNA concentration of $i$, and $p_i$ is the (scaled) protein concentration of $i$. The cyclic repressing behavior can be seen from these equations and the accompanying diagram; i.e., a large concentration of protein $cI$ will inhibit the transcription of $lacI$, and likewise proteins $lacI$ and $tetR$ inhibit transcription of $tetR$ and $cI$, respectively. The system parameters to be estimated from the noisy observational data are the Hill coefficient $n$, the ratio of protein-to-mRNA decay rate $\beta$, and the rates $\alpha_0$, $\alpha$ that govern transcription. Note that $\alpha_0$ can be interpreted as the transcription rate for $m_i$ when protein $j$ is saturated ($p_j \to \infty$), while $\alpha + \alpha_0$ is the rate when $p_j=0$.
	To mimic a realistic experimental scenario where typically either the protein or the mRNA is not measured \citep{maier2009correlation}, we assume in the simulation that noisy measurements of the mRNA levels are taken at 50 time points, while the protein concentrations are entirely unobserved. The black points in Fig \ref{fig:rep-infer-single-dataset} show an example of the noisy observations of mRNA, simulated from the system with parameter values and initial conditions that mimic the setup of Elowitz and Leibler\cite{elowitz2000synthetic}. The true trajectories are shown in red. The goal is to infer the system trajectories and parameter values based on the noisy observations of the mRNAs only. The panels of Fig \ref{fig:rep-infer-single-dataset} show that MAGI is able to infer the underlying trajectories (green curves) quite well, including the unobserved protein concentrations, without the use of any numerical integration (as the green curves largely agree with the red curves). The (Bayesian) posterior distributions of the parameters inferred from the data are shown in Fig \ref{fig:rep-param-single-dataset}, which well recover the true parameter values (represented by the red bars). 
	We will come back to this example in more detail in the Parameter Estimation and Inference section.
	
	The second general question considered by this article is model assessment in light of time-course data. It is often the case that multiple mechanisms (represented by different ODE models) are hypothesized for a process/reaction, and it is of interest to determine which mechanism is most compatible with the experimental data. This question could be difficult because (i) some system components may not be observed during the experiment, and (ii) different models might generate similar trajectories despite their different underlying mechanisms. The inference capability of MAGI leads to a natural method for assessing ODE models. The basic idea is to divide the observed time-course data into two parts: a training set (for example, data in the first time period) and a test set (for example, data in the second time period); then, apply MAGI only on the training data to estimate the model parameters, infer the system trajectories and generate predictions under the different model specifications. The models are then evaluated by comparing the model prediction obtained by MAGI against the test data. In the Model Assessment section, we will illustrate the method through an example that compares the original Michaelis-Menten model versus the Michaelis-Menten model with competitive inhibitor:
	\begin{equation*}
	E + S \underset{k_{-1}}{\stackrel{k_1}{\rightleftharpoons}} ES \, {\stackrel{k_2}{\longrightarrow}} \, E + P \quad\text{vs.}\quad 
	\begin{array}{l} E + S \underset{k_{-1}}{\stackrel{k_1}{\rightleftharpoons}} ES \, {\stackrel{k_2}{\longrightarrow}} \, E + P \\
	E + I \underset{k_{-3}}{\stackrel{k_3}{\rightleftharpoons}} EI 
	\end{array}
	\end{equation*}

	{{\bf Related work.}} 
	Another research direction parallel to ODE parameter inference and model assessment is the ODE identification problem \cite{bongard2007automated}, where the explicit form of the equations is unknown and needs to be inferred. One approach is to represent the unknown equations using basis function expansions. Subsequently, sparse regression techniques can be used to regularize the coefficients towards zero. As a result, identification methods are often limited to identifying ODEs or partial differential equations (PDEs) that are linear in the parameters \cite{rudy2017data,schaeffer2017learning,chen2021gaussian}. Deep learning methods have also been recently proposed to learn the governing equations; the idea is to represent the unknown ODE function as a neural network \cite{chen2018odenn,qin2019data}. However, (i) these methods still rely on numerical integration and are thus computationally intensive; (ii) the identified ODE functions are also neural networks, and while they might provide reasonable predictive performance, they still fall short in providing interpretable scientific models for the underlying mechanisms.
	
	A central ingredient of the MAGI method is the GP, which is a stochastic process with the property that the joint distribution of the process at any finite collection of time points is always multivariate Gaussian. Some well-known GPs include the Ornstein-Uhlenbeck process and the Brownian motion, both of which are also Markovian \cite{van2007stochastic}. In general, a GP is fully specified by its mean function and covariance function, and can be Markov or non-Markov.
	GPs have been previously explored in the parameter inference of differential equation models \cite{williams2006gaussian}, where the GP is employed as a prior for the solutions of ODEs or PDEs \cite{dondelinger2013ode,cockayne2017probabilistic,pmlr-v89-wenk19a}, but they all come with various limitations. For example, to infer the parameters of ODE models, Refs.\citenum{dondelinger2013ode,pmlr-v89-wenk19a} rely on an artificial parameter governing the mismatch between the GP derivatives and the ODE, which is \textit{ad hoc} and lacks rigorous justification. The approach adopted in Ref.\citenum{cockayne2017probabilistic} to study the PDE inverse problem solely considers linear PDEs, where the GP is explicitly conditional on the linear constraints. 
	
	Besides GPs, another popular surrogate-model approach for problems involving differential equations is the physics-informed neural network (PINN) \cite{raissi2019physics}, owing to the rapid development of deep learning. It has been proposed to solve the forward problem \cite{raissi2019physics}, the inverse problem \cite{depina2022application}, and PDE identification problems \cite{raissi2017physics}. However, PINNs for the inverse problem face challenges in computational efficiency, where the training of a complex neural network is necessary. Furthermore, deep learning methods, including PINNs, cannot properly address \textit{uncertainty quantification} for parameter estimation and ODE solutions without major Bayesian modifications \cite{yang2021b}. They may also be sensitive to the specific penalty terms used for the initial and boundary conditions \cite{lawal2022physics}, the neural network architecture \cite{krishnapriyan2021characterizing}, and the optimization algorithm \cite{berg2018unified}.
	
	{\bf Organization of the article. } The rest of the article is organized as follows. We start with a brief review of the MAGI method. We then illustrate the general applicability of the MAGI method on three examples of ODE models from chemical kinetics. We next discuss model assessment with MAGI, focusing on the comparison obtained via model prediction. We conclude this article with a general discussion. 
	
	\section{Methods} \label{sec2}
	
	The MAGI method, proposed in \citet{yang2021inference}, infers the trajectories and parameters of dynamic systems from noisy time-course data, without the need for any numerical integration. The method can work well even when some component trajectories of the dynamic system are entirely unobserved (as shown in Fig \ref{fig:rep-infer-single-dataset}, where all the protein concentrations are unobserved). MAGI accomplishes the inference goal by leveraging two key ideas: (i) a Gaussian process (GP) model for the trajectories of the dynamic system; (ii) constraining the GP on a manifold that satisfies the ODEs. This section provides a brief review of MAGI.
	
	Let $\bm{x}(t)$ denote the $D$-dimensional system of interest over time $t \in [0,T]$, whose dynamics are governed by the ODE
	\begin{equation}\label{eq:ode}
	\dot{\bm{x}}(t) = \frac{d \bm{x}(t)}{dt} = \mathbf{f}(\bm{x}(t),\bm{\theta}, t), 
	\end{equation}
	where $\dot{\bm{x}}(t)$ is shorthand for $d \bm{x}(t) / dt$ and $\mathbf{f}$ is a function that involves unknown parameters $\bm{\theta}$. Let $\bm{\tau}$ denote the vector of time points at which time-course data are available for some system components, and $\bm{y}(\bm \tau)$ the corresponding noisy measurements.
	
	MAGI is a Bayesian method. It begins by placing a prior distribution $\pi(\bm{\theta})$ on the unknown parameters $\bm{\theta}$ and a GP prior on $\bm{x}(t)$ (we thus view $\bm{\theta}$ as a realization from the distribution $\pi(\bm{\theta})$ and $\bm{x}(t)$ as a realization of a Gaussian process $\bm{X}(t)$). Under the GP, the conditional probability distribution of the derivative $\dot{\bm{X}}(t)$ given ${\bm{X}}(t)$ has a closed-form expression when the covariance function is twice-differentiable. This property allows a manifold constraint to be imposed on the GP such that $\bm{X}(t)$ satisfies the ODE equation \eqref{eq:ode}. Mathematically, this manifold constraint is defined by conditioning the GP on $W = 0$, where
	\begin{equation*}
	W = \sup_{t \in [0,T], d \in \{1,\ldots, D\}} |\dot X_d(t) - \mathbf{f}(\bm{X}(t), \bm{\theta}, t)_d|,
	\end{equation*}
	where the subscript $d$ refers to the $d$-th dimension of $\dot{\bm{X}}$ and $\mathbf{f}$. In actual computation, the constraint of $W =0$ is approximated by $W_{\bm{I}} =0$, where 
	\begin{equation}
	\label{eqn: dicretization}
	W_{\bm{I}} = \max_{t \in \bm{I}, d \in \{1,\ldots, D\}} |\dot X_d(t) - \mathbf{f}(\bm{X}(t), \bm{\theta}, t)_d|,
	\end{equation}
	and the maximum is over a set of discretization points $\bm{I} = \{ t_1, t_2, \ldots, t_n\}$ in  $[0,T]$. Following the Bayesian paradigm, MAGI then considers the  joint posterior distribution of $\bm{\theta}$ and $\bm{x}(\bm{I})$ (i.e., $\bm{x}$ at the points in $\bm{I}$) given the manifold constraint $W_{\bm{I}} =0$ and the noisy time-course data $\bm{y}(\bm \tau)$: $p(\bm{\theta}, \bm{x}(\bm{I}) |  W_{\bm{I}} = 0, \bm{y}(\bm \tau))$, which is
	\begin{align}
	&p(\bm{\theta}, \bm{x}(\bm{I}) |  W_{\bm{I}} = 0, \bm{y}(\bm \tau))  \notag \\
	& \propto 
	\pi(\bm{\theta}) \, p(\bm{X}(\bm{I}) = \bm{x}(\bm{I}) )\,  p(  \bm{y}(\bm \tau) | \bm{x}(\bm{I})) \, p(\bm{\dot X}(\bm{I}) = \mathbf{f}(\bm{x}(\bm{I}), \bm{\theta}, \bm{I}) | \bm{x}(\bm{I}) ). \label{eq:main}
	\end{align}
	Here, $\pi(\bm{\theta})$ is the prior density of the parameters, $p(\bm{X}(\bm{I})=\bm{x}(\bm{I}))$ is the multivariate Gaussian density from the GP prior on $\bm{X}(t)$ for $\bm{X}(t)$ to take the value $\bm{x}(\bm{I})$ at the time points $\bm{I}$, $ p(  \bm{y}(\bm \tau) | \bm{x}(\bm{I}))$ is the likelihood of the noisy observations, and $p(\bm{\dot X}(\bm{I}) = \mathbf{f}(\bm{x}(\bm{I}), \bm{\theta}, \bm{I}) | \bm{x}(\bm{I}) )$ is the multivariate Gaussian density for $\dot{\bm{X}}(t)$ conditioning on ${\bm{X}}(t)$, taking the value $\mathbf{f}(\bm{x}(\bm{I}), \bm{\theta}, \bm{I})$ at time points $\bm{I}$. Each of the terms on the right hand side of \eqref{eq:main} has a closed-form expression as described in the Appendix. 
	
	With the posterior distribution specified, MAGI uses Hamiltonian Monte Carlo (HMC) to draw samples of $\bm{\theta}$ and $\bm{x}(\bm{I})$ from  \eqref{eq:main}. After sampling convergence, inference of $\bm{\theta}$ and $\bm{x}(\bm{I})$ can be drawn from the Monte Carlo samples. For example, one can take the posterior means of $\bm{\theta}$ and $\bm{x}(\bm{I})$ as the parameter estimates and the inferred trajectories of $\bm{x}(t)$, respectively. It is the combination of GP and the introduction of manifold constraint in MAGI that leads to a principled statistical framework for inference of ODE systems, which completely bypasses the need of any numerical integration. For additional details of the MAGI method, such as the specification of the GP, we refer the reader to \citet{yang2021inference}. 
	
	\section{Results and Discussion}
	\subsection{Parameter estimation and the inference of system trajectories from time-course data}
	This section applies the MAGI method to three models. MAGI infers both the parameter values and the system trajectories (including those of the completely unobserved components) in each system. The three examples, ranging from the repressilator gene regulation network, the Michaelis-Menten model to a gene regulation network that contains more than 10 system components, illustrate the versatility of the MAGI method.
	
	\subsubsection{Repressilator gene regulation network} \label{sec:represillator}
	
	We continue the gene represillator example presented in the Introduction. \citet{elowitz2000synthetic} used the following parameter values for their theoretical study: $\alpha_0 = 0.24$, $\alpha = 240$, $n = 2$, and $\beta = 1/5$. In equation \eqref{eq:represillator}, the concentrations of $m_i$ and $p_i$ are respectively scaled such that they are unitless, and time is also unitless after scaling by mRNA lifetime \citep{elowitz2000synthetic}. As a consequence, all four parameters $\alpha_0$, $\alpha$, $n$ and $\beta$ are unitless. To mimic their system trajectories, the initial conditions for the mRNA concentrations of $lacI$, $tetR$, and $cI$ in our simulation were taken to be $0.4$, $20$, and $40$, respectively (in terms of the number of proteins produced per transcript), while the initial protein concentrations (in terms of the number of copies needed to achieve half the maximum repressor efficacy) are considered to be negligible (0.01). (These initial conditions are only used in simulating the data and are not known when applying MAGI, i.e., MAGI does not assume knowledge of the initial conditions.) The experimental examples of oscillatory behavior shown in \citet{elowitz2000synthetic} suggest that noise in fluorescence measurements is approximately 10-15\% of the mRNA levels. For our illustration in Fig \ref{fig:rep-infer-single-dataset} we used a larger simulation noise, namely 30\% multiplicative error, to demonstrate the inference capabilities of MAGI on noisy data. This was implemented by applying a log-transform to the ODEs and adding white Gaussian noise with SD 0.3 to the true mRNA trajectories at the times $t = 6, 12, \ldots, 300$ (mRNA lifetimes), yielding the black points in Fig \ref{fig:rep-infer-single-dataset}. We ran MAGI on the log-transformed ODEs, assuming the noise SD to be known based on knowledge about the experiment.
	
	\begin{figure}[t!]
		\includegraphics[width=\textwidth]{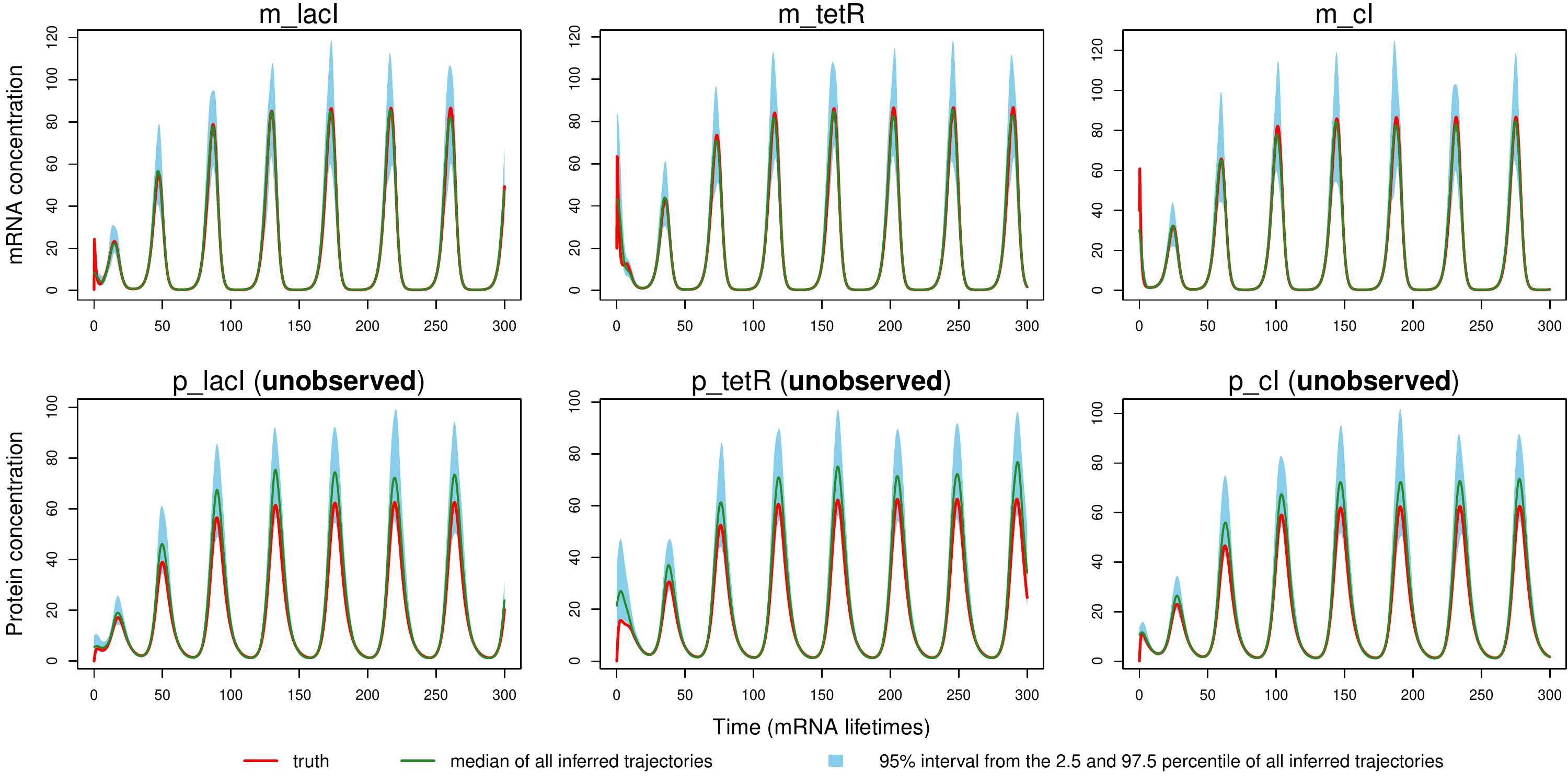}
		\caption{Inferred trajectories over 100 simulation repetitions from the repressilator gene regulation model. The red curves are the true trajectories. The green curves are the median of all inferred trajectories, with the blue shaded areas showing the 95\% intervals given by the 2.5 and 97.5 percentile of all inferred trajectories. Both mRNA and protein concentrations are normalized as in ref \citenum{elowitz2000synthetic}.}
		\label{fig:rep-infer-repeated-dataset}
	\end{figure}
	
	Fig \ref{fig:rep-infer-single-dataset} shows that the inferred trajectories for all six components, including the trajectories for the three protein concentrations, which are entirely unobserved, are largely recovered for the example dataset simulated from this setup: the true trajectory (red) is well-contained within the 95\% posterior bands (blue shaded areas in Fig \ref{fig:rep-infer-single-dataset}) for each component. Fig \ref{fig:rep-param-single-dataset}, plotting the posterior distributions of the parameters, shows that all the parameters are within the 95\% intervals (the gray shaded areas). Note that the prior distributions of the parameters in this case were the Lebesgue measure (i.e., uniform) over the positive real numbers; thus, the inference is informed entirely by the data. To assess the robustness of the inference, we generated 100 independently simulated datasets based on this same setup and ran the MAGI method for each. Fig \ref{fig:rep-infer-repeated-dataset} summarizes the results across these 100 datasets. The median inferred trajectories (green curves) capture the underlying system behavior for the unobserved protein components, and are very close to the truth (red curves) for the mRNA components. We summarize the parameter inference in Table \ref{tab:rep-param}, taking the posterior mean as the parameter estimate for each dataset. Over the 100 noisy datasets, we see that $\alpha_0$ is recovered very accurately, while $\alpha$, $n$, and $\beta$ have little to moderate errors in their recovery. (The intuitive reason for these results is that $\alpha$, $n$, and $\beta$ are closely tied to the behavior of the unobserved protein components rather than the observed mRNAs as seen in equation (\ref{eq:represillator})).
	
	\begin{table}[ht]
		\centering
		\caption{Parameter estimates over 100 simulated datasets from the repressilator gene regulation model. The average parameter estimate over the 100 simulation repetitions is shown with its SD after the $\pm$ sign.}
		\begin{tabular}{ccc}
			\hline
			Parameter & Truth & MAGI estimate \\
			\hline
			$\alpha_0$ & 0.24 & 0.239 $\pm$ 0.022 \\
			$\alpha$ & 240 &  186.7 $\pm$ 11.9 \\
			$n$ & 2 & 1.89 $\pm$ 0.02 \\
			$\beta$ & 0.2 & 0.194 $\pm$ 0.003 \\
			\hline
		\end{tabular} \label{tab:rep-param}
	\end{table}
	
	\subsubsection{The Michaelis-Menten model}\label{sec:MM}
	
	The Michaelis-Menten model, originated from the pioneering work of \citet{michaelis1913kinetik} in invertase experiments, has been foundational for studies of enzyme catalysis \citep{cornish2015one}. The corresponding Michaelis-Menten mechanism for enzymatic reactions can be depicted as
	
	\begin{minipage}{.2\textwidth}
		\begin{align*}
		E + S \underset{k_{-1}}{\stackrel{k_1}{\rightleftharpoons}} ES \, {\stackrel{k_2}{\longrightarrow}} \, E + P,
		\end{align*}
		\smallskip
	\end{minipage}%
	\begin{minipage}{.7\textwidth}
		\footnotesize{
			\begin{equation}\label{eq:mm}
			\begin{split}
			\frac{d[E]}{dt} &= -k_1 [E][S] + (k_{-1}+k_2) [ES] \\
			\frac{d[S]}{dt} &= -k_1 [E][S] + k_{-1} [ES]\\
			\frac{d[ES]}{dt} &=  k_1 [E][S] - (k_{-1}+k_2) [ES] \\
			\frac{d[P]}{dt} &=  k_2 [ES]
			\end{split}
			\end{equation} \smallskip}
	\end{minipage}
	
	\noindent where the enzyme $E$ binds reversibly with a substrate $S$, forming an intermediate complex $ES$ that decomposes into the product $P$ along with the original enzyme. The key quantities that summarize the kinetics are the Michaelis constant $K_M$ and the rate of catalysis $k_{cat}$, which are related to the rate parameters ($k_1$, $k_{-1}$ and $k_2$) through $K_M = (k_2+k_{-1})/k_1$ and $k_{cat} = k_2$.
	
	Time-course experiments can provide the data to estimate $K_M$ and $k_{cat}$ for a particular reaction, by measuring the product and substrate concentrations over time (known as \textit{progress curves}). There are two main approaches to utilizing progress curve data \citep{duggleby2001quantitative,choi2017beyond,aledo2022renz}: (i) extracting the initial reaction rate, as a function of substrate concentration; and (ii) using measurements at all available time points in the progress curve. The first approach has a long history beginning from the original \citet{michaelis1913kinetik} study, partly owing to its computational simplicity; e.g., simple linear regression of substrate and rate (\citep[Lineweaver-Burk plots][]{lineweaver1934determination}) can be used to estimate $K_M$ and $k_{cat}$. However, this approach typically requires multiple experiments over a range of substrate concentrations, and extracting the initial reaction rates does not fully utilize the experimental data collected in a time-course experiment. Furthermore, deviations from the assumed kinetics may only be evident when the full time-course is analyzed \citep{duggleby2001quantitative,lu1998single,xie1999single,xie2001single,sunney2002single,kou2005single,min2005fluctuating,english2006ever,min2006does}. In contrast, the kinetic parameters $K_M$ and $k_{cat}$ may potentially be estimated with only a single progress curve with the second approach. With advances in computational analyses of differential equations, analysis of full progress curves has become more commonly practiced and is the approach we take in this work using MAGI.
	
\begin{figure}[ht!]
		\includegraphics[width=\textwidth]{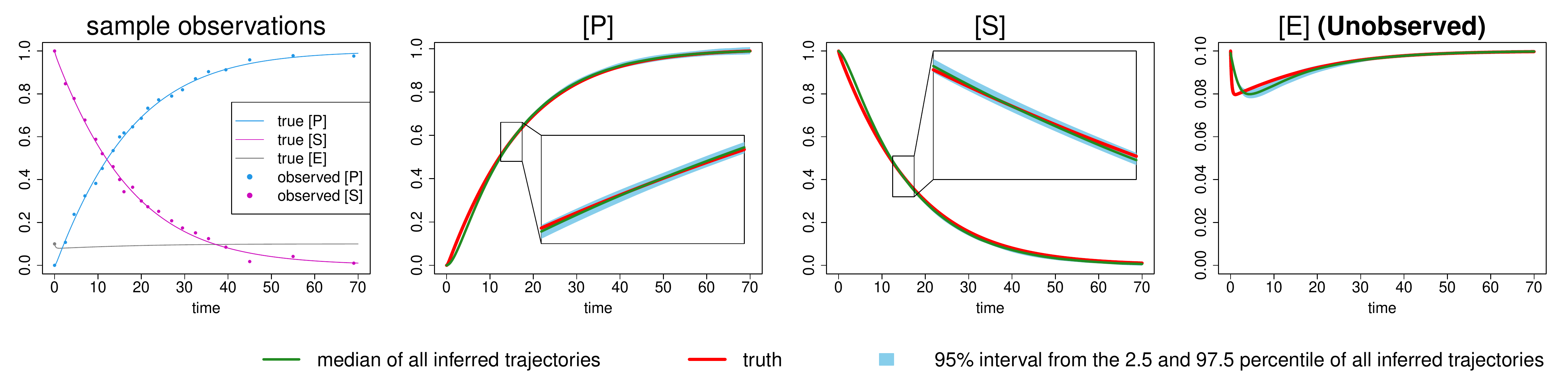}
		\caption{Inferred trajectories from the Michaelis-Menten model. \textbf{Left-most panel:} A sample dataset of observations, where only $P$ and $S$ components are observed at 20 time points \{2.5, 4.5, 7, 9.5, 11, 13.5, 15, 16, 18, 20, 21.5, 24, 27, 
			29.5, 32.5, 35.5, 39.5, 45, 55, 69\} (minute). The initial conditions $P = 0$ mM, $S = 1$ mM, $E = 0.1$ mM are also known. \textbf{Right three panels:} Inferred trajectories over 100 simulated datasets from the Michaelis-Menten model. The blue shaded area represents the 95\% interval, which is magnified in the figure inset.  }
		\label{fig:MM-vanilla}
	\end{figure}

	To numerically illustrate the MAGI method, we simulate the Michaelis-Menten model at true parameter values $k_1 = 0.9$ (min$\cdot$mM)$^{-1}$, $k_{-1} = 0.75$ (min)$^{-1}$, $k_2 = 2.54$ (min)$^{-1}$ and initial conditions $P = 0$ mM, $S = 1$ mM, and $E = 0.1$ mM. This setting closely follows the experimental data in the hydrolysis of phenylphosphate, catalyzed by prostate acid phosphatase considered in \citet{yun1977simple}, an early study with progress curve. To make the system more challenging, we suppose that only 20 unevenly spaced observations are available for the $P$ and $S$ components, which is more sparse than the experimental data in \citet{yun1977simple}. We further increase the inference difficulty by doubling the noise level of the experimental data \citep{yun1977simple} in our simulation study, which is implemented as additive Gaussian noise with SD 0.02 mM. The left-most panel of Fig \ref{fig:MM-vanilla} shows a sample dataset of sparse and noisy observations, where the figure caption lists the observation time points. We also assume that the initial conditions (i.e., $P$, $S$, $E$ at $t=0$) are known without noise, as is the case in most experimental settings. 
	
	In the implementation of MAGI, we used the relationship of $[ES] = [E]_0 - [E]$ to reduce the original 4-component system involving $P$, $S$, $E$ and $ES$ into an equivalent 3-component system involving only $P$, $S$ and $E$, where $E$ is completely unobserved beyond the initial condition. We set the discretization points $\bm I$ in equation \eqref{eqn: dicretization} to be evenly spaced from $t=0$ to $t=70$ minutes at 0.5 minute intervals, i.e., $\bm I = \{0, 0.5, 1,\ldots, 70\}$ (minute).
	
	Fig \ref{fig:MM-vanilla} shows MAGI's inferred trajectories over 100 simulated datasets. The $P$ and $S$ components are well recovered from the noisy observations across the 100 simulation repetitions: the 95\% interval band (the blue area) is so narrow around the truth that we can only see the band clearly after magnification (as shown in the figure inset). For the unobserved $E$ component, MAGI is able to recover it reasonably well, albeit with some estimation error in the early stages of the reaction, which could possibly be attributed to the initial rapid changes and the sparse observations. The unobserved $E$ component is recovered well after that initial period. 
	
	\begin{table}[ht]
		\centering
		\caption{Parameter estimates over 100 simulated datasets from the Michaelis-Menten model. The average parameter estimate is shown with its SD after the $\pm$ sign.}
		\begin{tabular}{ccc}
			\hline
			Parameter & Truth & MAGI estimate \\ 
			\hline
			$k_{cat}$ & 2.54 & 2.47 $\pm$ 0.17 \\
			$K_M$ & 3.66 &  3.43 $\pm$ 0.26  \\ 
			\hline
		\end{tabular} \label{tab:MM-vanilla-param}
	\end{table}
	
	\begin{figure}[ht!]
		\centering
		\includegraphics[width=0.5 \textwidth]{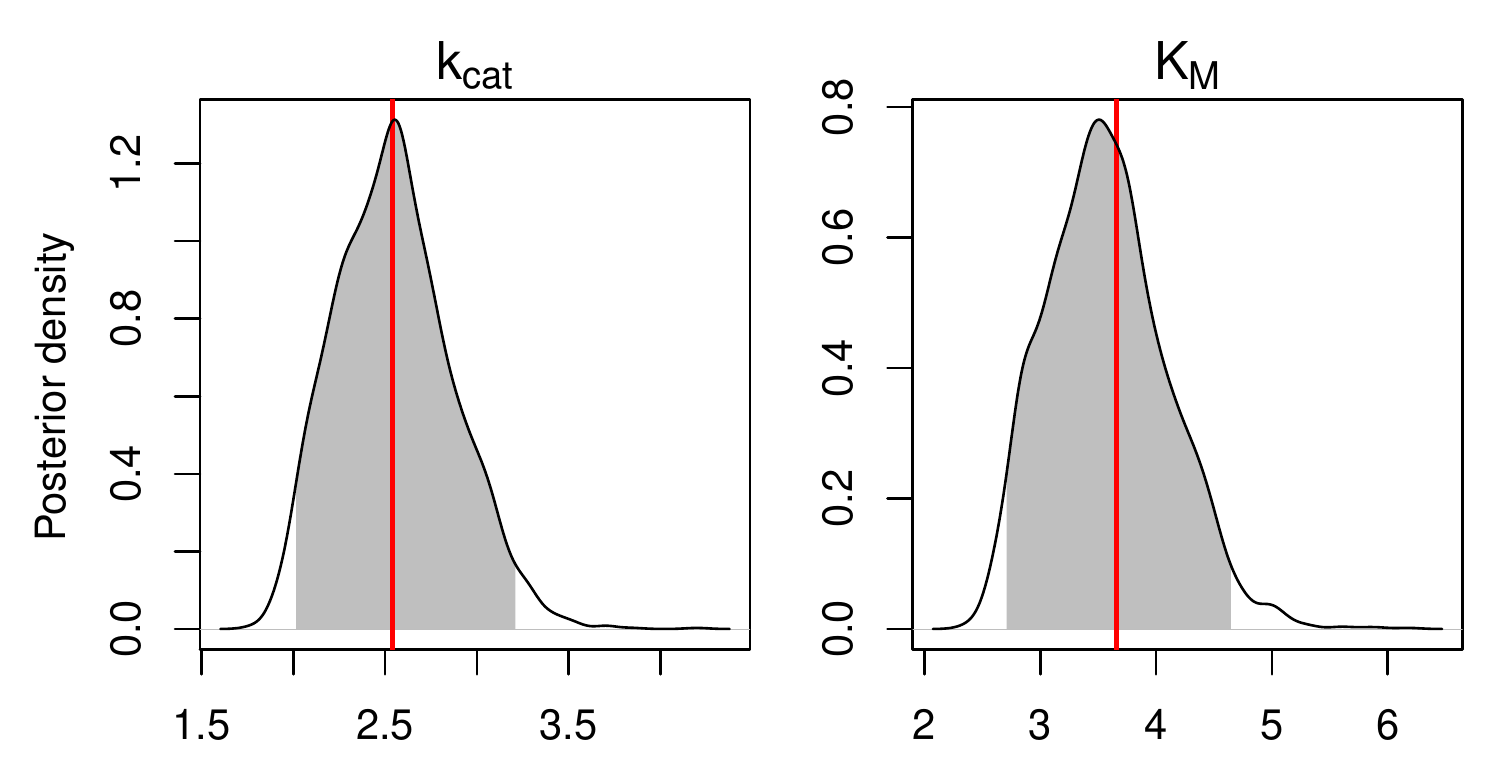}
		\caption{Bayesian posterior probability densities of $k_{cat}$ and $K_M$ for a sample dataset from the Michaelis-Menten model obtained by MAGI. The red vertical lines show the true parameter values used in the simulation. The shaded area represents the 95\% interval estimate of each parameter.}
		\label{fig:MM-vanilla-inferred-param-example}
	\end{figure}

	Fig \ref{fig:MM-vanilla-inferred-param-example} shows the posterior distribution of the kinetic parameters $k_{cat}$ and $K_M$ inferred by MAGI from one sample dataset (this dataset is presented in the left-most panel of Fig \ref{fig:MM-vanilla}). It is seen that the posterior distributions are well centered around the true values of $k_{cat}$ and $K_M$ (the red bars), where the 95\% interval represented by the shaded area in each panel provides the uncertainty quantification. Table \ref{tab:MM-vanilla-param} reports the estimation of $k_{cat}$ and $K_M$ across the 100 simulation datasets, where the MAGI estimate (which is the posterior mean) well recovers the true value with small SD. Here we focus on the estimation of $k_{cat}$ and $K_M$ for two reasons: (a) they are of the most scientific interest, and (b) they are identifiable from time-course data, whereas $k_1$, $k_{-1}$ and $k_2$ are not identifiable (meaning that there are multiple combinations of $k_1$, $k_{-1}$ and $k_2$ that can fit a time-course dataset equally well and yield the same $k_{cat}$ and $K_M$) \citep{johnson2013century}.
	
	\subsubsection{Larger reaction networks: a model for the \textit{lac} operon}
	
	The $lac$ operon has been studied for over a half-century, since the Nobel prize-winning work of \citet{jacob1961regulation} that established the concept of gene regulation. As a representative example of transcription negatively regulated by a repressor, the $lac$ operon only produces the enzymes necessary for the metabolism of lactose in the presence of an inducer (lactose); see \citet{muller2011lac} for a detailed historical account. Mechanistic models of varying complexity have been proposed to provide mathematical descriptions of the $lac$ operon, which range from modeling just a few key components \citep{novick1957enzyme,vilar2003modeling} to capturing the dynamics of a larger network, e.g., five to 10 system components \citep{yildirim2003feedback,stamatakis2009comparison,wilkinson2018stochastic}.
	
\begin{figure}[ht!]
		\includegraphics[width=\textwidth]{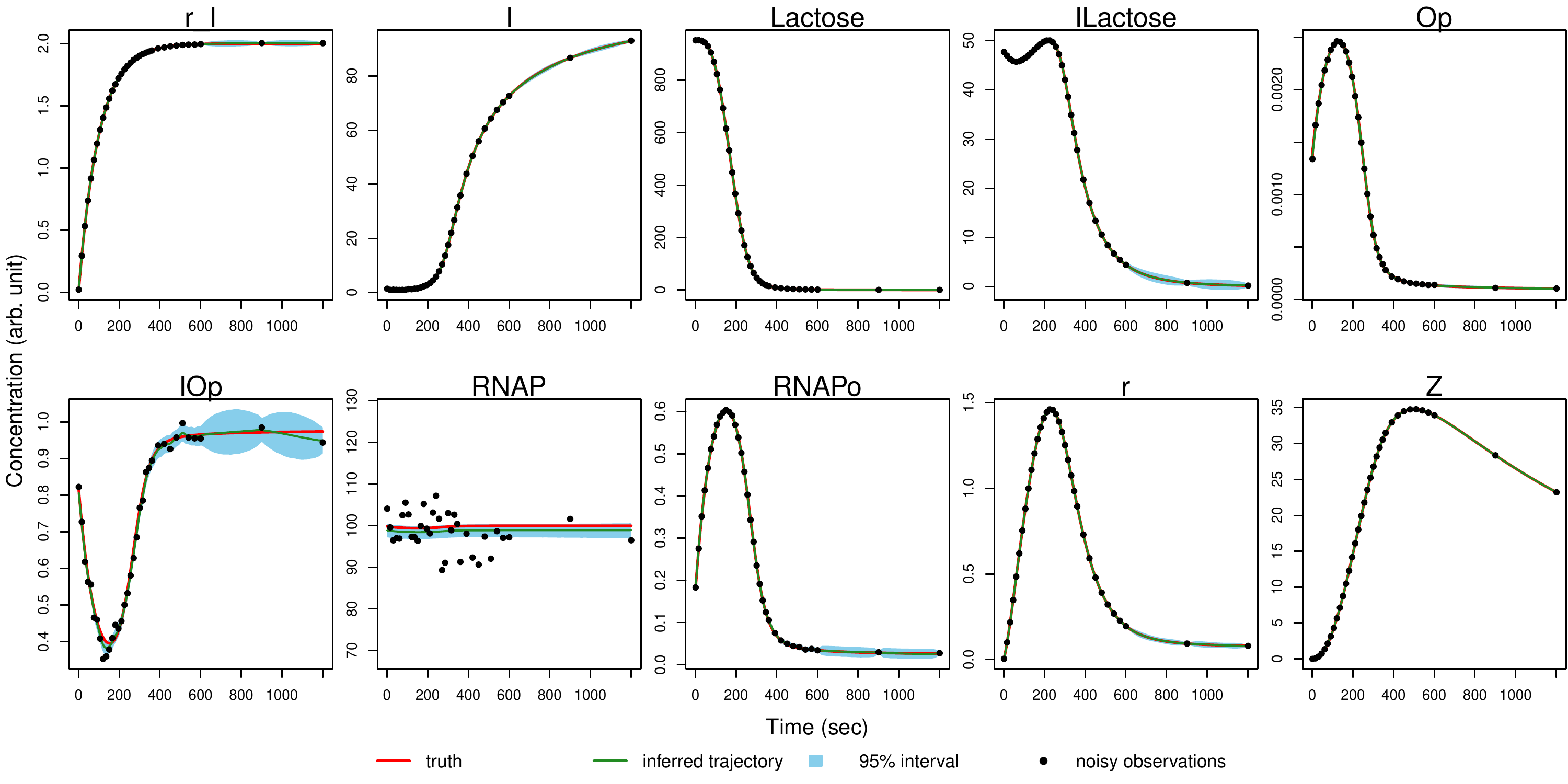}
		\caption{Inferred trajectories for a sample dataset from the $lac$ operon model. The black points are the measurements. The red curves are the true trajectories. The inferred trajectories are shown by the green curves, with the blue shaded areas representing 95\% intervals. Concentrations are given in arbitrary units, relative to $[i]$.}
		\label{fig:lac-infer-single-dataset}
	\end{figure}

	As an example of the application of MAGI to larger systems, we consider the inference of a 10-component $lac$ operon model \citep{wilkinson2018stochastic}, which would pose a serious challenge for any inference method due to its high dimensionality. A description of this model is as follows. In the absence of lactose, an inhibitor protein (denoted by $I$) binds to the $lac$ operon (denoted by $Op$), thereby blocking transcription of the operon by RNA polymerase (denoted by $RNAP$). In contrast when lactose is abundant, $I$ favors binding to lactose rather than $Op$, which enables $RNAP$ to act on $Op$ so that transcription proceeds. The mRNA transcripts from the operon (denoted by $r$) are translated into enzymes including $\beta$-galactosidase (denoted by $Z$) that then metabolizes lactose. To complete a mechanism that describes these steps, further let $r_I$ denote the inhibitor mRNA and $i$ its corresponding gene, $ILactose$ the inhibitor bound to lactose, $IOp$ the inhibitor bound to the $lac$ operon, and $RNAPo$ the $RNAP$-$Op$ complex. Even in this simplified form (e.g., degraded lactose and glucose mechanisms are not included), the model contains more than 10 system components and 16 rate parameters, with overall scheme given by:
	
	\begin{minipage}{.4\textwidth}
		\footnotesize{
			\begin{eqnarray*}
				i &{\stackrel{k_1}{\longrightarrow}}& i+r_I \\
				r_I &{\stackrel{k_2}{\longrightarrow}}& r_I + I \\
				I + Lactose  &\underset{k_{4}}{\stackrel{k_3}{\rightleftharpoons}}& ILactose \\
				I + Op &\underset{k_{6}}{\stackrel{k_5}{\rightleftharpoons}}& IOp \\
				Op + RNAP &\underset{k_{8}}{\stackrel{k_7}{\rightleftharpoons}}& RNAPo \\
				RNAPo &{\stackrel{k_9}{\longrightarrow}}& Op + RNAP + r \\
				r &{\stackrel{k_{10}}{\longrightarrow}}& r + Z \\
				Lactose + Z &{\stackrel{k_{11}}{\longrightarrow}}& Z \\
				r_I &{\stackrel{k_{12}}{\longrightarrow}}&  \emptyset\\
				I &{\stackrel{k_{13}}{\longrightarrow}}&  \emptyset\\
				ILactose &{\stackrel{k_{14}}{\longrightarrow}}&  Lactose \\
				r &{\stackrel{k_{15}}{\longrightarrow}}& \emptyset \\
				Z &{\stackrel{k_{16}}{\longrightarrow}}&  \emptyset\\
		\end{eqnarray*}}
	\end{minipage}
	\begin{minipage}{.6\textwidth}
		\tiny{
			\begin{eqnarray*}
				\frac{d[i]}{dt} &=& 0 \\
				\frac{d[r_I]}{dt} &=& k_{1}[i] - k_{12} [r_I] \\
				\frac{d[I]}{dt} &=& k_{2} [r_I] - k_{3} [I][Lactose] + k_{4} [ILactose] - k_{5} [I] [Op] + k_{6} [IOp] - k_{13} [I] \\
				\frac{d[Lactose]}{dt} &=& k_{4} [ILactose] - k_{3} [I] [Lactose] + k_{14} [ILactose] - k_{11} [Lactose] [Z] \\
				\frac{d[ILactose]}{dt} &=& k_{3} [I] [Lactose] - k_{4} [ILactose] - k_{14} [ILactose] \\
				\frac{d[Op]}{dt} &=& k_{6} [IOp] - k_{5} [I] [Op] - k_{7} [Op] [RNAP] + (k_{8} + k_{9}) [RNAPo] \\
				\frac{d[IOp]}{dt} &=& k_{5} [I] [Op] - k_{6} [IOp] \\
				\frac{d[RNAP]}{dt} &=& (k_{8} + k_{9}) [RNAPo] - k_{7} [Op] [RNAP] \\
				\frac{d[RNAPo]}{dt} &=& k_{7} [Op] [RNAP] - (k_{8}+k_{9}) [RNAPo] \\
				\frac{d[r]}{dt} &=& k_{9} [RNAPo] - k_{15} [r] \\
				\frac{d[Z]}{dt} &=& k_{10} [r] - k_{16} [Z] \\
			\end{eqnarray*}\bigskip}
	\end{minipage}%

	\noindent with $k_1, \ldots, k_{16}$ representing the parameters to be estimated, as adapted from ref \citenum{barbuti2020survey}, where $k_{12}, k_{13}, k_{15}, k_{16}$ are degradation rates. To simulate from this system, we follow the parameter values provided in ref \citenum{wilkinson2018stochastic} (shown in the second column of Table \ref{tab:lac-param}) and the initial conditions given in ref \citenum{barbuti2020survey}:  $[r_I] = 0$, $[I] = 50$, $[Op] = 1$, $[IOp] = 0$, $[RNAP] = 100$, $[RNAPo] = 0$, $[r] = 0$, $[Z] = 0$, $[Lactose] = 1000$, $[ILactose] = 0$. Each of these levels is specified relative to $[i]$, which is given a fixed unit concentration. These initial conditions are only used in simulating the data and are not known to MAGI. For illustration, suppose the system is observed every 15 seconds from $t=1$ sec to $t=361$ sec, and then every 30 seconds from $t=361$ sec to $t=601$ sec, and finally at $t=901$ sec and $t=1201$ sec, for a total of 35 observation times.  Measurement noise with known SD equal to 5\% of the minimum level of each component is added. A sample dataset generated with this setup is shown via the black points in Fig \ref{fig:lac-infer-single-dataset}. 
	
	MAGI's inferred trajectories for this dataset are shown via the green curves in Fig \ref{fig:lac-infer-single-dataset} together with blue 95\% intervals, where the true trajectories are plotted in red. There is larger uncertainty in the 95\% intervals when the observations are sparse. Overall, the inferred trajectory closely follows the true trajectory for each component. (To run MAGI on this example, we use an evenly-spaced discretization set $\bm{I}$ with interval 15 seconds that includes all of the observation times.)

	\begin{table}[ht]
		\centering
		\caption{Parameter estimates over 100 simulated datasets from the $lac$ operon model. The average parameter estimate is shown with its SD after the $\pm$ sign.}
		\begin{tabular}{ccc}
			\hline
			Parameter & Truth & MAGI estimate \\ 
			\hline
			\hline
			$k_1$ & 0.02 & 0.0199 $\pm$ 0.0000 \\
			$k_2$ & 1 & 0.0971 $\pm$ 0.0004\\
			$k_3$ & 0.005 & 0.0043 $\pm$ 0.0002 \\ 
			$k_4$ & 0.1 & 0.0857 $\pm$ 0.0031 \\ 
			$k_5$ & 1 & 0.9010 $\pm$ 0.0112 \\ 
			$k_6$ & 0.01 & 0.0090 $\pm$ 0.0001\\ 
			$k_7$ & 0.1 & 0.0958 $\pm$ 0.0012 \\ 
			$k_8$ & 0.01 & 0.0083 $\pm$ 0.0003\\ 
			$k_9$ & 0.03 & 0.0300 $\pm$ 0.0000  \\ 
			$k_{10}$ & 0.1 & 0.1000 $\pm$ 0.0000 \\
			$k_{11}$ & 0.001 & 0.0010 $\pm$ 0.0000 \\ 
			$k_{12}$ & 0.01 & 0.0100 $\pm$ 0.0000 \\ 
			$k_{13}$ & 0.002 & 0.0019 $\pm$ 0.0000 \\ 
			$k_{14}$ & 0.002 & 0.0019 $\pm$ 0.0000 \\ 
			$k_{15}$ & 0.01 & 0.0100 $\pm$ 0.0000 \\ 
			$k_{16}$ & 0.001 & 0.0010 $\pm$ 0.0000 \\
			\hline
		\end{tabular} \label{tab:lac-param}
	\end{table}
	
	\begin{figure}[ht!]
		\includegraphics[width=\textwidth]{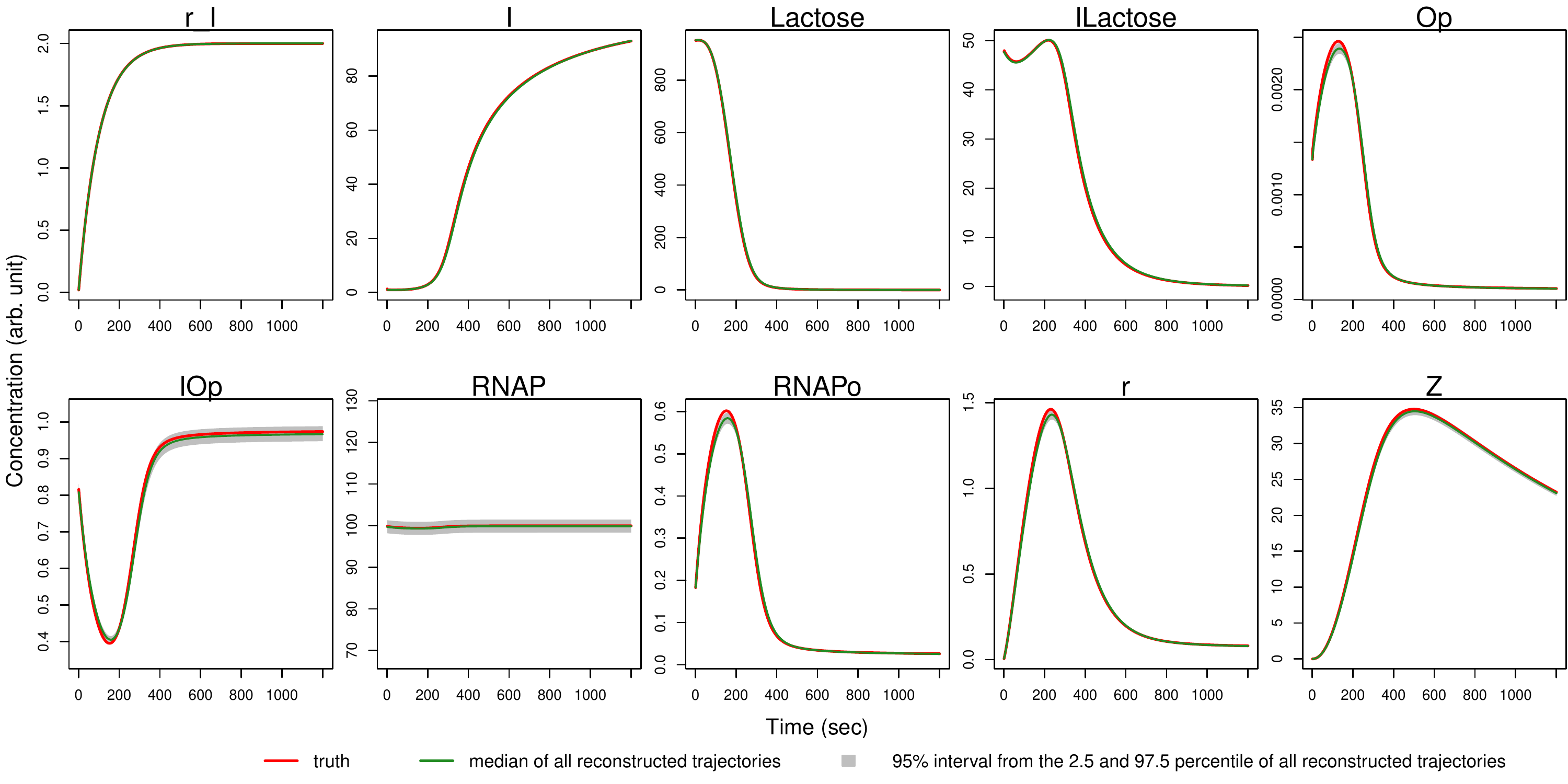}
		\caption{Reconstructed trajectories over 100 simulated datasets from the $lac$ operon model. The red curves are the true trajectories. The green curves are the median of all reconstructed trajectories, with the shaded areas showing the 95\% intervals given by the 2.5 and 97.5 percentile of all reconstructed trajectories. Concentrations are given in arbitrary units, relative to $[i]$.}
		\label{fig:lac-recon-repeated-dataset}
	\end{figure}
	
	Next, we assess the recovery of the system parameters $k_1, \ldots, k_{16}$. We generate 100 simulated datasets following the same setup, running MAGI on each and taking the posterior mean as the parameter estimate. The results across the 100 datasets are summarized in Table \ref{tab:lac-param}. The parameters are largely well recovered, with most parameters having small estimation error.  Some estimation error is apparent for the parameters $k_2$ to $k_8$; the system, however, might not be sensitive to all of these parameters as it is possible that different combinations of the parameters might give quite similar system trajectories. Thus, as a further check of the sensitivity of the system to parameter values, we calculated a \textit{reconstructed trajectory} for each dataset: taking MAGI's estimates of the parameters and the initial conditions, we used a numerical solver to reconstruct the system trajectories implied by those estimates. When the system is relatively insensitive to some parameters, different sets of parameter values and initial conditions will possibly lead to reconstructed trajectories that are close to each other and also close to the true trajectories. A graphical summary of the reconstructed trajectories across the 100 datasets is plotted in Fig \ref{fig:lac-recon-repeated-dataset}, which shows that (i) MAGI's parameter estimates recover the true trajectories (red curves) well and with only a small amount of uncertainty as evidenced by the narrow 95\% intervals (grey bands), and (ii) the system is indeed relatively insensitive to the parameter values as different combinations of parameter values give trajectories that are close to the truth. Note that this calculation of reconstructed trajectories is purely for additional verification; the MAGI method infers trajectories via posterior sampling (e.g., those plotted in Fig \ref{fig:lac-infer-single-dataset}) and does not use any numerical solver.
	
	\subsection{Model assessment with time-course data}\label{sec:model-comparison}
	In studying a chemical reaction or a biological process, it is often the case that different mechanisms are proposed. A natural question is to determine which mechanism best explains or fits the available time-course data. A proposed mechanism might be rejected on the basis of kinetic data or might be shown to be compatible with the mechanism \citep{huisgen1989kinetics}. Reciprocal plots are a simple tool to assess the compatibility of data with enzyme mechanisms \citep{dixon1953determination,cornish1974simple}. To use reciprocal plots, however, multiple experiments with varied substrate concentrations would be needed. This section demonstrates how MAGI can serve as a more powerful statistical approach to select the better model among mechanistic alternatives, on the basis of a \emph{single} (as opposed to multiple) time-course experiment.
	
	As an example, consider the competitive inhibition model of enzyme kinetics \citep{lineweaver1934determination}. Enzyme catalysis follows the Michaelis-Menten model, but is hindered by the presence of an inhibitor $I$ that binds reversibly with the enzyme $E$. When an enzyme molecule is bound to $I$, forming the enzyme-inhibitor complex $EI$, it can no longer bind to the substrate $S$ and can no longer facilitate the formation of product $P$. This competitive inhibition model of enzyme kinetics is denoted as \textit{Scheme A} below. Competitive inhibitors have wide applications \citep{alberts1980mevinolin,todd1989competitive,otton1984competitive,cushman1977design} and their identification often plays a key role in the drug discovery process \citep{holdgate2018mechanistic}. In the absence of inhibitor $I$, or if a proposed inhibitor is ineffectual, \textit{Scheme A} reduces to the Michaelis-Menten model in Eq.\eqref{eq:mm}, which we denote as \textit{Scheme B} in what follows. Suppose we have data from a single time-course experiment. Does the progress curve suggest the presence of an effectual competitive inhibitor, i.e., does Scheme A provide better agreement to the data than Scheme B when the two models are compared? 
	
	\paragraph{Scheme A} Michaelis-Menten model with competitive inhibitor 
	
	\begin{minipage}{.35 \textwidth}
		\begin{eqnarray*}
			&E + S \underset{k_{-1}}{\stackrel{k_1}{\rightleftharpoons}} ES \, {\stackrel{k_2}{\longrightarrow}} \, E + P \\
			&E + I \underset{k_{-3}}{\stackrel{k_3}{\rightleftharpoons}} EI 
		\end{eqnarray*}
	\end{minipage}%
	\begin{minipage}{.65 \textwidth}
		\footnotesize{
			\begin{eqnarray*}
				\frac{d[E]}{dt} &=& -k_1 [E][S] + (k_{-1}+k_2) [ES] - k_3 [I][E] + k_{-3} [EI]\\
				\frac{d[S]}{dt} &=& -k_1 [E][S] + k_{-1} [ES]\\
				\frac{d[ES]}{dt} &=&  k_1 [E][S] - (k_{-1}+k_2) [ES] \\
				\frac{d[P]}{dt} &=&  k_2 [ES]\\
				\frac{d[EI]}{dt} &=&  k_3 [I][E] - k_{-3}[EI] \\
				\frac{d[I]}{dt} &=&  -k_3 [I][E] + k_{-3}[EI] \\
		\end{eqnarray*}}
	\end{minipage}%
	
	\paragraph{Scheme B} Original Michaelis-Menten model in Eq.\eqref{eq:mm}.
	
	\subsubsection{Assessing models with MAGI}
	The statistical inference provided by MAGI leads to a natural method to assess a proposed model. We divide the observed time-course data into two parts: a training set (for example, data in the first time period) and a test set (for example, data in the second time period); then we apply MAGI only on the training set to estimate the model parameters and infer the system trajectories; finally we evaluate the model by comparing the model prediction on the test set. This idea of assessing a model based on its prediction of a ``future'' time period has been previously used in the context of numerical integration methods \citep{hasdemir2015validation}. Here the MAGI method offers a way to completely bypass numerical integrations for model assessment. 
	
	Operation-wise, we divide the observation time points $\bm\tau$ into the training part $\bm{\tau}^{\text{train}}$ and the testing part $\bm{\tau}^{\text{test}}$ in chronological order such that $\bm\tau = \bm{\tau}^{\text{train}} \cup \bm{\tau}^{\text{test}}$ and $\max\{ \bm{\tau}^{\text{train}} \} < \min\{ \bm{\tau}^{\text{test}} \}$. We choose the discretization points $\bm I$ in equation \eqref{eqn: dicretization} to cover the entire set of observation time points (i.e., $\bm\tau \subset \bm I$). Then inference of $\bm{x}(\bm{I})$ for both the training period (namely the fitting) and the testing period (namely the prediction) can be achieved in one integrated step through equation \eqref{eq:main} but using only the training data (i.e., conditioning only on the training data): 
	\begin{align*}
	& p(\bm{\theta}, \bm{x}(\bm{I}) |  W_{\bm{I}} = 0, \bm{y}(\bm{\tau}^{\text{train}}) ) \\
	&\propto 
	\pi(\bm{\theta}) \, p(\bm{X}(\bm{I}) = \bm{x}(\bm{I}) )\,  p(  \bm{y}(\bm{\tau}^{\text{train}}) | \bm{x}(\bm{I})) \, p(\bm{\dot X}(\bm{I}) = \mathbf{f}(\bm{x}(\bm{I}), \bm{\theta}, \bm{I}) | \bm{x}(\bm{I}) ).
	\end{align*}
	The prediction at $\bm{\tau}^{\text{test}}$, namely $\hat{\bm{x}}(\bm{\tau}^{\text{test}}) = \mathbb{E}({\bm{x}}(\bm{\tau}^{\text{test}}) | W_{\bm{I}} = 0, \bm{y}(\bm{\tau}^{\text{train}})  )$, is the corresponding posterior mean of $\bm{x}(\bm{\tau}^{\text{test}})$. Note that throughout the MAGI computation, no numerical integration is ever needed. The assessment of how compatible a proposed model is to the observed data can be quantified by measuring the discrepancy between the model prediction and the test data, for example, by using the sum of squared errors (SSE) of prediction: 
	\begin{equation}\label{eq:mse}
	\text{SSE} = \| \hat{\bm{x}}(\bm{\tau}^{\text{test}}) - {\bm{y}}(\bm{\tau}^{\text{test}}) \|^2_2.
	\end{equation}
	When two or more models are being compared, we will compute the SSE of prediction from each model. A smaller SSE indicates a better compatibility between the model and the data.

	\subsubsection{Comparing two models: Scheme A vs Scheme B}
	
\begin{figure}[ht!]
		\includegraphics[width=\textwidth]{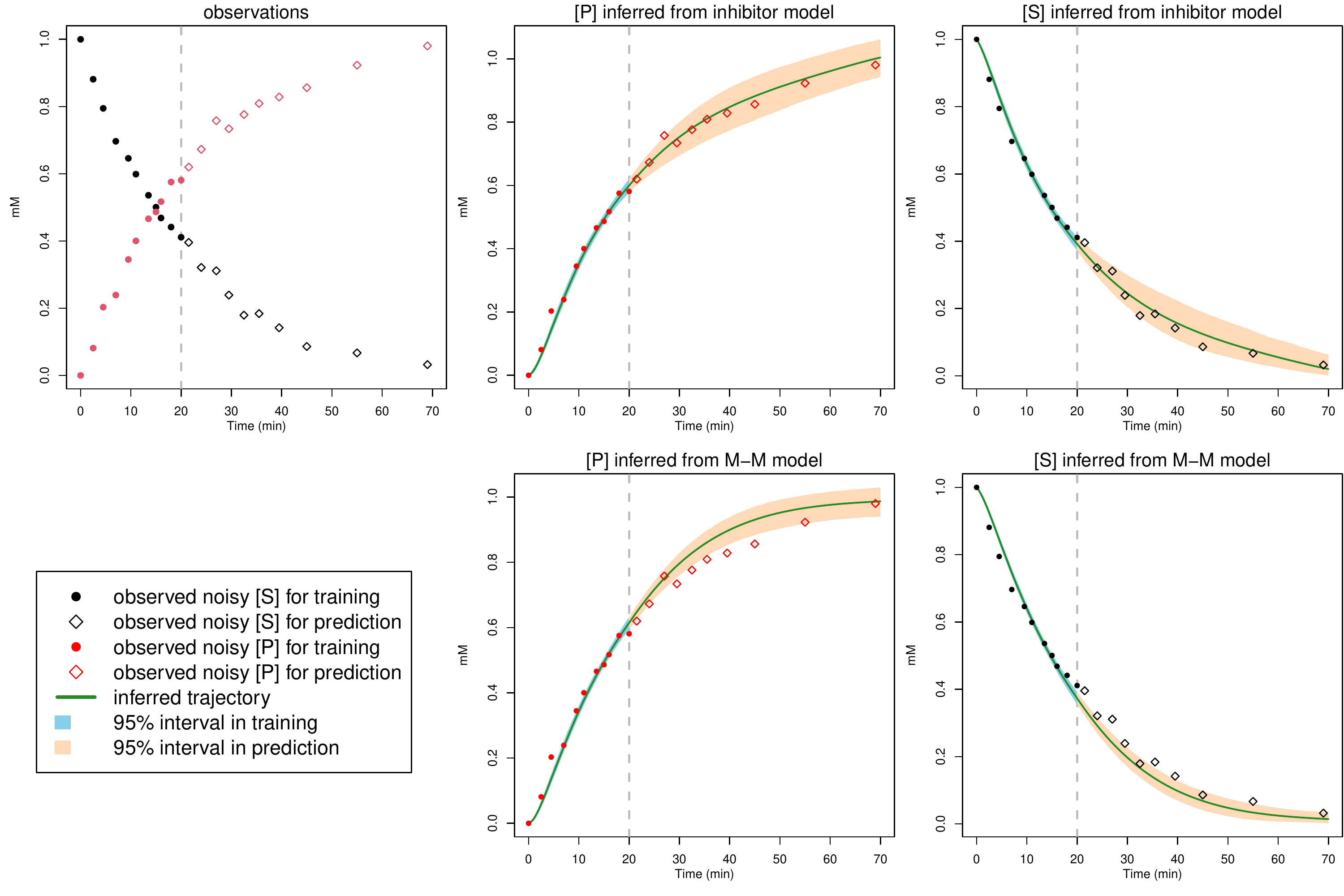}
		\caption{Michaelis-Menten model comparison for a sample dataset simulated with inhibitor. The observations are divided into two parts: the training data are marked with solid dots, and the testing data are marked with hollow dots (the vertical grey line separates the training period and the prediction period). \textbf{Top-left panel}: a sample dataset of observations. \textbf{Top-center and top-right panels}: inference results from Scheme A (with inhibitor). \textbf{Bottom panels}: inference results from Scheme B (original Michaelis-Menten model). The green curve is the inferred trajectory. The blue shaded area indicates the 95\% estimation interval for the training period, while the yellow area indicates 95\% prediction interval for the test period. Both Scheme A and Scheme B can fit the data well in the training period, but the predicted $S$ and $P$ from Scheme A is closer to the observations in the test period.}
		\label{fig:MM-model-comparison}
	\end{figure}

	For illustration, we compare Scheme A with Scheme B with simulated data: we simulate data from Scheme A (the Michaelis-Menten model with competitive inhibitor) and check if MAGI can correctly identify Scheme A as the right model (as opposed to Scheme B). The true parameter values used in the simulation are $k_1 = 0.9$ (min$\cdot$mM)$^{-1}$, $k_{-1} = 0.75$ (min)$^{-1}$, $k_2 = 2.54$ (min)$^{-1}$, $k_3 = 1$ (min$\cdot$mM)$^{-1}$, $k_{-3} = 0.5$ (min)$^{-1}$; and the true initial conditions are $P = 0$ mM, $S = 1$ mM, $E = 0.1$ mM, $I = 0.08$ mM. Only $P$ and $S$ are observed at 20 sparse time points listed in the legend of Fig \ref{fig:MM-vanilla}. The measurement noise of $P$ and $S$ is taken to be additive Gaussian noise with known SD equal to 0.02 mM. A visual illustration of the simulated observation data is presented in Fig \ref{fig:MM-model-comparison} (the upper left panel). We assume the initial conditions of $S$ and $E$ are known to MAGI without noise, along with the usual initial experimental settings that $P = ES = EI = 0$ (the initial condition of $I$ is unknown). 
	
	When applying MAGI, we took the discretization points $\bm I$ in equation \eqref{eqn: dicretization} to be evenly spaced from $t=0$ to $t=70$ (minute) at 0.5 minute intervals, i.e., $\bm{I} = \{0, 0.5, 1,\ldots, 70\}$ (minutes). To assessing a given model (i.e., Scheme A or Scheme B), we divide the time into two periods. The first time period (0 to 20 minutes) was used as the training period to fit a given model with MAGI, and then the model prediction of the system trajectories in the second time period (20 to 70 minutes), i.e., the test period, was compared against the data in the test period. The model with smaller prediction error (measured by SSE) is considered to be more compatible with the observations.  
	
	Fig \ref{fig:MM-model-comparison} shows the result obtained by MAGI for comparing Scheme A to Scheme B on a sample dataset. The two panels on the top (center and right) show the inference result based on this dataset under Scheme A (inhibitor model), while the two panels on the bottom show the inference result under Scheme B (original Michaelis-Menten). The inferred system trajectories are given by the green curves. The blue shaded area corresponds to the 95\% estimation interval within the training period (0 to 20 minutes). The yellow shaded area corresponds to the 95\% prediction interval in the test period (20 to 70 minutes). 
	
	Comparing the top two panels to the bottom two panels, we can visually see that Scheme A provides a better prediction than Scheme B: Scheme B underestimates the $S$ component (under Scheme B the substrate is consumed faster than seen in the real data) and overestimates the $P$ component (under Scheme B the product is generated faster than seen in the real data); in contrast, Scheme A gives prediction seen to be closer to the real data. The SSE under Scheme B is 0.035, which is more than 3 times as large as the SSE of 0.009 under Scheme A, quantitatively agreeing with the visual impression. As the data are generated from Scheme A, for this dataset MAGI is seen to correctly select Scheme A over Scheme B.
	
	It is also noticeable from Fig \ref{fig:MM-model-comparison} that while the two models appear to give similar quality of fit to the data within the training period, it is the prediction for the (future) test period that separates the two models, which highlights the importance of comparing models based on their prediction. For further assessment, we repeat this procedure 100 times, i.e., generate the datasets from Scheme A 100 times independently and apply MAGI to select Scheme A vs Scheme B on each dataset. For 100 times out of 100, MAGI correctly selected Scheme A over Scheme B. This result demonstrates the effectiveness of MAGI for (ODE) model comparison.

	\clearpage
	
	\section{Conclusion}
	
	This article considered the analysis of time-course data using the MAGI method. To illustrate the estimation of differential equation models from data, we presented three examples from chemistry and biology: the repressilator, the Michaelis-Menten model, and a gene regulation network with more than 10 system components.  The models and time-course setups pose various challenges for fitting to data: in the repressilator, half of the system components are entirely unobserved; for Michaelis-Menten, uncertainty quantification for the key $k_{cat}$ and $K_M$ parameters, using sparse and noisy measurements; in the \textit{lac} operon model, handling a larger number of system components and parameters. MAGI was shown to be a capable method for inferring the parameters (including interval estimates) and underlying system trajectories (including for unobserved components), without the need for any numerical integration. We then considered the problem of model assessment, to decipher whether one mechanism supports the observed data better than another. After dividing a single time-course dataset into training and testing parts, MAGI can simultaneously fit the data and generate predictions under a given model, again without any numerical integration. By comparing the prediction of each model against the test data, the best model can be identified. The approach was shown to be highly effective in the context of Michaelis-Menten kinetics with a competitive inhibitor.
	
	Overall, we believe MAGI can be a widely applicable method for analyses involving ODE models. The Supplement provides step-by-step code examples for each of the models discussed in this article. The MAGI software package for R is available from CRAN (\url{https://cran.r-project.org/package=magi}), and packages are also available for MATLAB and Python (\url{https://github.com/wongswk/magi}). A detailed usage guide for the software is provided in \citet{wong2022magi}.
	
	In the study of chemical and biological processes, ODE models are ubiquitous for describing the behavior of each system component at the aggregate level. In contrast, when interest lies in the behavior of single molecules, then models such as stochastic differential equations (SDEs) can provide a more realistic depiction of intrinsic noise \citep{kou2004generalized}. It would be interesting to explore how MAGI could be extended to the SDE setting in future research.

	\section{Appendix: Derivation of the MAGI method}
	This appendix describes the derivation of equation (\ref{eq:main}) and the detailed expressions of each term on the right-hand-side of equation (\ref{eq:main}).
	
	We first introduce the notations needed:
	$\|\bm v\|_A^2 = \bm{v}^\intercal A \bm{v}$, $|\bm{I}|$ is the cardinality of $\bm{I}$, $\mathbf{f}_{d, \bm{I}}^{\bm{x}, \bm{\theta}}$ is the $d$-th component of $\mathbf{f}(\bm{x}(\bm{I}), \bm{\theta}, \bm{I})$ evaluated at the time points $\bm{I}$, $N_d$ is the number of observations for component $d$, $\sigma_d$ is the noise level for component $d$, $\mu_d(t)$ is the mean function of the GP for component $d$, $\mathcal{K}_d(s, t)$ is the covariance function of the GP for component $d$, and the $|\bm{I}| \times |\bm{I}|$ matrices $C_d$, $m_d$ and $\Psi_d$ that govern the covariance and conditional covariances of the GP evaluated at $\bm{I}$ are given as follows for each component $d$:
	\begin{equation*}
	\begin{cases}
	C_d &= \mathcal{K}_d(\bm{{I}}, \bm{I}) \\
	m_d &= \mathcal{'K}_d(\bm{I}, \bm{I}) \mathcal{K}_d(\bm{I}, \bm{I})^{-1} \\
	\Psi_d &= \mathcal{K''}_d(\bm{I}, \bm{I}) - \mathcal{'K}_d(\bm{I}, \bm{I}) \mathcal{K}_d(\bm{I}, \bm{I})^{-1} \mathcal{K'}_d(\bm{I}, \bm{I})
	\end{cases}
	\end{equation*}
	where $\mathcal{'K}_d = \frac{\partial}{\partial s} \mathcal{K}_d(s, t)$, $\mathcal{K'}_d = \frac{\partial}{\partial t} \mathcal{K}_d(s, t)$, and $\mathcal{K''}_d = \frac{\partial^2}{\partial s\partial t} \mathcal{K}_d(s, t)$.  In our examples we use the Matern kernel: $\mathcal{K}_d(s, t) = \phi_1\frac{2^{1-\nu}}{\Gamma(\nu)}\left(\sqrt{2\nu}\frac{l}{\phi_2}\right)^\nu B_\nu\left(\sqrt{2\nu}\frac{l}{\phi_2}\right)$ where $l = |s-t|$, $\Gamma$ is the Gamma function and $B_\nu$ is the modified Bessel function of the second kind, and $\nu = 2.01$ is the degree of freedom. The values of $\phi_1$ and $\phi_2$ are tuned for each component by fitting to the data.
	
	To obtain the four terms on the right-hand-side of equation (\ref{eq:main}), we begin by applying Bayes' rule on the joint distribution of $\bm{\theta}$ and $\bm{x}(\bm{I})$, given the manifold constraint $W_{\bm{I}} = 0$ and the noisy observations $\bm{y}(\bm \tau))$,
	\begin{equation*}
	p(\bm{\theta}, \bm{x}(\bm{I}) |  W_{\bm{I}} = 0, \bm{Y}(\bm\tau) = \bm{y}(\bm \tau)) \propto  p({\bm{\Theta}} = \bm{\theta}, \bm{X}(\bm{I}) = \bm{x}(\bm{I}), W_{\bm{I}}=0, \bm{Y}(\bm\tau) = \bm{y}(\bm \tau)),
	\end{equation*}
	which then factorizes into
	\begin{align*}
	& p({\bm{\Theta}} = \bm{\theta}, \bm{X}(\bm{I}) = \bm{x}(\bm{I}), W_{\bm{I}}=0, \bm{Y}(\bm\tau) = \bm{y}(\bm \tau)) \\
	&= \pi(\bm{\theta}) \; \times \; p(\bm{X}(\bm{I}) = \bm{x({I})}  | {\bm{\Theta}} = \bm{\theta} ) \;
	\times \; p(\bm{Y}(\bm\tau) = \bm{y}(\bm\tau) | \bm{X}(\bm I) = \bm{x}(\bm I), {\bm{\Theta}} = \bm{\theta}) \\
	&\qquad \times p(W_{\bm{I}}=0 | \bm{Y}(\bm\tau) = \bm{y}(\bm\tau), \bm{X}(\bm I) = \bm{x}(\bm I), {\bm{\Theta}} = \bm{\theta}).
	\end{align*}
	Since the GP prior on $\bm{X}$ is independent of $\bm{\Theta}$, we have $p(\bm{X}(\bm{I}) = \bm{x}(\bm{I}) | {\bm{\Theta}} = \bm{\theta} ) = p(\bm{X}(\bm{I}) = \bm{x}(\bm{I}) )$. Likewise, the noisy observations do not depend on $\bm{\Theta}$, so $p(\bm{Y}(\bm\tau) = \bm{y}(\bm\tau) | \bm{X}(\bm I) = \bm{x}(\bm I), {\bm{\Theta}} = \bm{\theta})  = p(\bm{y}(\bm\tau) | \bm{x}(\bm I))$.
	Lastly, 
	\begin{align*}
	&  p(W_{\bm{I}}=0 | \bm{Y}(\bm\tau) = \bm{y}(\bm\tau), \bm{X}(\bm I) = \bm{x}(\bm I), {\bm{\Theta}} = \bm{\theta}) \\
	&= p(\bm{\dot X}(\bm{I}) = \mathbf{f}(\bm{x}(\bm{I}), \bm{\theta}, \bm{I}) | \bm{Y}(\bm\tau) = \bm{y}(\bm\tau), \bm{X}(\bm I) = \bm{x}(\bm I), {\bm{\Theta}} = \bm{\theta} ) \\
	&=  p(\bm{\dot X}(\bm{I}) = \mathbf{f}(\bm{x}(\bm{I}), \bm{\theta}, \bm{I}) | \bm{x}(\bm{I}) )
	\end{align*}
	by first substituting the definition of $W_{\bm{I}}=0$ and then noting that $\bm{\dot X}$ conditioning on $\bm{Y}(\bm\tau)$, $\bm{\Theta}$ and $\bm{X(\bm I)}$ only depends on $\bm{X(\bm I)}$.
	Therefore, the four terms on the right-hand-side of equation (\ref{eq:main}) can be expressed as
	\begin{enumerate}
		\item $\pi(\bm{\theta})$ is the prior density of the parameters; 
		\item  $p(\bm{X}(\bm{I})=\bm{x}(\bm{I})) \propto \exp\Big\{-\frac{1}{2}\sum_{d=1}^D \Big[
		|\bm{I}|\log(2\pi) + \log{\det(C_d)} + \left\|x_d(\bm{I}) - \mu_d(\bm{I})\right\|_{C_d^{-1}}^2 \Big] \Big\} $;
		\item  $ p(  \bm{y}(\bm \tau) | \bm{x}(\bm{I})) \propto \exp\Big\{-\frac{1}{2}\sum_{d=1}^D \Big[ N_d \log(2\pi \sigma_d^{2}) +  \left\|x_d(\bm{\tau}_d) - y_d(\bm{\tau}_d)\right\|_{\sigma_d^{-2} }^2
		\Big] \Big\} $;
		\item  $p(\bm{\dot X}(\bm{I}) = \mathbf{f}(\bm{x}(\bm{I}), \bm{\theta}, \bm{I}) | \bm{x}(\bm{I}) )  \propto $ \\
		$ ~~~ \exp\Big\{-\frac{1}{2}\sum_{d=1}^D \Big[ |\bm{I}|\log(2\pi) + \log{\det(\Psi_d)} + \left\|\mathbf{f}_{d, \bm{I}}^{\bm{x}, \bm{\theta}} - \dot{\mu}_d(\bm{I}) - m_d \{x_d(\bm{I}) - \mu_d(\bm{I})\}\right\|_{\Psi_d^{-1}}^2
		\Big] \Big\} $.
	\end{enumerate}

	\begin{acknowledgement}
		The authors thank Professor Sunney Xie for many groundbreaking contributions in single-molecule and single-cell studies and for fruitful collaborations. We are grateful for the inspirations and encouragement we received from Professor Xie over the years. S.W.K.W.~was partially supported by Discovery Grant RGPIN-2019-04771 from the Natural Sciences and Engineering Research Council of Canada.
	\end{acknowledgement}
	
	\begin{suppinfo}
		The supplement included with this article provides complete step-by-step code in R for running MAGI on each of the examples in the paper.
	\end{suppinfo}

\bibliographystyle{achemso}
\bibliography{physchem}

\clearpage
\newgeometry{margin=1in}

\makeatletter
\def\maxwidth{\ifdim\Gin@nat@width>\linewidth\linewidth\else\Gin@nat@width\fi}
\def\maxheight{\ifdim\Gin@nat@height>\textheight\textheight\else\Gin@nat@height\fi}
\makeatother
\setkeys{Gin}{width=\maxwidth,height=\maxheight,keepaspectratio}
\makeatletter
\def\fps@figure{htbp}
\makeatother
\setlength{\emergencystretch}{3em} 
\providecommand{\tightlist}{%
	\setlength{\itemsep}{0pt}\setlength{\parskip}{0pt}}
\setcounter{secnumdepth}{-\maxdimen} 
\setlength{\parindent}{0pt}
\urlstyle{same}

\renewcommand{\thefigure}{S\arabic{figure}}
\setcounter{figure}{0}

\section*{Supporting Information for ``Estimating and Assessing Differential Equation Models with Time-Course Data''}
\footnotesize{

\input{JPC-supplement.tex}
}
\end{document}

%% file: JPC-supplement.tex
This supporting information file provides complete step-by-step code in R for running MAGI on each of the examples in the main text. For further information on the software package, we refer the reader to the usage guide for MAGI at \url{https://arxiv.org/abs/2203.06066}.

\hypertarget{setup}{%
\subsection{Setup}\label{setup}}

Ensure that the \texttt{magi} R package is installed and loaded:

\begin{Shaded}
\begin{Highlighting}[]
\FunctionTok{install.packages}\NormalTok{(}\StringTok{"magi"}\NormalTok{)}
\end{Highlighting}
\end{Shaded}

\begin{Shaded}
\begin{Highlighting}[]
\FunctionTok{library}\NormalTok{(magi)}
\end{Highlighting}
\end{Shaded}

\hypertarget{repressilator-gene-regulation-network}{%
\subsection{Repressilator gene regulation network}\label{repressilator-gene-regulation-network}}

We begin by defining a function that codes the log-transformed ODEs:

\begin{Shaded}
\begin{Highlighting}[]
\NormalTok{RrepressilatorGeneRegulationLogODE }\OtherTok{\textless{}{-}} \ControlFlowTok{function}\NormalTok{(theta, x, tvec) \{}
\NormalTok{  m\_laci }\OtherTok{=} \FunctionTok{exp}\NormalTok{(x[,}\DecValTok{1}\NormalTok{])}
\NormalTok{  m\_tetr }\OtherTok{=} \FunctionTok{exp}\NormalTok{(x[,}\DecValTok{2}\NormalTok{])}
\NormalTok{  m\_ci }\OtherTok{=} \FunctionTok{exp}\NormalTok{(x[,}\DecValTok{3}\NormalTok{])}
\NormalTok{  p\_laci }\OtherTok{=} \FunctionTok{exp}\NormalTok{(x[,}\DecValTok{4}\NormalTok{])}
\NormalTok{  p\_tetr }\OtherTok{=} \FunctionTok{exp}\NormalTok{(x[,}\DecValTok{5}\NormalTok{])}
\NormalTok{  p\_ci }\OtherTok{=} \FunctionTok{exp}\NormalTok{(x[,}\DecValTok{6}\NormalTok{])}

\NormalTok{  alpha0 }\OtherTok{=}\NormalTok{ theta[}\DecValTok{1}\NormalTok{]}
\NormalTok{  alpha }\OtherTok{=}\NormalTok{ theta[}\DecValTok{2}\NormalTok{]}
\NormalTok{  n }\OtherTok{=}\NormalTok{ theta[}\DecValTok{3}\NormalTok{]}
\NormalTok{  beta }\OtherTok{=}\NormalTok{ theta[}\DecValTok{4}\NormalTok{]}

\NormalTok{  resultdt }\OtherTok{\textless{}{-}}  \FunctionTok{array}\NormalTok{(}\DecValTok{0}\NormalTok{, }\FunctionTok{c}\NormalTok{(}\FunctionTok{nrow}\NormalTok{(x),}\FunctionTok{ncol}\NormalTok{(x)))}

\NormalTok{  resultdt[,}\DecValTok{1}\NormalTok{] }\OtherTok{=} \SpecialCharTok{{-}}\DecValTok{1} \SpecialCharTok{+}\NormalTok{ (alpha }\SpecialCharTok{/}\NormalTok{ (}\DecValTok{1} \SpecialCharTok{+}\NormalTok{ p\_ci}\SpecialCharTok{\^{}}\NormalTok{n) }\SpecialCharTok{+}\NormalTok{ alpha0) }\SpecialCharTok{/}\NormalTok{ m\_laci}
\NormalTok{  resultdt[,}\DecValTok{2}\NormalTok{] }\OtherTok{=} \SpecialCharTok{{-}}\DecValTok{1} \SpecialCharTok{+}\NormalTok{ (alpha }\SpecialCharTok{/}\NormalTok{ (}\DecValTok{1} \SpecialCharTok{+}\NormalTok{ p\_laci}\SpecialCharTok{\^{}}\NormalTok{n) }\SpecialCharTok{+}\NormalTok{ alpha0) }\SpecialCharTok{/}\NormalTok{ m\_tetr}
\NormalTok{  resultdt[,}\DecValTok{3}\NormalTok{] }\OtherTok{=} \SpecialCharTok{{-}}\DecValTok{1} \SpecialCharTok{+}\NormalTok{ (alpha }\SpecialCharTok{/}\NormalTok{ (}\DecValTok{1} \SpecialCharTok{+}\NormalTok{ p\_tetr}\SpecialCharTok{\^{}}\NormalTok{n) }\SpecialCharTok{+}\NormalTok{ alpha0) }\SpecialCharTok{/}\NormalTok{ m\_ci}
\NormalTok{  resultdt[,}\DecValTok{4}\NormalTok{] }\OtherTok{=}\NormalTok{ (}\SpecialCharTok{{-}}\NormalTok{beta}\SpecialCharTok{*}\NormalTok{(}\DecValTok{1} \SpecialCharTok{{-}}\NormalTok{ m\_laci }\SpecialCharTok{/}\NormalTok{ p\_laci))}
\NormalTok{  resultdt[,}\DecValTok{5}\NormalTok{] }\OtherTok{=}\NormalTok{ (}\SpecialCharTok{{-}}\NormalTok{beta}\SpecialCharTok{*}\NormalTok{(}\DecValTok{1} \SpecialCharTok{{-}}\NormalTok{ m\_tetr }\SpecialCharTok{/}\NormalTok{ p\_tetr)) }
\NormalTok{  resultdt[,}\DecValTok{6}\NormalTok{] }\OtherTok{=}\NormalTok{ (}\SpecialCharTok{{-}}\NormalTok{beta}\SpecialCharTok{*}\NormalTok{(}\DecValTok{1} \SpecialCharTok{{-}}\NormalTok{ m\_ci }\SpecialCharTok{/}\NormalTok{ p\_ci))}

\NormalTok{  resultdt}
\NormalTok{\}}
\end{Highlighting}
\end{Shaded}

Next, we provide the gradients of the ODEs with respect to the system components \(X\) and the parameters \(\theta\).

\begin{Shaded}
\begin{Highlighting}[]
\NormalTok{RrepressilatorGeneRegulationLogDx }\OtherTok{\textless{}{-}} \ControlFlowTok{function}\NormalTok{(theta, x, tvec) \{}
\NormalTok{  resultDx }\OtherTok{\textless{}{-}} \FunctionTok{array}\NormalTok{(}\DecValTok{0}\NormalTok{, }\FunctionTok{c}\NormalTok{(}\FunctionTok{nrow}\NormalTok{(x), }\FunctionTok{ncol}\NormalTok{(x), }\FunctionTok{ncol}\NormalTok{(x)))}

\NormalTok{  tm\_laci }\OtherTok{=}\NormalTok{ x[,}\DecValTok{1}\NormalTok{]}
\NormalTok{  tm\_tetr }\OtherTok{=}\NormalTok{ x[,}\DecValTok{2}\NormalTok{]}
\NormalTok{  tm\_ci }\OtherTok{=}\NormalTok{ x[,}\DecValTok{3}\NormalTok{]}
\NormalTok{  tp\_laci }\OtherTok{=}\NormalTok{ x[,}\DecValTok{4}\NormalTok{]}
\NormalTok{  tp\_tetr }\OtherTok{=}\NormalTok{ x[,}\DecValTok{5}\NormalTok{]}
\NormalTok{  tp\_ci }\OtherTok{=}\NormalTok{ x[,}\DecValTok{6}\NormalTok{]}

\NormalTok{  alpha0 }\OtherTok{=}\NormalTok{ theta[}\DecValTok{1}\NormalTok{]}
\NormalTok{  alpha }\OtherTok{=}\NormalTok{ theta[}\DecValTok{2}\NormalTok{]}
\NormalTok{  n }\OtherTok{=}\NormalTok{ theta[}\DecValTok{3}\NormalTok{]}
\NormalTok{  beta }\OtherTok{=}\NormalTok{ theta[}\DecValTok{4}\NormalTok{]}

\NormalTok{  resultDx[,}\DecValTok{1}\NormalTok{,}\DecValTok{1}\NormalTok{] }\OtherTok{=} \SpecialCharTok{{-}}\NormalTok{(alpha }\SpecialCharTok{/}\NormalTok{ (}\DecValTok{1} \SpecialCharTok{+} \FunctionTok{exp}\NormalTok{(n }\SpecialCharTok{*}\NormalTok{ tp\_ci)) }\SpecialCharTok{+}\NormalTok{ alpha0) }\SpecialCharTok{*} \FunctionTok{exp}\NormalTok{(}\SpecialCharTok{{-}}\NormalTok{tm\_laci)}
\NormalTok{  resultDx[,}\DecValTok{6}\NormalTok{,}\DecValTok{1}\NormalTok{] }\OtherTok{=}\NormalTok{ alpha }\SpecialCharTok{*} \FunctionTok{exp}\NormalTok{(}\SpecialCharTok{{-}}\NormalTok{tm\_laci) }\SpecialCharTok{*}\NormalTok{ (}\SpecialCharTok{{-}}\DecValTok{1}\NormalTok{) }\SpecialCharTok{*} 
\NormalTok{    (}\DecValTok{1} \SpecialCharTok{+} \FunctionTok{exp}\NormalTok{(n }\SpecialCharTok{*}\NormalTok{ tp\_ci))}\SpecialCharTok{\^{}}\NormalTok{(}\SpecialCharTok{{-}}\DecValTok{2}\NormalTok{) }\SpecialCharTok{*}\NormalTok{ n }\SpecialCharTok{*} \FunctionTok{exp}\NormalTok{(n }\SpecialCharTok{*}\NormalTok{ tp\_ci)}
\NormalTok{  resultDx[,}\DecValTok{2}\NormalTok{,}\DecValTok{2}\NormalTok{] }\OtherTok{=} \SpecialCharTok{{-}}\NormalTok{(alpha }\SpecialCharTok{/}\NormalTok{ (}\DecValTok{1} \SpecialCharTok{+} \FunctionTok{exp}\NormalTok{(n }\SpecialCharTok{*}\NormalTok{ tp\_laci)) }\SpecialCharTok{+}\NormalTok{ alpha0) }\SpecialCharTok{*} \FunctionTok{exp}\NormalTok{(}\SpecialCharTok{{-}}\NormalTok{tm\_tetr)}
\NormalTok{  resultDx[,}\DecValTok{4}\NormalTok{,}\DecValTok{2}\NormalTok{] }\OtherTok{=}\NormalTok{ alpha }\SpecialCharTok{*} \FunctionTok{exp}\NormalTok{(}\SpecialCharTok{{-}}\NormalTok{tm\_tetr) }\SpecialCharTok{*}\NormalTok{ (}\SpecialCharTok{{-}}\DecValTok{1}\NormalTok{) }\SpecialCharTok{*} 
\NormalTok{    (}\DecValTok{1} \SpecialCharTok{+} \FunctionTok{exp}\NormalTok{(n }\SpecialCharTok{*}\NormalTok{ tp\_laci))}\SpecialCharTok{\^{}}\NormalTok{(}\SpecialCharTok{{-}}\DecValTok{2}\NormalTok{) }\SpecialCharTok{*}\NormalTok{ n }\SpecialCharTok{*} \FunctionTok{exp}\NormalTok{(n }\SpecialCharTok{*}\NormalTok{ tp\_laci)}
\NormalTok{  resultDx[,}\DecValTok{3}\NormalTok{,}\DecValTok{3}\NormalTok{] }\OtherTok{=} \SpecialCharTok{{-}}\NormalTok{(alpha }\SpecialCharTok{/}\NormalTok{ (}\DecValTok{1} \SpecialCharTok{+} \FunctionTok{exp}\NormalTok{(n }\SpecialCharTok{*}\NormalTok{ tp\_tetr)) }\SpecialCharTok{+}\NormalTok{ alpha0) }\SpecialCharTok{*} \FunctionTok{exp}\NormalTok{(}\SpecialCharTok{{-}}\NormalTok{tm\_ci)}
\NormalTok{  resultDx[,}\DecValTok{5}\NormalTok{,}\DecValTok{3}\NormalTok{] }\OtherTok{=}\NormalTok{ alpha }\SpecialCharTok{*} \FunctionTok{exp}\NormalTok{(}\SpecialCharTok{{-}}\NormalTok{tm\_ci) }\SpecialCharTok{*}\NormalTok{ (}\SpecialCharTok{{-}}\DecValTok{1}\NormalTok{) }\SpecialCharTok{*} 
\NormalTok{    (}\DecValTok{1} \SpecialCharTok{+} \FunctionTok{exp}\NormalTok{(n }\SpecialCharTok{*}\NormalTok{ tp\_tetr))}\SpecialCharTok{\^{}}\NormalTok{(}\SpecialCharTok{{-}}\DecValTok{2}\NormalTok{) }\SpecialCharTok{*}\NormalTok{ n }\SpecialCharTok{*} \FunctionTok{exp}\NormalTok{(n }\SpecialCharTok{*}\NormalTok{ tp\_tetr)}

\NormalTok{  resultDx[,}\DecValTok{1}\NormalTok{,}\DecValTok{4}\NormalTok{] }\OtherTok{=}\NormalTok{ beta }\SpecialCharTok{*} \FunctionTok{exp}\NormalTok{(tm\_laci }\SpecialCharTok{{-}}\NormalTok{ tp\_laci)}
\NormalTok{  resultDx[,}\DecValTok{4}\NormalTok{,}\DecValTok{4}\NormalTok{] }\OtherTok{=} \SpecialCharTok{{-}}\NormalTok{beta }\SpecialCharTok{*} \FunctionTok{exp}\NormalTok{(tm\_laci }\SpecialCharTok{{-}}\NormalTok{ tp\_laci)}
\NormalTok{  resultDx[,}\DecValTok{2}\NormalTok{,}\DecValTok{5}\NormalTok{] }\OtherTok{=}\NormalTok{ beta }\SpecialCharTok{*} \FunctionTok{exp}\NormalTok{(tm\_tetr }\SpecialCharTok{{-}}\NormalTok{ tp\_tetr)}
\NormalTok{  resultDx[,}\DecValTok{5}\NormalTok{,}\DecValTok{5}\NormalTok{] }\OtherTok{=} \SpecialCharTok{{-}}\NormalTok{beta }\SpecialCharTok{*} \FunctionTok{exp}\NormalTok{(tm\_tetr }\SpecialCharTok{{-}}\NormalTok{ tp\_tetr)}
\NormalTok{  resultDx[,}\DecValTok{3}\NormalTok{,}\DecValTok{6}\NormalTok{] }\OtherTok{=}\NormalTok{ beta }\SpecialCharTok{*} \FunctionTok{exp}\NormalTok{(tm\_ci }\SpecialCharTok{{-}}\NormalTok{ tp\_ci)}
\NormalTok{  resultDx[,}\DecValTok{6}\NormalTok{,}\DecValTok{6}\NormalTok{] }\OtherTok{=} \SpecialCharTok{{-}}\NormalTok{beta }\SpecialCharTok{*} \FunctionTok{exp}\NormalTok{(tm\_ci }\SpecialCharTok{{-}}\NormalTok{ tp\_ci)}

\NormalTok{  resultDx}
\NormalTok{\}}

\NormalTok{RrepressilatorGeneRegulationLogDtheta }\OtherTok{\textless{}{-}} \ControlFlowTok{function}\NormalTok{(theta, x, tvec) \{}
\NormalTok{  resultDtheta }\OtherTok{\textless{}{-}} \FunctionTok{array}\NormalTok{(}\DecValTok{0}\NormalTok{, }\FunctionTok{c}\NormalTok{(}\FunctionTok{nrow}\NormalTok{(x), }\FunctionTok{length}\NormalTok{(theta), }\FunctionTok{ncol}\NormalTok{(x)))}

\NormalTok{  tm\_laci }\OtherTok{=}\NormalTok{ x[,}\DecValTok{1}\NormalTok{]}
\NormalTok{  tm\_tetr }\OtherTok{=}\NormalTok{ x[,}\DecValTok{2}\NormalTok{]}
\NormalTok{  tm\_ci }\OtherTok{=}\NormalTok{ x[,}\DecValTok{3}\NormalTok{]}
\NormalTok{  tp\_laci }\OtherTok{=}\NormalTok{ x[,}\DecValTok{4}\NormalTok{]}
\NormalTok{  tp\_tetr }\OtherTok{=}\NormalTok{ x[,}\DecValTok{5}\NormalTok{]}
\NormalTok{  tp\_ci }\OtherTok{=}\NormalTok{ x[,}\DecValTok{6}\NormalTok{]}

\NormalTok{  p\_ci }\OtherTok{=} \FunctionTok{exp}\NormalTok{(tp\_ci)}
\NormalTok{  p\_laci }\OtherTok{=} \FunctionTok{exp}\NormalTok{(tp\_laci)}
\NormalTok{  p\_tetr }\OtherTok{=} \FunctionTok{exp}\NormalTok{(tp\_tetr)}

\NormalTok{  alpha0 }\OtherTok{=}\NormalTok{ theta[}\DecValTok{1}\NormalTok{]}
\NormalTok{  alpha }\OtherTok{=}\NormalTok{ theta[}\DecValTok{2}\NormalTok{]}
\NormalTok{  n }\OtherTok{=}\NormalTok{ theta[}\DecValTok{3}\NormalTok{]}
\NormalTok{  beta }\OtherTok{=}\NormalTok{ theta[}\DecValTok{4}\NormalTok{]}

\NormalTok{  resultDtheta[,}\DecValTok{1}\NormalTok{,}\DecValTok{1}\NormalTok{] }\OtherTok{=} \FunctionTok{exp}\NormalTok{(}\SpecialCharTok{{-}}\NormalTok{x[,}\DecValTok{1}\NormalTok{])}
\NormalTok{  resultDtheta[,}\DecValTok{2}\NormalTok{,}\DecValTok{1}\NormalTok{] }\OtherTok{=} \DecValTok{1} \SpecialCharTok{/}\NormalTok{ (}\DecValTok{1} \SpecialCharTok{+} \FunctionTok{exp}\NormalTok{(n }\SpecialCharTok{*}\NormalTok{ tp\_ci)) }\SpecialCharTok{*} \FunctionTok{exp}\NormalTok{(}\SpecialCharTok{{-}}\NormalTok{x[,}\DecValTok{1}\NormalTok{])}
\NormalTok{  resultDtheta[,}\DecValTok{3}\NormalTok{,}\DecValTok{1}\NormalTok{] }\OtherTok{=}\NormalTok{ alpha }\SpecialCharTok{*} \FunctionTok{exp}\NormalTok{(}\SpecialCharTok{{-}}\NormalTok{x[,}\DecValTok{1}\NormalTok{]) }\SpecialCharTok{*}\NormalTok{ (}\SpecialCharTok{{-}}\DecValTok{1}\NormalTok{) }\SpecialCharTok{*}
\NormalTok{    (}\DecValTok{1} \SpecialCharTok{+}\NormalTok{ p\_ci}\SpecialCharTok{\^{}}\NormalTok{n)}\SpecialCharTok{\^{}}\NormalTok{(}\SpecialCharTok{{-}}\DecValTok{2}\NormalTok{) }\SpecialCharTok{*}\NormalTok{ p\_ci}\SpecialCharTok{\^{}}\NormalTok{n }\SpecialCharTok{*} \FunctionTok{log}\NormalTok{(p\_ci)}
\NormalTok{  resultDtheta[,}\DecValTok{1}\NormalTok{,}\DecValTok{2}\NormalTok{] }\OtherTok{=} \FunctionTok{exp}\NormalTok{(}\SpecialCharTok{{-}}\NormalTok{x[,}\DecValTok{2}\NormalTok{])}
\NormalTok{  resultDtheta[,}\DecValTok{2}\NormalTok{,}\DecValTok{2}\NormalTok{] }\OtherTok{=} \DecValTok{1} \SpecialCharTok{/}\NormalTok{ (}\DecValTok{1} \SpecialCharTok{+} \FunctionTok{exp}\NormalTok{(n }\SpecialCharTok{*}\NormalTok{ tp\_laci)) }\SpecialCharTok{*} \FunctionTok{exp}\NormalTok{(}\SpecialCharTok{{-}}\NormalTok{x[,}\DecValTok{2}\NormalTok{])}
\NormalTok{  resultDtheta[,}\DecValTok{3}\NormalTok{,}\DecValTok{2}\NormalTok{] }\OtherTok{=}\NormalTok{ alpha }\SpecialCharTok{*} \FunctionTok{exp}\NormalTok{(}\SpecialCharTok{{-}}\NormalTok{x[,}\DecValTok{2}\NormalTok{]) }\SpecialCharTok{*}\NormalTok{ (}\SpecialCharTok{{-}}\DecValTok{1}\NormalTok{) }\SpecialCharTok{*} 
\NormalTok{    (}\DecValTok{1} \SpecialCharTok{+}\NormalTok{ p\_laci}\SpecialCharTok{\^{}}\NormalTok{n)}\SpecialCharTok{\^{}}\NormalTok{(}\SpecialCharTok{{-}}\DecValTok{2}\NormalTok{) }\SpecialCharTok{*}\NormalTok{ p\_laci}\SpecialCharTok{\^{}}\NormalTok{n }\SpecialCharTok{*} \FunctionTok{log}\NormalTok{(p\_laci)}
\NormalTok{  resultDtheta[,}\DecValTok{1}\NormalTok{,}\DecValTok{3}\NormalTok{] }\OtherTok{=} \FunctionTok{exp}\NormalTok{(}\SpecialCharTok{{-}}\NormalTok{x[,}\DecValTok{3}\NormalTok{])}
\NormalTok{  resultDtheta[,}\DecValTok{2}\NormalTok{,}\DecValTok{3}\NormalTok{] }\OtherTok{=} \DecValTok{1} \SpecialCharTok{/}\NormalTok{ (}\DecValTok{1} \SpecialCharTok{+} \FunctionTok{exp}\NormalTok{(n }\SpecialCharTok{*}\NormalTok{ tp\_tetr)) }\SpecialCharTok{*} \FunctionTok{exp}\NormalTok{(}\SpecialCharTok{{-}}\NormalTok{x[,}\DecValTok{3}\NormalTok{])}
\NormalTok{  resultDtheta[,}\DecValTok{3}\NormalTok{,}\DecValTok{3}\NormalTok{] }\OtherTok{=}\NormalTok{ alpha }\SpecialCharTok{*} \FunctionTok{exp}\NormalTok{(}\SpecialCharTok{{-}}\NormalTok{x[,}\DecValTok{3}\NormalTok{]) }\SpecialCharTok{*}\NormalTok{ (}\SpecialCharTok{{-}}\DecValTok{1}\NormalTok{) }\SpecialCharTok{*} 
\NormalTok{    (}\DecValTok{1} \SpecialCharTok{+}\NormalTok{ p\_tetr}\SpecialCharTok{\^{}}\NormalTok{n)}\SpecialCharTok{\^{}}\NormalTok{(}\SpecialCharTok{{-}}\DecValTok{2}\NormalTok{) }\SpecialCharTok{*}\NormalTok{ p\_tetr}\SpecialCharTok{\^{}}\NormalTok{n }\SpecialCharTok{*} \FunctionTok{log}\NormalTok{(p\_tetr)}

\NormalTok{  resultDtheta[,}\DecValTok{4}\NormalTok{,}\DecValTok{4}\NormalTok{] }\OtherTok{=} \FunctionTok{exp}\NormalTok{(x[,}\DecValTok{1}\NormalTok{] }\SpecialCharTok{{-}}\NormalTok{ x[,}\DecValTok{4}\NormalTok{]) }\SpecialCharTok{{-}} \DecValTok{1}
\NormalTok{  resultDtheta[,}\DecValTok{4}\NormalTok{,}\DecValTok{5}\NormalTok{] }\OtherTok{=} \FunctionTok{exp}\NormalTok{(x[,}\DecValTok{2}\NormalTok{] }\SpecialCharTok{{-}}\NormalTok{ x[,}\DecValTok{5}\NormalTok{]) }\SpecialCharTok{{-}} \DecValTok{1}
\NormalTok{  resultDtheta[,}\DecValTok{4}\NormalTok{,}\DecValTok{6}\NormalTok{] }\OtherTok{=} \FunctionTok{exp}\NormalTok{(x[,}\DecValTok{3}\NormalTok{] }\SpecialCharTok{{-}}\NormalTok{ x[,}\DecValTok{6}\NormalTok{]) }\SpecialCharTok{{-}} \DecValTok{1}

\NormalTok{  resultDtheta}
\NormalTok{\}}
\end{Highlighting}
\end{Shaded}

Define parameters and settings for the experiment and MAGI:

\begin{Shaded}
\begin{Highlighting}[]
\CommentTok{\# MAGI configuration}
\NormalTok{config }\OtherTok{\textless{}{-}} \FunctionTok{list}\NormalTok{(}
\AttributeTok{nobs =} \DecValTok{51}\NormalTok{,}
\AttributeTok{noise =} \FunctionTok{rep}\NormalTok{(}\FloatTok{0.3}\NormalTok{, }\DecValTok{6}\NormalTok{),}
\AttributeTok{kernel =} \StringTok{"generalMatern"}\NormalTok{,}
\AttributeTok{seed =} \DecValTok{142249801}\NormalTok{,  }\CommentTok{\# example seed, or choose a random seed}
\AttributeTok{niterHmc =} \DecValTok{10001}\NormalTok{,}
\AttributeTok{filllevel =} \DecValTok{1}\NormalTok{,}
\AttributeTok{t.end =} \DecValTok{300}\NormalTok{,}
\AttributeTok{modelName =} \StringTok{"repressilator{-}gene{-}regulation{-}log"}
\NormalTok{)}

\CommentTok{\# Parameters and initial conditions}
\NormalTok{alpha }\OtherTok{\textless{}{-}} \DecValTok{240} \CommentTok{\# obtain from Fig 1b in Elowitz and Leibler (2000)}
\NormalTok{KM }\OtherTok{\textless{}{-}} \DecValTok{40}     \CommentTok{\# scale factor only, to convert protein number to match Fig 1c in paper}
\NormalTok{pram.true }\OtherTok{\textless{}{-}} \FunctionTok{list}\NormalTok{(}
\AttributeTok{theta=}\FunctionTok{c}\NormalTok{(}\FloatTok{0.001}\SpecialCharTok{*}\NormalTok{alpha, alpha, }\DecValTok{2}\NormalTok{, }\DecValTok{1}\SpecialCharTok{/}\DecValTok{5}\NormalTok{),  }\CommentTok{\# alpha0/alpha = 0.001}
\AttributeTok{x0 =} \FunctionTok{log}\NormalTok{(}\FunctionTok{c}\NormalTok{(}\FloatTok{0.4}\NormalTok{, }\DecValTok{20}\NormalTok{, }\DecValTok{40}\NormalTok{, }\FloatTok{0.01}\NormalTok{, }\FloatTok{0.01}\NormalTok{, }\FloatTok{0.01}\NormalTok{)), }\CommentTok{\# initial conditions}
\AttributeTok{sigma=}\NormalTok{config}\SpecialCharTok{$}\NormalTok{noise}
\NormalTok{)}
\end{Highlighting}
\end{Shaded}

Use a numerical solver to generate the true trajectories to simulate data and to compare with inference from MAGI:

\begin{Shaded}
\begin{Highlighting}[]
\NormalTok{times }\OtherTok{\textless{}{-}} \FunctionTok{seq}\NormalTok{(}\DecValTok{0}\NormalTok{,config}\SpecialCharTok{$}\NormalTok{t.end,}\AttributeTok{length=}\DecValTok{1001}\NormalTok{)}

\NormalTok{modelODE }\OtherTok{\textless{}{-}} \ControlFlowTok{function}\NormalTok{(t, state, parameters) \{}
\FunctionTok{list}\NormalTok{(}\FunctionTok{as.vector}\NormalTok{(}\FunctionTok{RrepressilatorGeneRegulationLogODE}\NormalTok{(parameters, }\FunctionTok{t}\NormalTok{(state), t)))}
\NormalTok{\}}

\NormalTok{xtrue }\OtherTok{\textless{}{-}}\NormalTok{ deSolve}\SpecialCharTok{::}\FunctionTok{ode}\NormalTok{(}\AttributeTok{y =}\NormalTok{ pram.true}\SpecialCharTok{$}\NormalTok{x0, }\AttributeTok{times =}\NormalTok{ times,}
\AttributeTok{func =}\NormalTok{ modelODE, }\AttributeTok{parms =}\NormalTok{ pram.true}\SpecialCharTok{$}\NormalTok{theta)}
\NormalTok{xtrue }\OtherTok{\textless{}{-}} \FunctionTok{data.frame}\NormalTok{(xtrue)}
\end{Highlighting}
\end{Shaded}

Add multiplicative noise at the observation schedule to create simulated noisy data:

\begin{Shaded}
\begin{Highlighting}[]
\NormalTok{xtrueFunc }\OtherTok{\textless{}{-}} \FunctionTok{lapply}\NormalTok{(}\DecValTok{2}\SpecialCharTok{:}\FunctionTok{ncol}\NormalTok{(xtrue), }\ControlFlowTok{function}\NormalTok{(j)}
\FunctionTok{approxfun}\NormalTok{(xtrue[, }\StringTok{"time"}\NormalTok{], xtrue[, j]))}

\NormalTok{xsim }\OtherTok{\textless{}{-}} \FunctionTok{data.frame}\NormalTok{(}\AttributeTok{time =} \FunctionTok{seq}\NormalTok{(}\DecValTok{0}\NormalTok{,config}\SpecialCharTok{$}\NormalTok{t.end,}\AttributeTok{length=}\NormalTok{config}\SpecialCharTok{$}\NormalTok{nobs))}
\NormalTok{xsim }\OtherTok{\textless{}{-}} \FunctionTok{cbind}\NormalTok{(xsim, }\FunctionTok{sapply}\NormalTok{(xtrueFunc, }\ControlFlowTok{function}\NormalTok{(f) }\FunctionTok{f}\NormalTok{(xsim}\SpecialCharTok{$}\NormalTok{time)))}

\FunctionTok{set.seed}\NormalTok{(config}\SpecialCharTok{$}\NormalTok{seed)}
\ControlFlowTok{for}\NormalTok{(j }\ControlFlowTok{in} \DecValTok{1}\SpecialCharTok{:}\NormalTok{(}\FunctionTok{ncol}\NormalTok{(xsim)}\SpecialCharTok{{-}}\DecValTok{1}\NormalTok{))\{}
\NormalTok{  xsim[,}\DecValTok{1}\SpecialCharTok{+}\NormalTok{j] }\OtherTok{\textless{}{-}}\NormalTok{ xsim[,}\DecValTok{1}\SpecialCharTok{+}\NormalTok{j]}\SpecialCharTok{+}\FunctionTok{rnorm}\NormalTok{(}\FunctionTok{nrow}\NormalTok{(xsim), }\AttributeTok{sd=}\NormalTok{config}\SpecialCharTok{$}\NormalTok{noise[j])}
\NormalTok{\}}

\NormalTok{xsim.obs }\OtherTok{\textless{}{-}}\NormalTok{ xsim[}\FunctionTok{seq}\NormalTok{(}\DecValTok{1}\NormalTok{,}\FunctionTok{nrow}\NormalTok{(xsim), }\AttributeTok{length=}\NormalTok{config}\SpecialCharTok{$}\NormalTok{nobs),]}
\end{Highlighting}
\end{Shaded}

Create the \texttt{odeModel} list, then confirm ODEs and derivatives are correct:

\begin{Shaded}
\begin{Highlighting}[]
\NormalTok{xsim }\OtherTok{\textless{}{-}} \FunctionTok{setDiscretization}\NormalTok{(xsim.obs,config}\SpecialCharTok{$}\NormalTok{filllevel)}

\NormalTok{dynamicalModelList }\OtherTok{\textless{}{-}} \FunctionTok{list}\NormalTok{(}
\AttributeTok{fOde=}\NormalTok{RrepressilatorGeneRegulationLogODE,}
\AttributeTok{fOdeDx=}\NormalTok{RrepressilatorGeneRegulationLogDx,}
\AttributeTok{fOdeDtheta=}\NormalTok{RrepressilatorGeneRegulationLogDtheta,}
\AttributeTok{thetaLowerBound=}\FunctionTok{rep}\NormalTok{(}\DecValTok{0}\NormalTok{, }\DecValTok{4}\NormalTok{),}
\AttributeTok{thetaUpperBound=}\FunctionTok{rep}\NormalTok{(}\ConstantTok{Inf}\NormalTok{, }\DecValTok{4}\NormalTok{)}
\NormalTok{)}

\FunctionTok{testDynamicalModel}\NormalTok{(dynamicalModelList}\SpecialCharTok{$}\NormalTok{fOde, dynamicalModelList}\SpecialCharTok{$}\NormalTok{fOdeDx,}
\NormalTok{                   dynamicalModelList}\SpecialCharTok{$}\NormalTok{fOdeDtheta, }\StringTok{"dynamicalModelList"}\NormalTok{,}
\FunctionTok{data.matrix}\NormalTok{(xsim.obs[}\SpecialCharTok{{-}}\DecValTok{1}\NormalTok{,}\SpecialCharTok{{-}}\DecValTok{1}\NormalTok{]), pram.true}\SpecialCharTok{$}\NormalTok{theta, xsim.obs}\SpecialCharTok{$}\NormalTok{time[}\SpecialCharTok{{-}}\DecValTok{1}\NormalTok{])}
\end{Highlighting}
\end{Shaded}

\begin{verbatim}
## dynamicalModelList model, with derivatives
## Dx and Dtheta appear to be correct
\end{verbatim}

\begin{verbatim}
## $testDx
## [1] TRUE
## 
## $testDtheta
## [1] TRUE
\end{verbatim}

Create inputs for MAGI:

\begin{Shaded}
\begin{Highlighting}[]
\CommentTok{\# Set discretization level}
\NormalTok{xsim }\OtherTok{\textless{}{-}} \FunctionTok{setDiscretization}\NormalTok{(xsim.obs,config}\SpecialCharTok{$}\NormalTok{filllevel)}

\CommentTok{\# Set some reasonable hyperparameters}
\NormalTok{phiExogenous }\OtherTok{\textless{}{-}} \FunctionTok{rbind}\NormalTok{(}\FunctionTok{rep}\NormalTok{(}\DecValTok{6}\NormalTok{, }\DecValTok{6}\NormalTok{), }\FunctionTok{rep}\NormalTok{(}\DecValTok{10}\NormalTok{, }\DecValTok{6}\NormalTok{))}

\CommentTok{\# Known noise level for mRNA}
\NormalTok{sigmaInit }\OtherTok{\textless{}{-}}\NormalTok{ config}\SpecialCharTok{$}\NormalTok{noise}

\CommentTok{\# Remove initial conditions}
\NormalTok{xsim }\OtherTok{\textless{}{-}}\NormalTok{ xsim[}\SpecialCharTok{{-}}\DecValTok{1}\NormalTok{,]}

\CommentTok{\# Set protein levels missing}
\NormalTok{xsim[,}\DecValTok{5}\SpecialCharTok{:}\DecValTok{7}\NormalTok{] }\OtherTok{\textless{}{-}} \ConstantTok{NA}
\NormalTok{xsim.obs[,}\DecValTok{5}\SpecialCharTok{:}\DecValTok{7}\NormalTok{] }\OtherTok{\textless{}{-}} \ConstantTok{NA}
\end{Highlighting}
\end{Shaded}

Now we are ready to run the MAGI method:

\begin{Shaded}
\begin{Highlighting}[]
\NormalTok{gpode }\OtherTok{\textless{}{-}} \FunctionTok{MagiSolver}\NormalTok{(xsim, dynamicalModelList,}
\AttributeTok{control =} \FunctionTok{list}\NormalTok{(}\AttributeTok{niterHmc=}\NormalTok{config}\SpecialCharTok{$}\NormalTok{niterHmc, }\AttributeTok{phi=}\NormalTok{phiExogenous,}
\AttributeTok{sigma=}\NormalTok{sigmaInit, }\AttributeTok{useFixedSigma=}\ConstantTok{TRUE}\NormalTok{))}
\end{Highlighting}
\end{Shaded}

\begin{figure}
\centering
\includegraphics{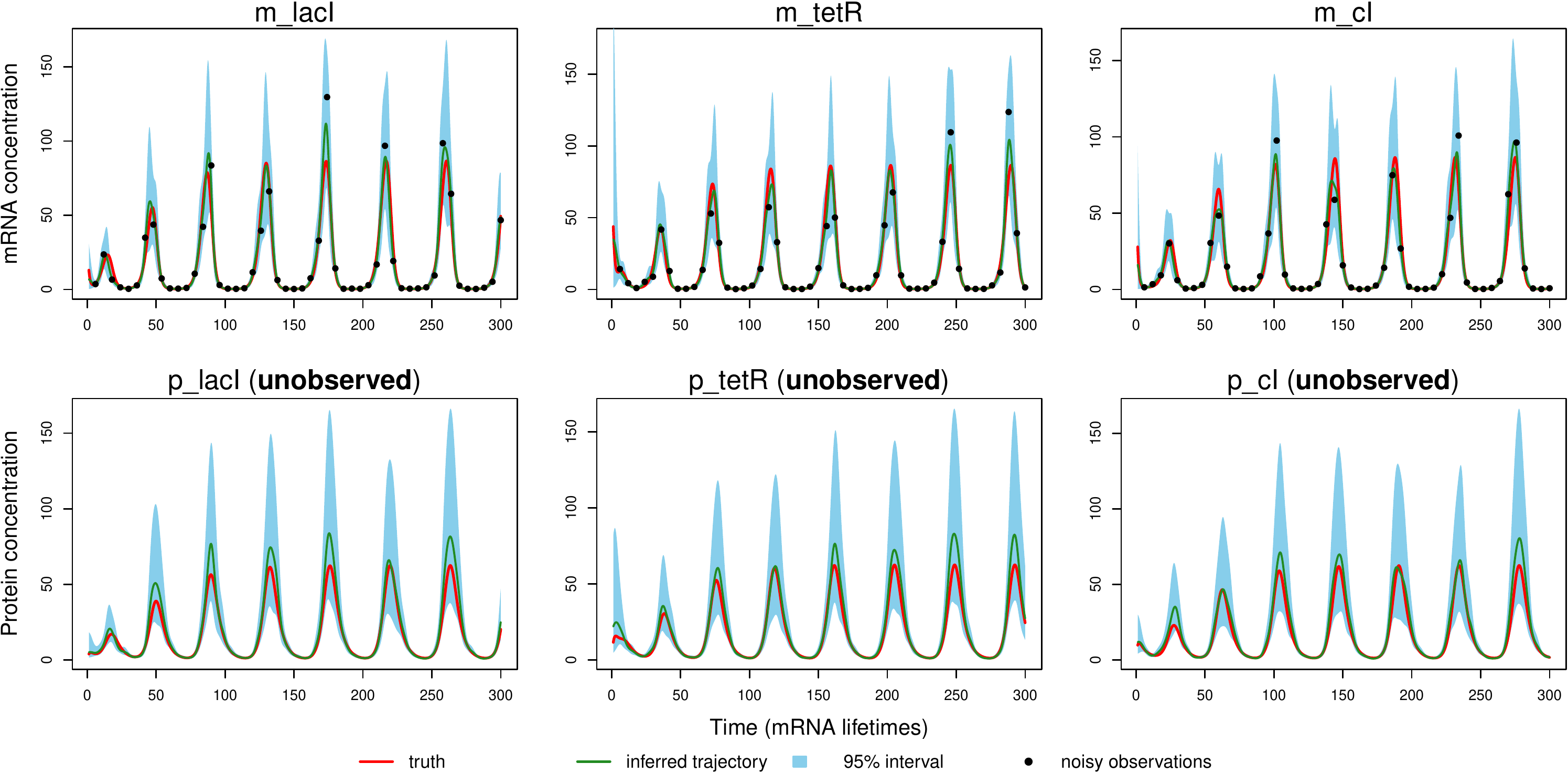}
\caption{\label{fig:rep-infer} \footnotesize Inferred trajectories for a sample dataset from the repressilator gene regulation model}
\end{figure}

Plot the noisy observations and inferred trajectories, to produce Fig \ref{fig:rep-infer}:

\begin{Shaded}
\begin{Highlighting}[]
\NormalTok{xtrue }\OtherTok{\textless{}{-}}\NormalTok{ xtrue[xtrue}\SpecialCharTok{$}\NormalTok{time }\SpecialCharTok{\textgreater{}=} \DecValTok{1}\NormalTok{,] }\CommentTok{\# remove initial conditions}

\NormalTok{xsampledexp }\OtherTok{\textless{}{-}} \FunctionTok{exp}\NormalTok{(gpode}\SpecialCharTok{$}\NormalTok{xsampled) }\CommentTok{\# exponentiate to original scale}
\NormalTok{oursPostExpX }\OtherTok{\textless{}{-}} \FunctionTok{cbind}\NormalTok{(}
\FunctionTok{apply}\NormalTok{(xsampledexp, }\DecValTok{2}\SpecialCharTok{:}\DecValTok{3}\NormalTok{, mean),}
\FunctionTok{apply}\NormalTok{(xsampledexp, }\DecValTok{2}\SpecialCharTok{:}\DecValTok{3}\NormalTok{, }\ControlFlowTok{function}\NormalTok{(x) }\FunctionTok{quantile}\NormalTok{(x, }\FloatTok{0.025}\NormalTok{)),}
\FunctionTok{apply}\NormalTok{(xsampledexp, }\DecValTok{2}\SpecialCharTok{:}\DecValTok{3}\NormalTok{, }\ControlFlowTok{function}\NormalTok{(x) }\FunctionTok{quantile}\NormalTok{(x, }\FloatTok{0.975}\NormalTok{)))}

\NormalTok{compnames }\OtherTok{\textless{}{-}} \FunctionTok{c}\NormalTok{(}\StringTok{"m\_lacI"}\NormalTok{, }\StringTok{"m\_tetR"}\NormalTok{, }\StringTok{"m\_cI"}\NormalTok{, }
\FunctionTok{expression}\NormalTok{(}\FunctionTok{paste}\NormalTok{(}\StringTok{"p\_lacI ("}\NormalTok{, }\FunctionTok{bold}\NormalTok{(}\StringTok{"unobserved"}\NormalTok{), }\StringTok{")"}\NormalTok{)),}
\FunctionTok{expression}\NormalTok{(}\FunctionTok{paste}\NormalTok{(}\StringTok{"p\_tetR ("}\NormalTok{, }\FunctionTok{bold}\NormalTok{(}\StringTok{"unobserved"}\NormalTok{), }\StringTok{")"}\NormalTok{)),}
\FunctionTok{expression}\NormalTok{(}\FunctionTok{paste}\NormalTok{(}\StringTok{"p\_cI ("}\NormalTok{, }\FunctionTok{bold}\NormalTok{(}\StringTok{"unobserved"}\NormalTok{), }\StringTok{")"}\NormalTok{)))}
\FunctionTok{layout}\NormalTok{(}\FunctionTok{rbind}\NormalTok{(}\FunctionTok{c}\NormalTok{(}\DecValTok{1}\NormalTok{,}\DecValTok{2}\NormalTok{,}\DecValTok{3}\NormalTok{), }\FunctionTok{c}\NormalTok{(}\DecValTok{4}\NormalTok{,}\DecValTok{5}\NormalTok{,}\DecValTok{6}\NormalTok{), }\FunctionTok{c}\NormalTok{(}\DecValTok{7}\NormalTok{,}\DecValTok{7}\NormalTok{,}\DecValTok{7}\NormalTok{)), }\AttributeTok{heights =} \FunctionTok{c}\NormalTok{(}\DecValTok{8}\NormalTok{,}\DecValTok{8}\NormalTok{,}\DecValTok{1}\NormalTok{))}
\ControlFlowTok{for}\NormalTok{ (ii }\ControlFlowTok{in} \DecValTok{1}\SpecialCharTok{:}\DecValTok{6}\NormalTok{) \{}

\FunctionTok{par}\NormalTok{(}\AttributeTok{mar =} \FunctionTok{c}\NormalTok{(}\DecValTok{4}\NormalTok{, }\FloatTok{4.5}\NormalTok{, }\FloatTok{1.75}\NormalTok{, }\FloatTok{0.1}\NormalTok{))}
\NormalTok{  ourEst }\OtherTok{\textless{}{-}}\NormalTok{ oursPostExpX[,ii]}
\NormalTok{  ourEst }\OtherTok{\textless{}{-}} \FunctionTok{exp}\NormalTok{(magi}\SpecialCharTok{:::}\FunctionTok{getMeanCurve}\NormalTok{(xsim}\SpecialCharTok{$}\NormalTok{time, }\FunctionTok{log}\NormalTok{(ourEst), xtrue[,}\DecValTok{1}\NormalTok{],}
\FunctionTok{t}\NormalTok{(phiExogenous[,ii]), }\DecValTok{0}\NormalTok{, }
\AttributeTok{kerneltype=}\NormalTok{config}\SpecialCharTok{$}\NormalTok{kernel, }\AttributeTok{deriv =} \ConstantTok{FALSE}\NormalTok{))}

\NormalTok{  ourUB }\OtherTok{\textless{}{-}}\NormalTok{ oursPostExpX[,}\DecValTok{12}\SpecialCharTok{+}\NormalTok{ii]}
\NormalTok{  ourUB }\OtherTok{\textless{}{-}} \FunctionTok{exp}\NormalTok{(magi}\SpecialCharTok{:::}\FunctionTok{getMeanCurve}\NormalTok{(xsim}\SpecialCharTok{$}\NormalTok{time, }\FunctionTok{log}\NormalTok{(ourUB), xtrue[,}\DecValTok{1}\NormalTok{],}
\FunctionTok{t}\NormalTok{(phiExogenous[,ii]), }\DecValTok{0}\NormalTok{,}
\AttributeTok{kerneltype=}\NormalTok{config}\SpecialCharTok{$}\NormalTok{kernel, }\AttributeTok{deriv =} \ConstantTok{FALSE}\NormalTok{))}

\NormalTok{  ourLB }\OtherTok{\textless{}{-}}\NormalTok{ oursPostExpX[,}\DecValTok{6}\SpecialCharTok{+}\NormalTok{ii]}
\NormalTok{  ourLB }\OtherTok{\textless{}{-}} \FunctionTok{exp}\NormalTok{(magi}\SpecialCharTok{:::}\FunctionTok{getMeanCurve}\NormalTok{(xsim}\SpecialCharTok{$}\NormalTok{time, }\FunctionTok{log}\NormalTok{(ourLB), xtrue[,}\DecValTok{1}\NormalTok{],}
\FunctionTok{t}\NormalTok{(phiExogenous[,ii]), }\DecValTok{0}\NormalTok{,}
\AttributeTok{kerneltype=}\NormalTok{config}\SpecialCharTok{$}\NormalTok{kernel, }\AttributeTok{deriv =} \ConstantTok{FALSE}\NormalTok{))}
\FunctionTok{plot}\NormalTok{( }\FunctionTok{c}\NormalTok{(}\FunctionTok{min}\NormalTok{(xtrue}\SpecialCharTok{$}\NormalTok{time),}\FunctionTok{max}\NormalTok{(xtrue}\SpecialCharTok{$}\NormalTok{time)), }\FunctionTok{c}\NormalTok{(}\FunctionTok{min}\NormalTok{(ourLB), }\FunctionTok{min}\NormalTok{(}\FunctionTok{max}\NormalTok{(ourUB),}\DecValTok{175}\NormalTok{)),}
\AttributeTok{type=}\StringTok{\textquotesingle{}n\textquotesingle{}}\NormalTok{, }\AttributeTok{xlab=}\StringTok{\textquotesingle{}\textquotesingle{}}\NormalTok{, }\AttributeTok{ylab=}\StringTok{\textquotesingle{}\textquotesingle{}}\NormalTok{)}

\FunctionTok{polygon}\NormalTok{(}\FunctionTok{c}\NormalTok{(xtrue[,}\DecValTok{1}\NormalTok{], }\FunctionTok{rev}\NormalTok{(xtrue[,}\DecValTok{1}\NormalTok{])), }\FunctionTok{c}\NormalTok{(ourUB, }\FunctionTok{rev}\NormalTok{(ourLB)),}
\AttributeTok{col =} \StringTok{"skyblue"}\NormalTok{, }\AttributeTok{border =} \ConstantTok{NA}\NormalTok{)    }

\ControlFlowTok{if}\NormalTok{ (ii }\SpecialCharTok{==} \DecValTok{1}\NormalTok{)}
\FunctionTok{title}\NormalTok{(}\AttributeTok{ylab=}\StringTok{\textquotesingle{}mRNA concentration\textquotesingle{}}\NormalTok{, }\AttributeTok{cex.lab =} \FloatTok{1.5}\NormalTok{)}

\ControlFlowTok{if}\NormalTok{ (ii }\SpecialCharTok{==} \DecValTok{4}\NormalTok{)}
\FunctionTok{title}\NormalTok{(}\AttributeTok{ylab=}\StringTok{\textquotesingle{}Protein concentration\textquotesingle{}}\NormalTok{, }\AttributeTok{cex.lab =} \FloatTok{1.5}\NormalTok{)}

\ControlFlowTok{if}\NormalTok{ (ii }\SpecialCharTok{==} \DecValTok{5}\NormalTok{)}
\FunctionTok{title}\NormalTok{(}\AttributeTok{xlab=}\StringTok{\textquotesingle{}Time (mRNA lifetimes)\textquotesingle{}}\NormalTok{, }\AttributeTok{cex.lab =} \FloatTok{1.5}\NormalTok{)}

\FunctionTok{lines}\NormalTok{(xtrue[, }\StringTok{"time"}\NormalTok{], }\FunctionTok{exp}\NormalTok{(xtrue[,ii}\SpecialCharTok{+}\DecValTok{1}\NormalTok{]),}\AttributeTok{col=}\StringTok{\textquotesingle{}red\textquotesingle{}}\NormalTok{, }\AttributeTok{lwd=}\DecValTok{2}\NormalTok{)}
\FunctionTok{lines}\NormalTok{(xtrue[,}\DecValTok{1}\NormalTok{], ourEst, }\AttributeTok{col=}\StringTok{\textquotesingle{}forestgreen\textquotesingle{}}\NormalTok{, }\AttributeTok{lwd=}\FloatTok{1.5}\NormalTok{)}
\FunctionTok{mtext}\NormalTok{(compnames[ii], }\AttributeTok{cex=}\FloatTok{1.25}\NormalTok{)}
\ControlFlowTok{if}\NormalTok{ (ii }\SpecialCharTok{\textless{}=} \DecValTok{3}\NormalTok{) }\FunctionTok{points}\NormalTok{(xsim.obs}\SpecialCharTok{$}\NormalTok{time[}\SpecialCharTok{{-}}\DecValTok{1}\NormalTok{], }\FunctionTok{exp}\NormalTok{(xsim.obs[}\SpecialCharTok{{-}}\DecValTok{1}\NormalTok{,ii}\SpecialCharTok{+}\DecValTok{1}\NormalTok{]), }\AttributeTok{col=}\StringTok{\textquotesingle{}black\textquotesingle{}}\NormalTok{, }\AttributeTok{pch=}\DecValTok{16}\NormalTok{)}
\NormalTok{\}}

\FunctionTok{par}\NormalTok{(}\AttributeTok{mar =} \FunctionTok{rep}\NormalTok{(}\DecValTok{0}\NormalTok{, }\DecValTok{4}\NormalTok{))}
\FunctionTok{plot}\NormalTok{(}\DecValTok{1}\NormalTok{, }\AttributeTok{type =} \StringTok{\textquotesingle{}n\textquotesingle{}}\NormalTok{, }\AttributeTok{xaxt =} \StringTok{\textquotesingle{}n\textquotesingle{}}\NormalTok{, }\AttributeTok{yaxt =} \StringTok{\textquotesingle{}n\textquotesingle{}}\NormalTok{,}
\AttributeTok{xlab =} \ConstantTok{NA}\NormalTok{, }\AttributeTok{ylab =} \ConstantTok{NA}\NormalTok{, }\AttributeTok{frame.plot =} \ConstantTok{FALSE}\NormalTok{)}

\FunctionTok{legend}\NormalTok{(}\StringTok{"center"}\NormalTok{, }\FunctionTok{c}\NormalTok{(}\StringTok{"truth"}\NormalTok{, }\StringTok{"inferred trajectory"}\NormalTok{,}
\StringTok{"95\% interval"}\NormalTok{, }\StringTok{"noisy observations"}\NormalTok{),}
\AttributeTok{lty =} \FunctionTok{c}\NormalTok{(}\DecValTok{1}\NormalTok{, }\DecValTok{1}\NormalTok{, }\DecValTok{0}\NormalTok{, }\DecValTok{0}\NormalTok{), }\AttributeTok{lwd =} \FunctionTok{c}\NormalTok{(}\DecValTok{2}\NormalTok{, }\DecValTok{2}\NormalTok{, }\DecValTok{0}\NormalTok{, }\DecValTok{1}\NormalTok{), }\AttributeTok{bty =} \StringTok{"n"}\NormalTok{,}
\AttributeTok{col =} \FunctionTok{c}\NormalTok{(}\StringTok{"red"}\NormalTok{, }\StringTok{"forestgreen"}\NormalTok{, }\ConstantTok{NA}\NormalTok{, }\StringTok{"black"}\NormalTok{), }\AttributeTok{fill =} \FunctionTok{c}\NormalTok{(}\DecValTok{0}\NormalTok{, }\DecValTok{0}\NormalTok{, }\StringTok{"skyblue"}\NormalTok{, }\DecValTok{0}\NormalTok{),}
\AttributeTok{border =} \FunctionTok{c}\NormalTok{(}\DecValTok{0}\NormalTok{, }\DecValTok{0}\NormalTok{, }\StringTok{"skyblue"}\NormalTok{, }\DecValTok{0}\NormalTok{), }\AttributeTok{pch =} \FunctionTok{c}\NormalTok{(}\ConstantTok{NA}\NormalTok{, }\ConstantTok{NA}\NormalTok{, }\DecValTok{15}\NormalTok{, }\DecValTok{16}\NormalTok{), }\AttributeTok{horiz =} \ConstantTok{TRUE}\NormalTok{, }\AttributeTok{cex=}\FloatTok{1.25}\NormalTok{)}
\end{Highlighting}
\end{Shaded}

\begin{figure}
\centering
\includegraphics{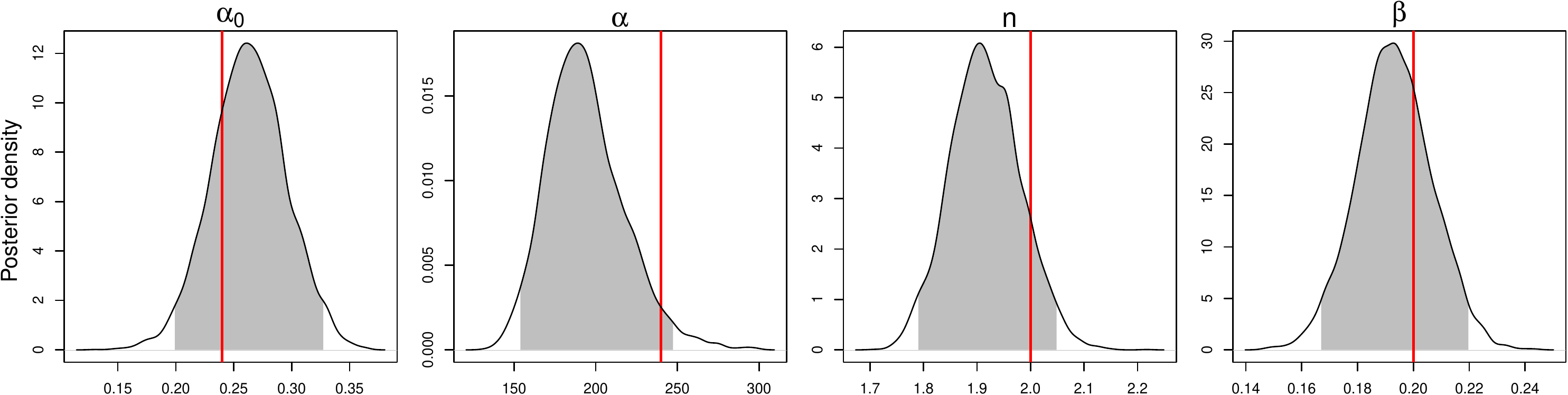}
\caption{\label{fig:rep-param} \footnotesize Bayesian posterior probability densities of system parameters for a sample dataset from the repressilator gene regulation model obtained by MAGI. The red vertical lines show the true parameter values used in the simulation. The shaded area represents the 95\% interval estimate of each parameter.}
\end{figure}

Plot the posterior densities of the parameters, to produce Fig \ref{fig:rep-param}:

\begin{Shaded}
\begin{Highlighting}[]
\NormalTok{par.names }\OtherTok{\textless{}{-}} \FunctionTok{c}\NormalTok{( }\FunctionTok{expression}\NormalTok{(alpha[}\DecValTok{0}\NormalTok{]), }\FunctionTok{expression}\NormalTok{(alpha), }\StringTok{"n"}\NormalTok{, }\FunctionTok{expression}\NormalTok{(beta))}
\FunctionTok{par}\NormalTok{(}\AttributeTok{mfrow=}\FunctionTok{c}\NormalTok{(}\DecValTok{1}\NormalTok{,}\DecValTok{4}\NormalTok{))}
\ControlFlowTok{for}\NormalTok{ (ii }\ControlFlowTok{in} \DecValTok{1}\SpecialCharTok{:}\DecValTok{4}\NormalTok{) \{}
\ControlFlowTok{if}\NormalTok{ (ii }\SpecialCharTok{==} \DecValTok{1}\NormalTok{) }\FunctionTok{par}\NormalTok{(}\AttributeTok{oma=}\FunctionTok{c}\NormalTok{(}\DecValTok{0}\NormalTok{,}\FloatTok{1.5}\NormalTok{,}\DecValTok{0}\NormalTok{,}\DecValTok{0}\NormalTok{))}
\FunctionTok{par}\NormalTok{(}\AttributeTok{mar =} \FunctionTok{c}\NormalTok{(}\FloatTok{2.5}\NormalTok{, }\FloatTok{2.5}\NormalTok{, }\DecValTok{2}\NormalTok{, }\FloatTok{0.75}\NormalTok{))}
\NormalTok{  den }\OtherTok{\textless{}{-}} \FunctionTok{density}\NormalTok{(gpode}\SpecialCharTok{$}\NormalTok{theta[,ii])}
\FunctionTok{plot}\NormalTok{(den, }\AttributeTok{main=}\StringTok{\textquotesingle{}\textquotesingle{}}\NormalTok{, }\AttributeTok{xlab =} \StringTok{\textquotesingle{}\textquotesingle{}}\NormalTok{, }\AttributeTok{ylab =} \StringTok{\textquotesingle{}\textquotesingle{}}\NormalTok{, }\AttributeTok{type=}\StringTok{\textquotesingle{}n\textquotesingle{}}\NormalTok{)}

\NormalTok{  value1 }\OtherTok{\textless{}{-}} \FunctionTok{quantile}\NormalTok{(gpode}\SpecialCharTok{$}\NormalTok{theta[,ii], }\FloatTok{0.025}\NormalTok{)}
\NormalTok{  value2 }\OtherTok{\textless{}{-}} \FunctionTok{quantile}\NormalTok{(gpode}\SpecialCharTok{$}\NormalTok{theta[,ii], }\FloatTok{0.975}\NormalTok{)}

\NormalTok{  l }\OtherTok{\textless{}{-}} \FunctionTok{min}\NormalTok{(}\FunctionTok{which}\NormalTok{(den}\SpecialCharTok{$}\NormalTok{x }\SpecialCharTok{\textgreater{}=}\NormalTok{ value1))}
\NormalTok{  h }\OtherTok{\textless{}{-}} \FunctionTok{max}\NormalTok{(}\FunctionTok{which}\NormalTok{(den}\SpecialCharTok{$}\NormalTok{x }\SpecialCharTok{\textless{}}\NormalTok{ value2))}

\FunctionTok{polygon}\NormalTok{(}\FunctionTok{c}\NormalTok{(den}\SpecialCharTok{$}\NormalTok{x[}\FunctionTok{c}\NormalTok{(l, l}\SpecialCharTok{:}\NormalTok{h, h)]),}
\FunctionTok{c}\NormalTok{(}\DecValTok{0}\NormalTok{, den}\SpecialCharTok{$}\NormalTok{y[l}\SpecialCharTok{:}\NormalTok{h], }\DecValTok{0}\NormalTok{),}
\AttributeTok{col =} \StringTok{"grey75"}\NormalTok{, }\AttributeTok{border=}\ConstantTok{NA}\NormalTok{)}
\FunctionTok{abline}\NormalTok{(}\AttributeTok{v=}\NormalTok{pram.true}\SpecialCharTok{$}\NormalTok{theta[ii], }\AttributeTok{col=}\StringTok{\textquotesingle{}red\textquotesingle{}}\NormalTok{, }\AttributeTok{lwd =}\DecValTok{2}\NormalTok{)}

\FunctionTok{lines}\NormalTok{(den)}

\ControlFlowTok{if}\NormalTok{ (ii }\SpecialCharTok{==} \DecValTok{1}\NormalTok{) }\FunctionTok{mtext}\NormalTok{(}\AttributeTok{text=}\StringTok{\textquotesingle{}Posterior density\textquotesingle{}}\NormalTok{,}\AttributeTok{side=}\DecValTok{2}\NormalTok{,}\AttributeTok{line=}\DecValTok{0}\NormalTok{,}\AttributeTok{outer=}\ConstantTok{TRUE}\NormalTok{)}
\FunctionTok{mtext}\NormalTok{(par.names[ii], }\AttributeTok{cex=}\FloatTok{1.25}\NormalTok{)}
\NormalTok{\}}
\end{Highlighting}
\end{Shaded}

\hypertarget{michaelis-menten-model}{%
\subsection{Michaelis-Menten model}\label{michaelis-menten-model}}

We begin by defining a function that codes the ODEs. Since \([E]_0 = [E]+[ES]\) is a constant, the model can be reduced to three equations.

\begin{Shaded}
\begin{Highlighting}[]
\NormalTok{RMichaelisMentenReducedODE }\OtherTok{\textless{}{-}} \ControlFlowTok{function}\NormalTok{(theta, x, tvec) \{}
\NormalTok{  e0 }\OtherTok{=} \FloatTok{0.1}
\NormalTok{  e }\OtherTok{=}\NormalTok{ x[,}\DecValTok{1}\NormalTok{]}
\NormalTok{  s }\OtherTok{=}\NormalTok{ x[,}\DecValTok{2}\NormalTok{]}
\NormalTok{  es }\OtherTok{=}\NormalTok{ e0 }\SpecialCharTok{{-}}\NormalTok{ e}
\NormalTok{  p }\OtherTok{=}\NormalTok{ x[,}\DecValTok{3}\NormalTok{]}

\NormalTok{  resultdt }\OtherTok{\textless{}{-}}  \FunctionTok{array}\NormalTok{(}\DecValTok{0}\NormalTok{, }\FunctionTok{c}\NormalTok{(}\FunctionTok{nrow}\NormalTok{(x),}\FunctionTok{ncol}\NormalTok{(x)))}

\NormalTok{  resultdt[,}\DecValTok{1}\NormalTok{] }\OtherTok{=} \SpecialCharTok{{-}}\NormalTok{theta[}\DecValTok{1}\NormalTok{] }\SpecialCharTok{*}\NormalTok{ e }\SpecialCharTok{*}\NormalTok{ s }\SpecialCharTok{+}\NormalTok{ (theta[}\DecValTok{2}\NormalTok{]}\SpecialCharTok{+}\NormalTok{theta[}\DecValTok{3}\NormalTok{]) }\SpecialCharTok{*}\NormalTok{ es}
\NormalTok{  resultdt[,}\DecValTok{2}\NormalTok{] }\OtherTok{=} \SpecialCharTok{{-}}\NormalTok{theta[}\DecValTok{1}\NormalTok{] }\SpecialCharTok{*}\NormalTok{ e }\SpecialCharTok{*}\NormalTok{ s }\SpecialCharTok{+}\NormalTok{ (theta[}\DecValTok{2}\NormalTok{]) }\SpecialCharTok{*}\NormalTok{ es}
\NormalTok{  resultdt[,}\DecValTok{3}\NormalTok{] }\OtherTok{=}\NormalTok{ theta[}\DecValTok{3}\NormalTok{] }\SpecialCharTok{*}\NormalTok{ es}

\NormalTok{  resultdt}
\NormalTok{\}}
\end{Highlighting}
\end{Shaded}

Next, we provide the gradients of the ODEs with respect to the system components \(X\) and the parameters \(\theta\).

\begin{Shaded}
\begin{Highlighting}[]
\NormalTok{RMichaelisMentenReducedDx }\OtherTok{\textless{}{-}} \ControlFlowTok{function}\NormalTok{(theta, x, tvec) \{}
\NormalTok{  resultDx }\OtherTok{\textless{}{-}} \FunctionTok{array}\NormalTok{(}\DecValTok{0}\NormalTok{, }\FunctionTok{c}\NormalTok{(}\FunctionTok{nrow}\NormalTok{(x), }\FunctionTok{ncol}\NormalTok{(x), }\FunctionTok{ncol}\NormalTok{(x)))}

\NormalTok{  e0 }\OtherTok{=} \FloatTok{0.1}
\NormalTok{  e }\OtherTok{=}\NormalTok{ x[,}\DecValTok{1}\NormalTok{]}
\NormalTok{  s }\OtherTok{=}\NormalTok{ x[,}\DecValTok{2}\NormalTok{]}
\NormalTok{  es }\OtherTok{=}\NormalTok{ e0 }\SpecialCharTok{{-}}\NormalTok{ e}
\NormalTok{  p }\OtherTok{=}\NormalTok{ x[,}\DecValTok{3}\NormalTok{]}

\NormalTok{  resultDx[,}\DecValTok{1}\NormalTok{,}\DecValTok{1}\NormalTok{] }\OtherTok{=} \SpecialCharTok{{-}}\NormalTok{theta[}\DecValTok{1}\NormalTok{] }\SpecialCharTok{*}\NormalTok{ s }\SpecialCharTok{{-}}\NormalTok{ (theta[}\DecValTok{2}\NormalTok{] }\SpecialCharTok{+}\NormalTok{ theta[}\DecValTok{3}\NormalTok{])}
\NormalTok{  resultDx[,}\DecValTok{2}\NormalTok{,}\DecValTok{1}\NormalTok{] }\OtherTok{=} \SpecialCharTok{{-}}\NormalTok{theta[}\DecValTok{1}\NormalTok{] }\SpecialCharTok{*}\NormalTok{ e}

\NormalTok{  resultDx[,}\DecValTok{1}\NormalTok{,}\DecValTok{2}\NormalTok{] }\OtherTok{=} \SpecialCharTok{{-}}\NormalTok{theta[}\DecValTok{1}\NormalTok{] }\SpecialCharTok{*}\NormalTok{ s }\SpecialCharTok{{-}}\NormalTok{ theta[}\DecValTok{2}\NormalTok{]}
\NormalTok{  resultDx[,}\DecValTok{2}\NormalTok{,}\DecValTok{2}\NormalTok{] }\OtherTok{=} \SpecialCharTok{{-}}\NormalTok{theta[}\DecValTok{1}\NormalTok{] }\SpecialCharTok{*}\NormalTok{ e}

\NormalTok{  resultDx[,}\DecValTok{1}\NormalTok{,}\DecValTok{3}\NormalTok{] }\OtherTok{=}\NormalTok{ (}\SpecialCharTok{{-}}\NormalTok{theta[}\DecValTok{3}\NormalTok{])}

\NormalTok{  resultDx}
\NormalTok{\}}

\NormalTok{RMichaelisMentenReducedDtheta }\OtherTok{\textless{}{-}} \ControlFlowTok{function}\NormalTok{(theta, x, tvec) \{}
\NormalTok{  resultDtheta }\OtherTok{\textless{}{-}} \FunctionTok{array}\NormalTok{(}\DecValTok{0}\NormalTok{, }\FunctionTok{c}\NormalTok{(}\FunctionTok{nrow}\NormalTok{(x), }\FunctionTok{length}\NormalTok{(theta), }\FunctionTok{ncol}\NormalTok{(x)))}

\NormalTok{  e0 }\OtherTok{=} \FloatTok{0.1}
\NormalTok{  e }\OtherTok{=}\NormalTok{ x[,}\DecValTok{1}\NormalTok{]}
\NormalTok{  s }\OtherTok{=}\NormalTok{ x[,}\DecValTok{2}\NormalTok{]}
\NormalTok{  es }\OtherTok{=}\NormalTok{ e0 }\SpecialCharTok{{-}}\NormalTok{ e}
\NormalTok{  p }\OtherTok{=}\NormalTok{ x[,}\DecValTok{3}\NormalTok{]}

\NormalTok{  resultDtheta[,}\DecValTok{1}\NormalTok{,}\DecValTok{1}\NormalTok{] }\OtherTok{=} \SpecialCharTok{{-}}\NormalTok{e }\SpecialCharTok{*}\NormalTok{ s}
\NormalTok{  resultDtheta[,}\DecValTok{2}\NormalTok{,}\DecValTok{1}\NormalTok{] }\OtherTok{=}\NormalTok{ es}
\NormalTok{  resultDtheta[,}\DecValTok{3}\NormalTok{,}\DecValTok{1}\NormalTok{] }\OtherTok{=}\NormalTok{ es}

\NormalTok{  resultDtheta[,}\DecValTok{1}\NormalTok{,}\DecValTok{2}\NormalTok{] }\OtherTok{=} \SpecialCharTok{{-}}\NormalTok{e }\SpecialCharTok{*}\NormalTok{ s}
\NormalTok{  resultDtheta[,}\DecValTok{2}\NormalTok{,}\DecValTok{2}\NormalTok{] }\OtherTok{=}\NormalTok{ es}

\NormalTok{  resultDtheta[,}\DecValTok{3}\NormalTok{,}\DecValTok{3}\NormalTok{] }\OtherTok{=}\NormalTok{ es}

\NormalTok{  resultDtheta}
\NormalTok{\}}
\end{Highlighting}
\end{Shaded}

Define parameters and settings for the experiment and MAGI:

\begin{Shaded}
\begin{Highlighting}[]
\CommentTok{\# Observation times}
\NormalTok{obs.times }\OtherTok{\textless{}{-}} \FunctionTok{c}\NormalTok{(}\FloatTok{2.5}\NormalTok{, }\FloatTok{4.5}\NormalTok{, }\DecValTok{7}\NormalTok{, }\FloatTok{9.5}\NormalTok{, }\DecValTok{11}\NormalTok{, }\FloatTok{13.5}\NormalTok{, }\DecValTok{15}\NormalTok{, }\DecValTok{16}\NormalTok{, }\DecValTok{18}\NormalTok{, }\DecValTok{20}\NormalTok{,}
\FloatTok{21.5}\NormalTok{, }\DecValTok{24}\NormalTok{, }\DecValTok{27}\NormalTok{, }\FloatTok{29.5}\NormalTok{, }\FloatTok{32.5}\NormalTok{, }\FloatTok{35.5}\NormalTok{, }\FloatTok{39.5}\NormalTok{, }\DecValTok{45}\NormalTok{, }\DecValTok{55}\NormalTok{, }\DecValTok{69}\NormalTok{)}

\NormalTok{config }\OtherTok{\textless{}{-}} \FunctionTok{list}\NormalTok{(}
\AttributeTok{nobs =} \FunctionTok{length}\NormalTok{(obs.times),}
\AttributeTok{noise =} \FunctionTok{c}\NormalTok{(}\ConstantTok{NA}\NormalTok{, }\FloatTok{0.02}\NormalTok{, }\FloatTok{0.02}\NormalTok{), }
\AttributeTok{kernel =} \StringTok{"generalMatern"}\NormalTok{,}
\AttributeTok{seed =} \DecValTok{1}\NormalTok{,}
\AttributeTok{n.iter =} \DecValTok{5001}\NormalTok{,}
\AttributeTok{linfillspace =} \FloatTok{0.5}\NormalTok{, }
\AttributeTok{t.start =} \DecValTok{0}\NormalTok{,}
\AttributeTok{t.end =} \DecValTok{70}\NormalTok{,  }
\AttributeTok{phi =} \FunctionTok{cbind}\NormalTok{(}\FunctionTok{c}\NormalTok{(}\FloatTok{0.1}\NormalTok{, }\DecValTok{70}\NormalTok{), }\FunctionTok{c}\NormalTok{(}\DecValTok{1}\NormalTok{, }\DecValTok{30}\NormalTok{), }\FunctionTok{c}\NormalTok{(}\DecValTok{1}\NormalTok{, }\DecValTok{30}\NormalTok{)),}
\AttributeTok{modelName =} \StringTok{"Michaelis{-}Menten{-}Reduced"}
\NormalTok{)}

\NormalTok{pram.true }\OtherTok{\textless{}{-}} \FunctionTok{list}\NormalTok{(}
\AttributeTok{theta=}\FunctionTok{c}\NormalTok{(}\FloatTok{0.9}\NormalTok{, }\FloatTok{0.75}\NormalTok{, }\FloatTok{2.54}\NormalTok{),}
\AttributeTok{x0 =} \FunctionTok{c}\NormalTok{(}\FloatTok{0.1}\NormalTok{, }\DecValTok{1}\NormalTok{, }\DecValTok{0}\NormalTok{),}
\AttributeTok{phi =}\NormalTok{ config}\SpecialCharTok{$}\NormalTok{phi,}
\AttributeTok{sigma=}\NormalTok{config}\SpecialCharTok{$}\NormalTok{noise}
\NormalTok{)}
\end{Highlighting}
\end{Shaded}

Use a numerical solver to generate the true trajectories to simulate data and to compare with inference from MAGI:

\begin{Shaded}
\begin{Highlighting}[]
\NormalTok{times }\OtherTok{\textless{}{-}} \FunctionTok{seq}\NormalTok{(}\DecValTok{0}\NormalTok{,config}\SpecialCharTok{$}\NormalTok{t.end,}\AttributeTok{length=}\DecValTok{1001}\NormalTok{)}

\NormalTok{modelODE }\OtherTok{\textless{}{-}} \ControlFlowTok{function}\NormalTok{(t, state, parameters) \{}
\FunctionTok{list}\NormalTok{(}\FunctionTok{as.vector}\NormalTok{(}\FunctionTok{RMichaelisMentenReducedODE}\NormalTok{(parameters, }\FunctionTok{t}\NormalTok{(state), t)))}
\NormalTok{\}}

\NormalTok{xtrue }\OtherTok{\textless{}{-}}\NormalTok{ deSolve}\SpecialCharTok{::}\FunctionTok{ode}\NormalTok{(}\AttributeTok{y =}\NormalTok{ pram.true}\SpecialCharTok{$}\NormalTok{x0, }\AttributeTok{times =}\NormalTok{ times,}
\AttributeTok{func =}\NormalTok{ modelODE, }\AttributeTok{parms =}\NormalTok{ pram.true}\SpecialCharTok{$}\NormalTok{theta)}
\NormalTok{xtrue }\OtherTok{\textless{}{-}} \FunctionTok{data.frame}\NormalTok{(xtrue)}
\end{Highlighting}
\end{Shaded}

Additive measurement noise at the observation schedule to create simulated noisy data:

\begin{Shaded}
\begin{Highlighting}[]
\NormalTok{xtrueFunc }\OtherTok{\textless{}{-}} \FunctionTok{lapply}\NormalTok{(}\DecValTok{2}\SpecialCharTok{:}\FunctionTok{ncol}\NormalTok{(xtrue), }\ControlFlowTok{function}\NormalTok{(j)}
\FunctionTok{approxfun}\NormalTok{(xtrue[, }\StringTok{"time"}\NormalTok{], xtrue[, j]))}

\NormalTok{xsim }\OtherTok{\textless{}{-}} \FunctionTok{data.frame}\NormalTok{(}\AttributeTok{time =} \FunctionTok{round}\NormalTok{(obs.times }\SpecialCharTok{/}\NormalTok{ config}\SpecialCharTok{$}\NormalTok{linfillspace) }\SpecialCharTok{*}\NormalTok{ config}\SpecialCharTok{$}\NormalTok{linfillspace)}
\NormalTok{xsim }\OtherTok{\textless{}{-}} \FunctionTok{cbind}\NormalTok{(xsim, }\FunctionTok{sapply}\NormalTok{(xtrueFunc, }\ControlFlowTok{function}\NormalTok{(f) }\FunctionTok{f}\NormalTok{(xsim}\SpecialCharTok{$}\NormalTok{time)))}
\NormalTok{xtest }\OtherTok{\textless{}{-}}\NormalTok{ xsim}

\FunctionTok{set.seed}\NormalTok{(config}\SpecialCharTok{$}\NormalTok{seed)}
\ControlFlowTok{for}\NormalTok{(j }\ControlFlowTok{in} \DecValTok{1}\SpecialCharTok{:}\NormalTok{(}\FunctionTok{ncol}\NormalTok{(xsim)}\SpecialCharTok{{-}}\DecValTok{1}\NormalTok{))\{}
\NormalTok{  xsim[,}\DecValTok{1}\SpecialCharTok{+}\NormalTok{j] }\OtherTok{\textless{}{-}}\NormalTok{ xsim[,}\DecValTok{1}\SpecialCharTok{+}\NormalTok{j]}\SpecialCharTok{+}\FunctionTok{rnorm}\NormalTok{(}\FunctionTok{nrow}\NormalTok{(xsim), }\AttributeTok{sd=}\NormalTok{config}\SpecialCharTok{$}\NormalTok{noise[j])}
\NormalTok{\}}

\NormalTok{xsim.obs }\OtherTok{\textless{}{-}}\NormalTok{ xsim[}\FunctionTok{seq}\NormalTok{(}\DecValTok{1}\NormalTok{,}\FunctionTok{nrow}\NormalTok{(xsim), }\AttributeTok{length=}\NormalTok{config}\SpecialCharTok{$}\NormalTok{nobs),]}
\NormalTok{xsim.obs }\OtherTok{\textless{}{-}} \FunctionTok{rbind}\NormalTok{(}\FunctionTok{c}\NormalTok{(}\DecValTok{0}\NormalTok{, pram.true}\SpecialCharTok{$}\NormalTok{x0), xsim.obs)}
\end{Highlighting}
\end{Shaded}

Create the \texttt{odeModel} list, then confirm ODEs and derivatives are correct:

\begin{Shaded}
\begin{Highlighting}[]
\NormalTok{dynamicalModelList }\OtherTok{\textless{}{-}} \FunctionTok{list}\NormalTok{(}
\AttributeTok{fOde=}\NormalTok{RMichaelisMentenReducedODE,}
\AttributeTok{fOdeDx=}\NormalTok{RMichaelisMentenReducedDx,}
\AttributeTok{fOdeDtheta=}\NormalTok{RMichaelisMentenReducedDtheta,}
\AttributeTok{thetaLowerBound=}\FunctionTok{c}\NormalTok{(}\DecValTok{0}\NormalTok{,}\SpecialCharTok{{-}}\DecValTok{100}\NormalTok{,}\DecValTok{0}\NormalTok{),}
\AttributeTok{thetaUpperBound=}\FunctionTok{c}\NormalTok{(}\ConstantTok{Inf}\NormalTok{,}\ConstantTok{Inf}\NormalTok{,}\ConstantTok{Inf}\NormalTok{)}
\NormalTok{)}

\FunctionTok{testDynamicalModel}\NormalTok{(dynamicalModelList}\SpecialCharTok{$}\NormalTok{fOde, dynamicalModelList}\SpecialCharTok{$}\NormalTok{fOdeDx,}
\NormalTok{                   dynamicalModelList}\SpecialCharTok{$}\NormalTok{fOdeDtheta, }\StringTok{"dynamicalModelList"}\NormalTok{,}
\FunctionTok{data.matrix}\NormalTok{(xtest[,}\SpecialCharTok{{-}}\DecValTok{1}\NormalTok{]), pram.true}\SpecialCharTok{$}\NormalTok{theta, xtest}\SpecialCharTok{$}\NormalTok{time)}
\end{Highlighting}
\end{Shaded}

\begin{verbatim}
## dynamicalModelList model, with derivatives
## Dx and Dtheta appear to be correct
\end{verbatim}

\begin{verbatim}
## $testDx
## [1] TRUE
## 
## $testDtheta
## [1] TRUE
\end{verbatim}

Create inputs for MAGI:

\begin{Shaded}
\begin{Highlighting}[]
\CommentTok{\# Discretization set}
\NormalTok{xsim }\OtherTok{\textless{}{-}} \FunctionTok{setDiscretization}\NormalTok{(xsim.obs, }\AttributeTok{by =}\NormalTok{ config}\SpecialCharTok{$}\NormalTok{linfillspace)}

\CommentTok{\# Linearly interpolate to initialize X, use known initial conditions}
\NormalTok{xInitExogenous }\OtherTok{\textless{}{-}} \FunctionTok{data.matrix}\NormalTok{(xsim[,}\SpecialCharTok{{-}}\DecValTok{1}\NormalTok{])}
\ControlFlowTok{for}\NormalTok{ (j }\ControlFlowTok{in} \FunctionTok{c}\NormalTok{(}\DecValTok{2}\NormalTok{,}\DecValTok{3}\NormalTok{))\{}
\NormalTok{  xInitExogenous[, j] }\OtherTok{\textless{}{-}} \FunctionTok{approx}\NormalTok{(xsim.obs}\SpecialCharTok{$}\NormalTok{time, xsim.obs[,j}\SpecialCharTok{+}\DecValTok{1}\NormalTok{], xsim}\SpecialCharTok{$}\NormalTok{time)}\SpecialCharTok{$}\NormalTok{y}
\NormalTok{  idx }\OtherTok{\textless{}{-}} \FunctionTok{which}\NormalTok{(}\FunctionTok{is.na}\NormalTok{(xInitExogenous[, j]))}
\NormalTok{  xInitExogenous[idx, j] }\OtherTok{\textless{}{-}}\NormalTok{ xInitExogenous[idx[}\DecValTok{1}\NormalTok{] }\SpecialCharTok{{-}} \DecValTok{1}\NormalTok{, j]}
\NormalTok{\}}
\NormalTok{xInitExogenous[}\SpecialCharTok{{-}}\DecValTok{1}\NormalTok{, }\DecValTok{1}\NormalTok{] }\OtherTok{\textless{}{-}} \FloatTok{0.1}  \CommentTok{\# fill missing E component with 0.1}

\CommentTok{\# Use setSizeFactor=0 to fix initial conditions [E]=0.1, [S]=1, [P]=0}
\NormalTok{stepSizeFactor }\OtherTok{\textless{}{-}} \FunctionTok{rep}\NormalTok{(}\FloatTok{0.01}\NormalTok{, }\FunctionTok{nrow}\NormalTok{(xsim)}\SpecialCharTok{*}\FunctionTok{length}\NormalTok{(pram.true}\SpecialCharTok{$}\NormalTok{x0) }\SpecialCharTok{+}
\FunctionTok{length}\NormalTok{(dynamicalModelList}\SpecialCharTok{$}\NormalTok{thetaLowerBound) }\SpecialCharTok{+} \FunctionTok{length}\NormalTok{(pram.true}\SpecialCharTok{$}\NormalTok{x0))}
\ControlFlowTok{for}\NormalTok{(j }\ControlFlowTok{in} \DecValTok{1}\SpecialCharTok{:}\DecValTok{3}\NormalTok{)\{}
\ControlFlowTok{for}\NormalTok{(incre }\ControlFlowTok{in} \DecValTok{1}\SpecialCharTok{:}\DecValTok{1}\NormalTok{)\{}
\NormalTok{    stepSizeFactor[(j}\DecValTok{{-}1}\NormalTok{)}\SpecialCharTok{*}\FunctionTok{nrow}\NormalTok{(xsim) }\SpecialCharTok{+}\NormalTok{ incre] }\OtherTok{\textless{}{-}} \DecValTok{0}  
\NormalTok{  \}}
\NormalTok{\}}
\end{Highlighting}
\end{Shaded}

Run the MAGI method:

\begin{Shaded}
\begin{Highlighting}[]
\CommentTok{\# Since we have manually initialized the components, skip initial optimization}
\NormalTok{gpode }\OtherTok{\textless{}{-}} \FunctionTok{MagiSolver}\NormalTok{(xsim, dynamicalModelList, }
\AttributeTok{control =} \FunctionTok{list}\NormalTok{(}\AttributeTok{niterHmc=}\NormalTok{config}\SpecialCharTok{$}\NormalTok{n.iter, }\AttributeTok{stepSizeFactor =}\NormalTok{ stepSizeFactor,}
\AttributeTok{xInit =}\NormalTok{ xInitExogenous, }\AttributeTok{phi =}\NormalTok{ pram.true}\SpecialCharTok{$}\NormalTok{phi,}
\AttributeTok{sigma=}\NormalTok{config}\SpecialCharTok{$}\NormalTok{noise, }\AttributeTok{useFixedSigma=}\ConstantTok{TRUE}\NormalTok{,}
\AttributeTok{skipMissingComponentOptimization=}\ConstantTok{TRUE}\NormalTok{))}
\end{Highlighting}
\end{Shaded}

Inference for \(k_{cat}\) and \(K_M\) (posterior mean, 2.5 and 97.5 percentiles):

\begin{Shaded}
\begin{Highlighting}[]
\CommentTok{\# Add KM to parameters as a function of k1, k\_\{{-}1\}, k2}
\NormalTok{gpode}\SpecialCharTok{$}\NormalTok{theta }\OtherTok{\textless{}{-}} \FunctionTok{cbind}\NormalTok{(gpode}\SpecialCharTok{$}\NormalTok{theta, (gpode}\SpecialCharTok{$}\NormalTok{theta[,}\DecValTok{2}\NormalTok{]}\SpecialCharTok{+}\NormalTok{gpode}\SpecialCharTok{$}\NormalTok{theta[,}\DecValTok{3}\NormalTok{])}\SpecialCharTok{/}\NormalTok{gpode}\SpecialCharTok{$}\NormalTok{theta[,}\DecValTok{1}\NormalTok{])}
\NormalTok{pram.true}\SpecialCharTok{$}\NormalTok{theta }\OtherTok{\textless{}{-}} \FunctionTok{c}\NormalTok{(pram.true}\SpecialCharTok{$}\NormalTok{theta,}
\NormalTok{                     (pram.true}\SpecialCharTok{$}\NormalTok{theta[}\DecValTok{2}\NormalTok{]}\SpecialCharTok{+}\NormalTok{pram.true}\SpecialCharTok{$}\NormalTok{theta[}\DecValTok{3}\NormalTok{])}\SpecialCharTok{/}\NormalTok{pram.true}\SpecialCharTok{$}\NormalTok{theta[}\DecValTok{1}\NormalTok{])}

\NormalTok{par.table }\OtherTok{\textless{}{-}} \ControlFlowTok{function}\NormalTok{(res) \{}
\NormalTok{  par.est }\OtherTok{\textless{}{-}} \FunctionTok{apply}\NormalTok{(}\FunctionTok{cbind}\NormalTok{(res}\SpecialCharTok{$}\NormalTok{theta[,}\SpecialCharTok{{-}}\FunctionTok{c}\NormalTok{(}\DecValTok{1}\SpecialCharTok{:}\DecValTok{2}\NormalTok{)]), }\DecValTok{2}\NormalTok{,}
\ControlFlowTok{function}\NormalTok{(x) }\FunctionTok{c}\NormalTok{(}\FunctionTok{mean}\NormalTok{(x), }\FunctionTok{quantile}\NormalTok{(x, }\FloatTok{0.025}\NormalTok{), }\FunctionTok{quantile}\NormalTok{(x, }\FloatTok{0.975}\NormalTok{)))}
\FunctionTok{colnames}\NormalTok{(par.est) }\OtherTok{\textless{}{-}} \FunctionTok{c}\NormalTok{(}\StringTok{"k\_cat"}\NormalTok{, }\StringTok{"KM"}\NormalTok{)}
\FunctionTok{rownames}\NormalTok{(par.est) }\OtherTok{\textless{}{-}} \FunctionTok{c}\NormalTok{(}\StringTok{"Mean"}\NormalTok{, }\StringTok{"2.5\%"}\NormalTok{, }\StringTok{"97.5\%"}\NormalTok{)}
\FunctionTok{signif}\NormalTok{(par.est, }\DecValTok{3}\NormalTok{)}
\NormalTok{\}}

\FunctionTok{par.table}\NormalTok{(gpode)}
\end{Highlighting}
\end{Shaded}

\begin{verbatim}
##       k_cat   KM
## Mean   2.47 3.49
## 2.5%   1.93 2.58
## 97.5%  3.13 4.60
\end{verbatim}

\begin{figure}
\centering
\includegraphics{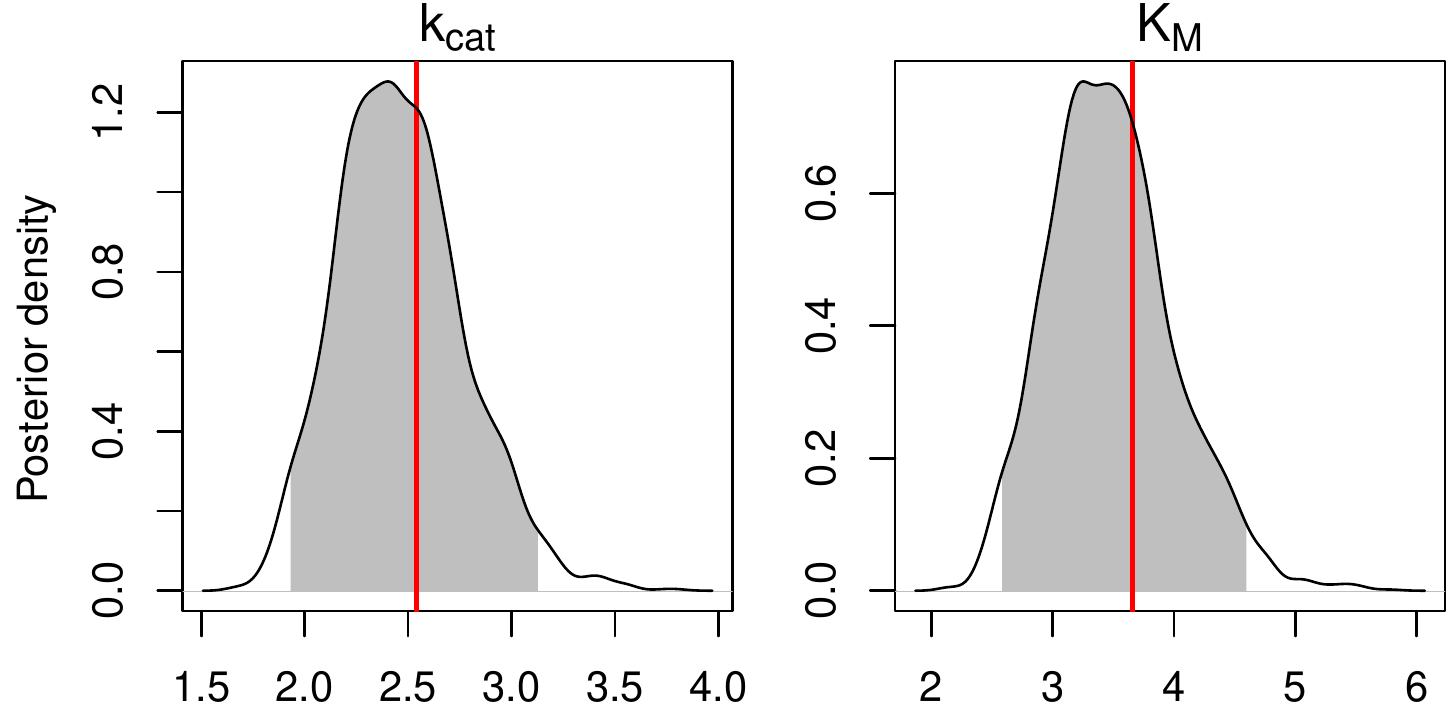}
\caption{\label{fig:mm-param} \footnotesize Bayesian posterior probability densities of \(k_{cat}\) and \(K_M\) for a sample dataset from the Michaelis-Menten model obtained by MAGI. The red vertical lines show the true parameter values used in the simulation. The shaded area represents the 95\% interval estimate of each parameter.}
\end{figure}

Plot the posterior densities of the parameters \(k_{cat}\) and \(K_M\), to produce Fig \ref{fig:mm-param}:

\begin{Shaded}
\begin{Highlighting}[]
\NormalTok{par.names }\OtherTok{\textless{}{-}} \FunctionTok{c}\NormalTok{( }\FunctionTok{expression}\NormalTok{(}\StringTok{\textquotesingle{}k\textquotesingle{}}\NormalTok{[}\StringTok{\textquotesingle{}cat\textquotesingle{}}\NormalTok{]), }\FunctionTok{expression}\NormalTok{(}\StringTok{\textquotesingle{}K\textquotesingle{}}\NormalTok{[}\StringTok{\textquotesingle{}M\textquotesingle{}}\NormalTok{]))}

\FunctionTok{par}\NormalTok{(}\AttributeTok{mfrow=}\FunctionTok{c}\NormalTok{(}\DecValTok{1}\NormalTok{,}\DecValTok{2}\NormalTok{))}
\ControlFlowTok{for}\NormalTok{ (ii }\ControlFlowTok{in} \DecValTok{3}\SpecialCharTok{:}\DecValTok{4}\NormalTok{) \{}
\ControlFlowTok{if}\NormalTok{ (ii }\SpecialCharTok{==} \DecValTok{3}\NormalTok{) }\FunctionTok{par}\NormalTok{(}\AttributeTok{oma=}\FunctionTok{c}\NormalTok{(}\DecValTok{0}\NormalTok{,}\FloatTok{1.5}\NormalTok{,}\DecValTok{0}\NormalTok{,}\DecValTok{0}\NormalTok{))}
\FunctionTok{par}\NormalTok{(}\AttributeTok{mar =} \FunctionTok{c}\NormalTok{(}\FloatTok{2.5}\NormalTok{, }\FloatTok{2.5}\NormalTok{, }\DecValTok{2}\NormalTok{, }\FloatTok{0.75}\NormalTok{))}
\NormalTok{  den }\OtherTok{\textless{}{-}} \FunctionTok{density}\NormalTok{(gpode}\SpecialCharTok{$}\NormalTok{theta[,ii])}
\FunctionTok{plot}\NormalTok{(den, }\AttributeTok{main=}\StringTok{\textquotesingle{}\textquotesingle{}}\NormalTok{, }\AttributeTok{xlab =} \StringTok{\textquotesingle{}\textquotesingle{}}\NormalTok{, }\AttributeTok{ylab =} \StringTok{\textquotesingle{}\textquotesingle{}}\NormalTok{, }\AttributeTok{type=}\StringTok{\textquotesingle{}n\textquotesingle{}}\NormalTok{)}

\NormalTok{  value1 }\OtherTok{\textless{}{-}} \FunctionTok{quantile}\NormalTok{(gpode}\SpecialCharTok{$}\NormalTok{theta[,ii], }\FloatTok{0.025}\NormalTok{)}
\NormalTok{  value2 }\OtherTok{\textless{}{-}} \FunctionTok{quantile}\NormalTok{(gpode}\SpecialCharTok{$}\NormalTok{theta[,ii], }\FloatTok{0.975}\NormalTok{)}

\NormalTok{  l }\OtherTok{\textless{}{-}} \FunctionTok{min}\NormalTok{(}\FunctionTok{which}\NormalTok{(den}\SpecialCharTok{$}\NormalTok{x }\SpecialCharTok{\textgreater{}=}\NormalTok{ value1))}
\NormalTok{  h }\OtherTok{\textless{}{-}} \FunctionTok{max}\NormalTok{(}\FunctionTok{which}\NormalTok{(den}\SpecialCharTok{$}\NormalTok{x }\SpecialCharTok{\textless{}}\NormalTok{ value2))}

\FunctionTok{polygon}\NormalTok{(}\FunctionTok{c}\NormalTok{(den}\SpecialCharTok{$}\NormalTok{x[}\FunctionTok{c}\NormalTok{(l, l}\SpecialCharTok{:}\NormalTok{h, h)]),}
\FunctionTok{c}\NormalTok{(}\DecValTok{0}\NormalTok{, den}\SpecialCharTok{$}\NormalTok{y[l}\SpecialCharTok{:}\NormalTok{h], }\DecValTok{0}\NormalTok{),}
\AttributeTok{col =} \StringTok{"grey75"}\NormalTok{, }\AttributeTok{border=}\ConstantTok{NA}\NormalTok{)}
\FunctionTok{abline}\NormalTok{(}\AttributeTok{v=}\NormalTok{pram.true}\SpecialCharTok{$}\NormalTok{theta[ii], }\AttributeTok{col=}\StringTok{\textquotesingle{}red\textquotesingle{}}\NormalTok{, }\AttributeTok{lwd =}\DecValTok{2}\NormalTok{)}

\FunctionTok{lines}\NormalTok{(den)}

\ControlFlowTok{if}\NormalTok{ (ii }\SpecialCharTok{==} \DecValTok{3}\NormalTok{) }\FunctionTok{mtext}\NormalTok{(}\AttributeTok{text=}\StringTok{\textquotesingle{}Posterior density\textquotesingle{}}\NormalTok{,}\AttributeTok{side=}\DecValTok{2}\NormalTok{,}\AttributeTok{line=}\DecValTok{0}\NormalTok{,}\AttributeTok{outer=}\ConstantTok{TRUE}\NormalTok{)}
\FunctionTok{mtext}\NormalTok{(par.names[ii}\DecValTok{{-}2}\NormalTok{], }\AttributeTok{cex=}\FloatTok{1.25}\NormalTok{)}
\NormalTok{\}}
\end{Highlighting}
\end{Shaded}

\hypertarget{model-for-lac-operon}{%
\subsection{\texorpdfstring{Model for \emph{lac} operon}{Model for lac operon}}\label{model-for-lac-operon}}

We begin by defining a function that codes the ODEs:

\begin{Shaded}
\begin{Highlighting}[]
\NormalTok{RlacOperonODE }\OtherTok{\textless{}{-}} \ControlFlowTok{function}\NormalTok{(theta, x, tvec) \{}
\NormalTok{  ri }\OtherTok{=}\NormalTok{ x[,}\DecValTok{1}\NormalTok{]}
\NormalTok{  i }\OtherTok{=}\NormalTok{ x[,}\DecValTok{2}\NormalTok{]}
\NormalTok{  lactose }\OtherTok{=}\NormalTok{ x[,}\DecValTok{3}\NormalTok{]}
\NormalTok{  ilactose }\OtherTok{=}\NormalTok{ x[,}\DecValTok{4}\NormalTok{]}
\NormalTok{  op }\OtherTok{=}\NormalTok{ x[,}\DecValTok{5}\NormalTok{]}
\NormalTok{  iop }\OtherTok{=}\NormalTok{ x[,}\DecValTok{6}\NormalTok{]}
\NormalTok{  rnap }\OtherTok{=}\NormalTok{ x[,}\DecValTok{7}\NormalTok{]}
\NormalTok{  rnapo }\OtherTok{=}\NormalTok{ x[,}\DecValTok{8}\NormalTok{]}
\NormalTok{  r }\OtherTok{=}\NormalTok{ x[,}\DecValTok{9}\NormalTok{]}
\NormalTok{  z }\OtherTok{=}\NormalTok{ x[,}\DecValTok{10}\NormalTok{]}

\NormalTok{  iconstant }\OtherTok{=} \FloatTok{1.0}
\NormalTok{  k }\OtherTok{=}\NormalTok{ theta}
\NormalTok{  resultdt }\OtherTok{\textless{}{-}}  \FunctionTok{array}\NormalTok{(}\DecValTok{0}\NormalTok{, }\FunctionTok{c}\NormalTok{(}\FunctionTok{nrow}\NormalTok{(x),}\FunctionTok{ncol}\NormalTok{(x)))}

\NormalTok{  resultdt[,}\DecValTok{1}\NormalTok{] }\OtherTok{=}\NormalTok{ k[}\DecValTok{2}\NormalTok{] }\SpecialCharTok{*}\NormalTok{ iconstant }\SpecialCharTok{{-}}\NormalTok{ k[}\DecValTok{13}\NormalTok{] }\SpecialCharTok{*}\NormalTok{ ri}
\NormalTok{  resultdt[,}\DecValTok{2}\NormalTok{] }\OtherTok{=}\NormalTok{ k[}\DecValTok{3}\NormalTok{] }\SpecialCharTok{*}\NormalTok{ ri }\SpecialCharTok{{-}}\NormalTok{ k[}\DecValTok{4}\NormalTok{] }\SpecialCharTok{*}\NormalTok{ i }\SpecialCharTok{*}\NormalTok{ lactose }\SpecialCharTok{+}\NormalTok{ k[}\DecValTok{5}\NormalTok{] }\SpecialCharTok{*}\NormalTok{ ilactose }\SpecialCharTok{{-}} 
\NormalTok{    k[}\DecValTok{6}\NormalTok{] }\SpecialCharTok{*}\NormalTok{ i }\SpecialCharTok{*}\NormalTok{ op }\SpecialCharTok{+}\NormalTok{ k[}\DecValTok{7}\NormalTok{] }\SpecialCharTok{*}\NormalTok{ iop }\SpecialCharTok{{-}}\NormalTok{ k[}\DecValTok{14}\NormalTok{] }\SpecialCharTok{*}\NormalTok{ i}
\NormalTok{  resultdt[,}\DecValTok{3}\NormalTok{] }\OtherTok{=}\NormalTok{ k[}\DecValTok{5}\NormalTok{] }\SpecialCharTok{*}\NormalTok{ ilactose }\SpecialCharTok{{-}}\NormalTok{ k[}\DecValTok{4}\NormalTok{] }\SpecialCharTok{*}\NormalTok{ i }\SpecialCharTok{*}\NormalTok{ lactose }\SpecialCharTok{+}\NormalTok{ k[}\DecValTok{15}\NormalTok{] }\SpecialCharTok{*}\NormalTok{ ilactose }\SpecialCharTok{{-}} 
\NormalTok{    k[}\DecValTok{12}\NormalTok{] }\SpecialCharTok{*}\NormalTok{ lactose }\SpecialCharTok{*}\NormalTok{ z}
\NormalTok{  resultdt[,}\DecValTok{4}\NormalTok{] }\OtherTok{=}\NormalTok{ k[}\DecValTok{4}\NormalTok{] }\SpecialCharTok{*}\NormalTok{ i }\SpecialCharTok{*}\NormalTok{ lactose }\SpecialCharTok{{-}}\NormalTok{ k[}\DecValTok{5}\NormalTok{] }\SpecialCharTok{*}\NormalTok{ ilactose }\SpecialCharTok{{-}}\NormalTok{ k[}\DecValTok{15}\NormalTok{] }\SpecialCharTok{*}\NormalTok{ ilactose}
\NormalTok{  resultdt[,}\DecValTok{5}\NormalTok{] }\OtherTok{=}\NormalTok{ k[}\DecValTok{7}\NormalTok{] }\SpecialCharTok{*}\NormalTok{ iop }\SpecialCharTok{{-}}\NormalTok{ k[}\DecValTok{6}\NormalTok{] }\SpecialCharTok{*}\NormalTok{ i }\SpecialCharTok{*}\NormalTok{ op }\SpecialCharTok{{-}}\NormalTok{ k[}\DecValTok{8}\NormalTok{] }\SpecialCharTok{*}\NormalTok{ op }\SpecialCharTok{*}\NormalTok{ rnap }\SpecialCharTok{+}\NormalTok{ (k[}\DecValTok{9}\NormalTok{] }\SpecialCharTok{+}\NormalTok{ k[}\DecValTok{10}\NormalTok{]) }\SpecialCharTok{*}\NormalTok{ rnapo}
\NormalTok{  resultdt[,}\DecValTok{6}\NormalTok{] }\OtherTok{=}\NormalTok{ k[}\DecValTok{6}\NormalTok{] }\SpecialCharTok{*}\NormalTok{ i }\SpecialCharTok{*}\NormalTok{ op }\SpecialCharTok{{-}}\NormalTok{ k[}\DecValTok{7}\NormalTok{] }\SpecialCharTok{*}\NormalTok{ iop}
\NormalTok{  resultdt[,}\DecValTok{7}\NormalTok{] }\OtherTok{=}\NormalTok{ (k[}\DecValTok{9}\NormalTok{] }\SpecialCharTok{+}\NormalTok{ k[}\DecValTok{10}\NormalTok{]) }\SpecialCharTok{*}\NormalTok{ rnapo }\SpecialCharTok{{-}}\NormalTok{ k[}\DecValTok{8}\NormalTok{] }\SpecialCharTok{*}\NormalTok{ op }\SpecialCharTok{*}\NormalTok{ rnap}
\NormalTok{  resultdt[,}\DecValTok{8}\NormalTok{] }\OtherTok{=}\NormalTok{ k[}\DecValTok{8}\NormalTok{] }\SpecialCharTok{*}\NormalTok{ op }\SpecialCharTok{*}\NormalTok{ rnap }\SpecialCharTok{{-}}\NormalTok{ (k[}\DecValTok{9}\NormalTok{]}\SpecialCharTok{+}\NormalTok{k[}\DecValTok{10}\NormalTok{]) }\SpecialCharTok{*}\NormalTok{ rnapo}
\NormalTok{  resultdt[,}\DecValTok{9}\NormalTok{] }\OtherTok{=}\NormalTok{ k[}\DecValTok{10}\NormalTok{] }\SpecialCharTok{*}\NormalTok{ rnapo }\SpecialCharTok{{-}}\NormalTok{ k[}\DecValTok{16}\NormalTok{] }\SpecialCharTok{*}\NormalTok{ r}
\NormalTok{  resultdt[,}\DecValTok{10}\NormalTok{] }\OtherTok{=}\NormalTok{ k[}\DecValTok{11}\NormalTok{]}\SpecialCharTok{*}\NormalTok{r }\SpecialCharTok{{-}}\NormalTok{ k[}\DecValTok{17}\NormalTok{] }\SpecialCharTok{*}\NormalTok{ z}

\NormalTok{  resultdt}
\NormalTok{\}}
\end{Highlighting}
\end{Shaded}

Next, we provide the gradients of the ODEs with respect to the system components \(X\) and the parameters \(\theta\).

\begin{Shaded}
\begin{Highlighting}[]
\NormalTok{RlacOperonDx }\OtherTok{\textless{}{-}} \ControlFlowTok{function}\NormalTok{(theta, x, tvec) \{}
\NormalTok{  resultDx }\OtherTok{\textless{}{-}} \FunctionTok{array}\NormalTok{(}\DecValTok{0}\NormalTok{, }\FunctionTok{c}\NormalTok{(}\FunctionTok{nrow}\NormalTok{(x), }\FunctionTok{ncol}\NormalTok{(x), }\FunctionTok{ncol}\NormalTok{(x)))}

\NormalTok{  ri }\OtherTok{=}\NormalTok{ x[,}\DecValTok{1}\NormalTok{]}
\NormalTok{  i }\OtherTok{=}\NormalTok{ x[,}\DecValTok{2}\NormalTok{]}
\NormalTok{  lactose }\OtherTok{=}\NormalTok{ x[,}\DecValTok{3}\NormalTok{]}
\NormalTok{  ilactose }\OtherTok{=}\NormalTok{ x[,}\DecValTok{4}\NormalTok{]}
\NormalTok{  op }\OtherTok{=}\NormalTok{ x[,}\DecValTok{5}\NormalTok{]}
\NormalTok{  iop }\OtherTok{=}\NormalTok{ x[,}\DecValTok{6}\NormalTok{]}
\NormalTok{  rnap }\OtherTok{=}\NormalTok{ x[,}\DecValTok{7}\NormalTok{]}
\NormalTok{  rnapo }\OtherTok{=}\NormalTok{ x[,}\DecValTok{8}\NormalTok{]}
\NormalTok{  r }\OtherTok{=}\NormalTok{ x[,}\DecValTok{9}\NormalTok{]}
\NormalTok{  z }\OtherTok{=}\NormalTok{ x[,}\DecValTok{10}\NormalTok{]}

\NormalTok{  iconstant }\OtherTok{=} \FloatTok{1.0}
\NormalTok{  k }\OtherTok{=}\NormalTok{ theta}

\NormalTok{  resultDx[,}\DecValTok{1}\NormalTok{,}\DecValTok{1}\NormalTok{] }\OtherTok{=}\NormalTok{ (}\SpecialCharTok{{-}}\NormalTok{k[}\DecValTok{13}\NormalTok{])}

\NormalTok{  resultDx[,}\DecValTok{1}\NormalTok{,}\DecValTok{2}\NormalTok{] }\OtherTok{=}\NormalTok{ (k[}\DecValTok{3}\NormalTok{])}
\NormalTok{  resultDx[,}\DecValTok{2}\NormalTok{,}\DecValTok{2}\NormalTok{] }\OtherTok{=} \SpecialCharTok{{-}}\NormalTok{k[}\DecValTok{4}\NormalTok{] }\SpecialCharTok{*}\NormalTok{ lactose }\SpecialCharTok{{-}}\NormalTok{ k[}\DecValTok{6}\NormalTok{] }\SpecialCharTok{*}\NormalTok{ op }\SpecialCharTok{{-}}\NormalTok{ k[}\DecValTok{14}\NormalTok{]}
\NormalTok{  resultDx[,}\DecValTok{3}\NormalTok{,}\DecValTok{2}\NormalTok{] }\OtherTok{=} \SpecialCharTok{{-}}\NormalTok{k[}\DecValTok{4}\NormalTok{] }\SpecialCharTok{*}\NormalTok{ i}
\NormalTok{  resultDx[,}\DecValTok{4}\NormalTok{,}\DecValTok{2}\NormalTok{] }\OtherTok{=}\NormalTok{ (k[}\DecValTok{5}\NormalTok{])}
\NormalTok{  resultDx[,}\DecValTok{5}\NormalTok{,}\DecValTok{2}\NormalTok{] }\OtherTok{=} \SpecialCharTok{{-}}\NormalTok{k[}\DecValTok{6}\NormalTok{] }\SpecialCharTok{*}\NormalTok{ i}
\NormalTok{  resultDx[,}\DecValTok{6}\NormalTok{,}\DecValTok{2}\NormalTok{] }\OtherTok{=}\NormalTok{ (k[}\DecValTok{7}\NormalTok{])}

\NormalTok{  resultDx[,}\DecValTok{2}\NormalTok{,}\DecValTok{3}\NormalTok{] }\OtherTok{=} \SpecialCharTok{{-}}\NormalTok{k[}\DecValTok{4}\NormalTok{]}\SpecialCharTok{*}\NormalTok{lactose}
\NormalTok{  resultDx[,}\DecValTok{3}\NormalTok{,}\DecValTok{3}\NormalTok{] }\OtherTok{=} \SpecialCharTok{{-}}\NormalTok{k[}\DecValTok{4}\NormalTok{]}\SpecialCharTok{*}\NormalTok{i }\SpecialCharTok{{-}}\NormalTok{ k[}\DecValTok{12}\NormalTok{]}\SpecialCharTok{*}\NormalTok{z}
\NormalTok{  resultDx[,}\DecValTok{4}\NormalTok{,}\DecValTok{3}\NormalTok{] }\OtherTok{=}\NormalTok{ (k[}\DecValTok{5}\NormalTok{] }\SpecialCharTok{+}\NormalTok{ k[}\DecValTok{15}\NormalTok{])}
\NormalTok{  resultDx[,}\DecValTok{10}\NormalTok{,}\DecValTok{3}\NormalTok{] }\OtherTok{=} \SpecialCharTok{{-}}\NormalTok{k[}\DecValTok{12}\NormalTok{]}\SpecialCharTok{*}\NormalTok{lactose}

\NormalTok{  resultDx[,}\DecValTok{2}\NormalTok{,}\DecValTok{4}\NormalTok{] }\OtherTok{=}\NormalTok{ k[}\DecValTok{4}\NormalTok{]}\SpecialCharTok{*}\NormalTok{lactose}
\NormalTok{  resultDx[,}\DecValTok{3}\NormalTok{,}\DecValTok{4}\NormalTok{] }\OtherTok{=}\NormalTok{ k[}\DecValTok{4}\NormalTok{]}\SpecialCharTok{*}\NormalTok{i}
\NormalTok{  resultDx[,}\DecValTok{4}\NormalTok{,}\DecValTok{4}\NormalTok{] }\OtherTok{=}\NormalTok{ (}\SpecialCharTok{{-}}\NormalTok{k[}\DecValTok{5}\NormalTok{] }\SpecialCharTok{{-}}\NormalTok{ k[}\DecValTok{15}\NormalTok{])}

\NormalTok{  resultDx[,}\DecValTok{2}\NormalTok{,}\DecValTok{5}\NormalTok{] }\OtherTok{=} \SpecialCharTok{{-}}\NormalTok{k[}\DecValTok{6}\NormalTok{] }\SpecialCharTok{*}\NormalTok{ op}
\NormalTok{  resultDx[,}\DecValTok{5}\NormalTok{,}\DecValTok{5}\NormalTok{] }\OtherTok{=} \SpecialCharTok{{-}}\NormalTok{k[}\DecValTok{6}\NormalTok{] }\SpecialCharTok{*}\NormalTok{ i }\SpecialCharTok{{-}}\NormalTok{ k[}\DecValTok{8}\NormalTok{]}\SpecialCharTok{*}\NormalTok{rnap}
\NormalTok{  resultDx[,}\DecValTok{6}\NormalTok{,}\DecValTok{5}\NormalTok{] }\OtherTok{=}\NormalTok{ (k[}\DecValTok{7}\NormalTok{])}
\NormalTok{  resultDx[,}\DecValTok{7}\NormalTok{,}\DecValTok{5}\NormalTok{] }\OtherTok{=} \SpecialCharTok{{-}}\NormalTok{k[}\DecValTok{8}\NormalTok{]}\SpecialCharTok{*}\NormalTok{op}
\NormalTok{  resultDx[,}\DecValTok{8}\NormalTok{,}\DecValTok{5}\NormalTok{] }\OtherTok{=}\NormalTok{ (k[}\DecValTok{9}\NormalTok{] }\SpecialCharTok{+}\NormalTok{ k[}\DecValTok{10}\NormalTok{])}

\NormalTok{  resultDx[,}\DecValTok{2}\NormalTok{,}\DecValTok{6}\NormalTok{] }\OtherTok{=}\NormalTok{ k[}\DecValTok{6}\NormalTok{] }\SpecialCharTok{*}\NormalTok{ op}
\NormalTok{  resultDx[,}\DecValTok{5}\NormalTok{,}\DecValTok{6}\NormalTok{] }\OtherTok{=}\NormalTok{ k[}\DecValTok{6}\NormalTok{] }\SpecialCharTok{*}\NormalTok{ i}
\NormalTok{  resultDx[,}\DecValTok{6}\NormalTok{,}\DecValTok{6}\NormalTok{] }\OtherTok{=}\NormalTok{ (}\SpecialCharTok{{-}}\NormalTok{k[}\DecValTok{7}\NormalTok{])}

\NormalTok{  resultDx[,}\DecValTok{5}\NormalTok{,}\DecValTok{7}\NormalTok{] }\OtherTok{=} \SpecialCharTok{{-}}\NormalTok{k[}\DecValTok{8}\NormalTok{]}\SpecialCharTok{*}\NormalTok{rnap}
\NormalTok{  resultDx[,}\DecValTok{7}\NormalTok{,}\DecValTok{7}\NormalTok{] }\OtherTok{=} \SpecialCharTok{{-}}\NormalTok{k[}\DecValTok{8}\NormalTok{]}\SpecialCharTok{*}\NormalTok{op}
\NormalTok{  resultDx[,}\DecValTok{8}\NormalTok{,}\DecValTok{7}\NormalTok{] }\OtherTok{=}\NormalTok{ (k[}\DecValTok{9}\NormalTok{] }\SpecialCharTok{+}\NormalTok{ k[}\DecValTok{10}\NormalTok{])}

\NormalTok{  resultDx[,}\DecValTok{5}\NormalTok{,}\DecValTok{8}\NormalTok{] }\OtherTok{=}\NormalTok{ k[}\DecValTok{8}\NormalTok{] }\SpecialCharTok{*}\NormalTok{ rnap}
\NormalTok{  resultDx[,}\DecValTok{7}\NormalTok{,}\DecValTok{8}\NormalTok{] }\OtherTok{=}\NormalTok{ k[}\DecValTok{8}\NormalTok{] }\SpecialCharTok{*}\NormalTok{ op}
\NormalTok{  resultDx[,}\DecValTok{8}\NormalTok{,}\DecValTok{8}\NormalTok{] }\OtherTok{=}\NormalTok{ (}\SpecialCharTok{{-}}\NormalTok{(k[}\DecValTok{9}\NormalTok{] }\SpecialCharTok{+}\NormalTok{ k[}\DecValTok{10}\NormalTok{]))}

\NormalTok{  resultDx[,}\DecValTok{8}\NormalTok{,}\DecValTok{9}\NormalTok{] }\OtherTok{=}\NormalTok{ (k[}\DecValTok{10}\NormalTok{])}
\NormalTok{  resultDx[,}\DecValTok{9}\NormalTok{,}\DecValTok{9}\NormalTok{] }\OtherTok{=}\NormalTok{ (}\SpecialCharTok{{-}}\NormalTok{k[}\DecValTok{16}\NormalTok{])}

\NormalTok{  resultDx[,}\DecValTok{9}\NormalTok{,}\DecValTok{10}\NormalTok{] }\OtherTok{=}\NormalTok{ (k[}\DecValTok{11}\NormalTok{])}
\NormalTok{  resultDx[,}\DecValTok{10}\NormalTok{,}\DecValTok{10}\NormalTok{] }\OtherTok{=}\NormalTok{ (}\SpecialCharTok{{-}}\NormalTok{k[}\DecValTok{17}\NormalTok{])}

\NormalTok{  resultDx}
\NormalTok{\}}

\NormalTok{RlacOperonDtheta }\OtherTok{\textless{}{-}} \ControlFlowTok{function}\NormalTok{(theta, x, tvec) \{}
\NormalTok{  resultDtheta }\OtherTok{\textless{}{-}} \FunctionTok{array}\NormalTok{(}\DecValTok{0}\NormalTok{, }\FunctionTok{c}\NormalTok{(}\FunctionTok{nrow}\NormalTok{(x), }\FunctionTok{length}\NormalTok{(theta), }\FunctionTok{ncol}\NormalTok{(x)))}

\NormalTok{  ri }\OtherTok{=}\NormalTok{ x[,}\DecValTok{1}\NormalTok{]}
\NormalTok{  i }\OtherTok{=}\NormalTok{ x[,}\DecValTok{2}\NormalTok{]}
\NormalTok{  lactose }\OtherTok{=}\NormalTok{ x[,}\DecValTok{3}\NormalTok{]}
\NormalTok{  ilactose }\OtherTok{=}\NormalTok{ x[,}\DecValTok{4}\NormalTok{]}
\NormalTok{  op }\OtherTok{=}\NormalTok{ x[,}\DecValTok{5}\NormalTok{]}
\NormalTok{  iop }\OtherTok{=}\NormalTok{ x[,}\DecValTok{6}\NormalTok{]}
\NormalTok{  rnap }\OtherTok{=}\NormalTok{ x[,}\DecValTok{7}\NormalTok{]}
\NormalTok{  rnapo }\OtherTok{=}\NormalTok{ x[,}\DecValTok{8}\NormalTok{]}
\NormalTok{  r }\OtherTok{=}\NormalTok{ x[,}\DecValTok{9}\NormalTok{]}
\NormalTok{  z }\OtherTok{=}\NormalTok{ x[,}\DecValTok{10}\NormalTok{]}

\NormalTok{  iconstant }\OtherTok{=} \FloatTok{1.0}
\NormalTok{  k }\OtherTok{=}\NormalTok{ theta}

\NormalTok{  resultDtheta[,}\DecValTok{1}\NormalTok{,}\DecValTok{1}\NormalTok{] }\OtherTok{=}\NormalTok{ (}\DecValTok{0}\NormalTok{)}
\NormalTok{  resultDtheta[,}\DecValTok{2}\NormalTok{,}\DecValTok{1}\NormalTok{] }\OtherTok{=}\NormalTok{ (iconstant)}
\NormalTok{  resultDtheta[,}\DecValTok{13}\NormalTok{,}\DecValTok{1}\NormalTok{] }\OtherTok{=} \SpecialCharTok{{-}}\NormalTok{ri}

\NormalTok{  resultDtheta[,}\DecValTok{3}\NormalTok{,}\DecValTok{2}\NormalTok{] }\OtherTok{=}\NormalTok{ ri}
\NormalTok{  resultDtheta[,}\DecValTok{4}\NormalTok{,}\DecValTok{2}\NormalTok{] }\OtherTok{=} \SpecialCharTok{{-}}\NormalTok{i}\SpecialCharTok{*}\NormalTok{lactose}
\NormalTok{  resultDtheta[,}\DecValTok{5}\NormalTok{,}\DecValTok{2}\NormalTok{] }\OtherTok{=}\NormalTok{ ilactose}
\NormalTok{  resultDtheta[,}\DecValTok{6}\NormalTok{,}\DecValTok{2}\NormalTok{] }\OtherTok{=} \SpecialCharTok{{-}}\NormalTok{ i}\SpecialCharTok{*}\NormalTok{op}
\NormalTok{  resultDtheta[,}\DecValTok{7}\NormalTok{,}\DecValTok{2}\NormalTok{] }\OtherTok{=}\NormalTok{ iop}
\NormalTok{  resultDtheta[,}\DecValTok{14}\NormalTok{,}\DecValTok{2}\NormalTok{] }\OtherTok{=} \SpecialCharTok{{-}}\NormalTok{i}

\NormalTok{  resultDtheta[,}\DecValTok{5}\NormalTok{,}\DecValTok{3}\NormalTok{] }\OtherTok{=}\NormalTok{ ilactose}
\NormalTok{  resultDtheta[,}\DecValTok{4}\NormalTok{,}\DecValTok{3}\NormalTok{] }\OtherTok{=} \SpecialCharTok{{-}}\NormalTok{i}\SpecialCharTok{*}\NormalTok{lactose}
\NormalTok{  resultDtheta[,}\DecValTok{15}\NormalTok{,}\DecValTok{3}\NormalTok{] }\OtherTok{=}\NormalTok{ ilactose}
\NormalTok{  resultDtheta[,}\DecValTok{12}\NormalTok{,}\DecValTok{3}\NormalTok{] }\OtherTok{=} \SpecialCharTok{{-}}\NormalTok{lactose}\SpecialCharTok{*}\NormalTok{z}

\NormalTok{  resultDtheta[,}\DecValTok{4}\NormalTok{,}\DecValTok{4}\NormalTok{] }\OtherTok{=}\NormalTok{ i}\SpecialCharTok{*}\NormalTok{lactose}
\NormalTok{  resultDtheta[,}\DecValTok{5}\NormalTok{,}\DecValTok{4}\NormalTok{] }\OtherTok{=} \SpecialCharTok{{-}}\NormalTok{ilactose}
\NormalTok{  resultDtheta[,}\DecValTok{15}\NormalTok{,}\DecValTok{4}\NormalTok{] }\OtherTok{=} \SpecialCharTok{{-}}\NormalTok{ilactose}

\NormalTok{  resultDtheta[,}\DecValTok{7}\NormalTok{,}\DecValTok{5}\NormalTok{] }\OtherTok{=}\NormalTok{ iop}
\NormalTok{  resultDtheta[,}\DecValTok{6}\NormalTok{,}\DecValTok{5}\NormalTok{] }\OtherTok{=} \SpecialCharTok{{-}}\NormalTok{i}\SpecialCharTok{*}\NormalTok{op}
\NormalTok{  resultDtheta[,}\DecValTok{8}\NormalTok{,}\DecValTok{5}\NormalTok{] }\OtherTok{=} \SpecialCharTok{{-}}\NormalTok{op}\SpecialCharTok{*}\NormalTok{rnap}
\NormalTok{  resultDtheta[,}\DecValTok{9}\NormalTok{,}\DecValTok{5}\NormalTok{] }\OtherTok{=}\NormalTok{ rnapo}
\NormalTok{  resultDtheta[,}\DecValTok{10}\NormalTok{,}\DecValTok{5}\NormalTok{] }\OtherTok{=}\NormalTok{ rnapo}

\NormalTok{  resultDtheta[,}\DecValTok{6}\NormalTok{,}\DecValTok{6}\NormalTok{] }\OtherTok{=}\NormalTok{ i}\SpecialCharTok{*}\NormalTok{op}
\NormalTok{  resultDtheta[,}\DecValTok{7}\NormalTok{,}\DecValTok{6}\NormalTok{] }\OtherTok{=} \SpecialCharTok{{-}}\NormalTok{iop}

\NormalTok{  resultDtheta[,}\DecValTok{9}\NormalTok{,}\DecValTok{7}\NormalTok{] }\OtherTok{=}\NormalTok{ rnapo}
\NormalTok{  resultDtheta[,}\DecValTok{10}\NormalTok{,}\DecValTok{7}\NormalTok{] }\OtherTok{=}\NormalTok{ rnapo}
\NormalTok{  resultDtheta[,}\DecValTok{8}\NormalTok{,}\DecValTok{7}\NormalTok{] }\OtherTok{=} \SpecialCharTok{{-}}\NormalTok{op}\SpecialCharTok{*}\NormalTok{rnap}

\NormalTok{  resultDtheta[,}\DecValTok{8}\NormalTok{,}\DecValTok{8}\NormalTok{] }\OtherTok{=}\NormalTok{ op}\SpecialCharTok{*}\NormalTok{rnap}
\NormalTok{  resultDtheta[,}\DecValTok{9}\NormalTok{,}\DecValTok{8}\NormalTok{] }\OtherTok{=} \SpecialCharTok{{-}}\NormalTok{rnapo}
\NormalTok{  resultDtheta[,}\DecValTok{10}\NormalTok{,}\DecValTok{8}\NormalTok{] }\OtherTok{=} \SpecialCharTok{{-}}\NormalTok{rnapo}

\NormalTok{  resultDtheta[,}\DecValTok{10}\NormalTok{,}\DecValTok{9}\NormalTok{] }\OtherTok{=}\NormalTok{ rnapo}
\NormalTok{  resultDtheta[,}\DecValTok{16}\NormalTok{,}\DecValTok{9}\NormalTok{] }\OtherTok{=} \SpecialCharTok{{-}}\NormalTok{r}

\NormalTok{  resultDtheta[,}\DecValTok{11}\NormalTok{,}\DecValTok{10}\NormalTok{] }\OtherTok{=}\NormalTok{ r}
\NormalTok{  resultDtheta[,}\DecValTok{17}\NormalTok{,}\DecValTok{10}\NormalTok{] }\OtherTok{=} \SpecialCharTok{{-}}\NormalTok{z}

\NormalTok{  resultDtheta}
\NormalTok{\}}
\end{Highlighting}
\end{Shaded}

Define parameters and settings for the experiment and MAGI:

\begin{Shaded}
\begin{Highlighting}[]
\NormalTok{noisefac }\OtherTok{\textless{}{-}} \FloatTok{0.05}
\NormalTok{config }\OtherTok{\textless{}{-}} \FunctionTok{list}\NormalTok{(}
\AttributeTok{noise =} \FunctionTok{c}\NormalTok{(}\FloatTok{0.0236194228362474}\NormalTok{, }\FloatTok{0.984379168607761}\NormalTok{, }\FloatTok{0.040561105491602}\NormalTok{, }\FloatTok{0.19224800630503}\NormalTok{,}
\FloatTok{0.000104872420893985}\NormalTok{, }\FloatTok{0.395628293932429}\NormalTok{, }\FloatTok{99.3980508395468}\NormalTok{, }\FloatTok{0.0263189487352539}\NormalTok{,}
\FloatTok{0.00631925202022132}\NormalTok{, }\FloatTok{0.00036869187211123}\NormalTok{)}\SpecialCharTok{*}\NormalTok{noisefac,}
\CommentTok{\# noise level is minimum of each component * noisefac}
\AttributeTok{kernel =} \StringTok{"generalMatern"}\NormalTok{,}
\AttributeTok{seed =} \DecValTok{115767108}\NormalTok{,}
\AttributeTok{hmcSteps =} \DecValTok{500}\NormalTok{,}
\AttributeTok{niterHmc =} \DecValTok{20001}\NormalTok{,}
\AttributeTok{fillinterval =} \DecValTok{15}\NormalTok{,}
\AttributeTok{t.start =} \DecValTok{1}\NormalTok{,}
\AttributeTok{t.end =} \DecValTok{1201}\NormalTok{,}
\AttributeTok{modelName =} \StringTok{"lac{-}operon"}\NormalTok{,}
\AttributeTok{obs.times =} \FunctionTok{c}\NormalTok{(}\FunctionTok{seq}\NormalTok{(}\DecValTok{1}\NormalTok{,}\DecValTok{361}\NormalTok{,}\AttributeTok{by=}\DecValTok{15}\NormalTok{), }\FunctionTok{seq}\NormalTok{(}\DecValTok{391}\NormalTok{,}\DecValTok{601}\NormalTok{,}\AttributeTok{by=}\DecValTok{30}\NormalTok{), }\DecValTok{901}\NormalTok{, }\DecValTok{1201}\NormalTok{)}
\NormalTok{)}
\NormalTok{config}\SpecialCharTok{$}\NormalTok{nobs }\OtherTok{=} \FunctionTok{length}\NormalTok{(config}\SpecialCharTok{$}\NormalTok{obs.times)}

\NormalTok{pram.true }\OtherTok{\textless{}{-}} \FunctionTok{list}\NormalTok{(}
\AttributeTok{theta=}\FunctionTok{c}\NormalTok{(}\DecValTok{1}\NormalTok{, }\FloatTok{0.02}\NormalTok{, }\FloatTok{0.1}\NormalTok{, }\FloatTok{0.005}\NormalTok{, }\FloatTok{0.1}\NormalTok{, }\DecValTok{1}\NormalTok{, }\FloatTok{0.01}\NormalTok{, }\FloatTok{0.1}\NormalTok{, }\FloatTok{0.01}\NormalTok{, }\FloatTok{0.03}\NormalTok{,}
\FloatTok{0.1}\NormalTok{, }\FloatTok{0.001}\NormalTok{, }\FloatTok{0.01}\NormalTok{, }\FloatTok{0.002}\NormalTok{, }\FloatTok{0.002}\NormalTok{, }\FloatTok{0.01}\NormalTok{, }\FloatTok{0.001}\NormalTok{),}
\AttributeTok{x0 =} \FunctionTok{c}\NormalTok{(}\DecValTok{0}\NormalTok{, }\DecValTok{50}\NormalTok{, }\DecValTok{1000}\NormalTok{, }\DecValTok{0}\NormalTok{, }\DecValTok{1}\NormalTok{, }\DecValTok{0}\NormalTok{, }\DecValTok{100}\NormalTok{, }\DecValTok{0}\NormalTok{, }\DecValTok{0}\NormalTok{, }\DecValTok{0}\NormalTok{),}
\AttributeTok{sigma=}\NormalTok{config}\SpecialCharTok{$}\NormalTok{noise}
\NormalTok{)}
\end{Highlighting}
\end{Shaded}

Use a numerical solver to generate the true trajectories to simulate data and to compare with inference from MAGI:

\begin{Shaded}
\begin{Highlighting}[]
\NormalTok{times }\OtherTok{\textless{}{-}} \FunctionTok{seq}\NormalTok{(}\DecValTok{0}\NormalTok{,config}\SpecialCharTok{$}\NormalTok{t.end,}\AttributeTok{by =} \FloatTok{0.01}\NormalTok{)}

\NormalTok{modelODE }\OtherTok{\textless{}{-}} \ControlFlowTok{function}\NormalTok{(t, state, parameters) \{}
\FunctionTok{list}\NormalTok{(}\FunctionTok{as.vector}\NormalTok{(}\FunctionTok{RlacOperonODE}\NormalTok{(parameters, }\FunctionTok{t}\NormalTok{(state), t)))}
\NormalTok{\}}

\NormalTok{xtrue }\OtherTok{\textless{}{-}}\NormalTok{ deSolve}\SpecialCharTok{::}\FunctionTok{ode}\NormalTok{(}\AttributeTok{y =}\NormalTok{ pram.true}\SpecialCharTok{$}\NormalTok{x0, }\AttributeTok{times =}\NormalTok{ times,}
\AttributeTok{func =}\NormalTok{ modelODE, }\AttributeTok{parms =}\NormalTok{ pram.true}\SpecialCharTok{$}\NormalTok{theta)}
\NormalTok{xtrue }\OtherTok{\textless{}{-}} \FunctionTok{data.frame}\NormalTok{(xtrue)}
\end{Highlighting}
\end{Shaded}

Additive measurement noise at the observation schedule to create simulated noisy data:

\begin{Shaded}
\begin{Highlighting}[]
\NormalTok{xtrueFunc }\OtherTok{\textless{}{-}} \FunctionTok{lapply}\NormalTok{(}\DecValTok{2}\SpecialCharTok{:}\FunctionTok{ncol}\NormalTok{(xtrue), }\ControlFlowTok{function}\NormalTok{(j)}
\FunctionTok{approxfun}\NormalTok{(xtrue[, }\StringTok{"time"}\NormalTok{], xtrue[, j]))}

\NormalTok{xsim }\OtherTok{\textless{}{-}} \FunctionTok{data.frame}\NormalTok{(}\AttributeTok{time =}\NormalTok{ config}\SpecialCharTok{$}\NormalTok{obs.times)}
\NormalTok{xsim }\OtherTok{\textless{}{-}} \FunctionTok{cbind}\NormalTok{(xsim, }\FunctionTok{sapply}\NormalTok{(xtrueFunc, }\ControlFlowTok{function}\NormalTok{(f) }\FunctionTok{f}\NormalTok{(xsim}\SpecialCharTok{$}\NormalTok{time)))}

\FunctionTok{set.seed}\NormalTok{(config}\SpecialCharTok{$}\NormalTok{seed)}
\ControlFlowTok{for}\NormalTok{(j }\ControlFlowTok{in} \DecValTok{1}\SpecialCharTok{:}\NormalTok{(}\FunctionTok{ncol}\NormalTok{(xsim)}\SpecialCharTok{{-}}\DecValTok{1}\NormalTok{))\{}
\NormalTok{  xsim[,}\DecValTok{1}\SpecialCharTok{+}\NormalTok{j] }\OtherTok{\textless{}{-}}\NormalTok{ xsim[,}\DecValTok{1}\SpecialCharTok{+}\NormalTok{j]}\SpecialCharTok{+}\FunctionTok{rnorm}\NormalTok{(}\FunctionTok{nrow}\NormalTok{(xsim), }\AttributeTok{sd=}\NormalTok{config}\SpecialCharTok{$}\NormalTok{noise[j])}
\NormalTok{\}}
\NormalTok{xsim.obs }\OtherTok{\textless{}{-}}\NormalTok{ xsim}
\end{Highlighting}
\end{Shaded}

Create the \texttt{odeModel} list, then confirm ODEs and derivatives are correct:

\begin{Shaded}
\begin{Highlighting}[]
\NormalTok{dynamicalModelList }\OtherTok{\textless{}{-}} \FunctionTok{list}\NormalTok{(}
\AttributeTok{fOde=}\NormalTok{RlacOperonODE,}
\AttributeTok{fOdeDx=}\NormalTok{RlacOperonDx,}
\AttributeTok{fOdeDtheta=}\NormalTok{RlacOperonDtheta,}
\AttributeTok{thetaLowerBound=}\FunctionTok{rep}\NormalTok{(}\DecValTok{0}\NormalTok{, }\DecValTok{17}\NormalTok{),}
\AttributeTok{thetaUpperBound=}\FunctionTok{rep}\NormalTok{(}\ConstantTok{Inf}\NormalTok{, }\DecValTok{17}\NormalTok{),}
\AttributeTok{name=}\StringTok{"lac{-}operon"}
\NormalTok{)}

\FunctionTok{testDynamicalModel}\NormalTok{(dynamicalModelList}\SpecialCharTok{$}\NormalTok{fOde, dynamicalModelList}\SpecialCharTok{$}\NormalTok{fOdeDx,}
\NormalTok{                   dynamicalModelList}\SpecialCharTok{$}\NormalTok{fOdeDtheta, }\StringTok{"dynamicalModelList"}\NormalTok{,}
\FunctionTok{data.matrix}\NormalTok{(xsim.obs[,}\SpecialCharTok{{-}}\DecValTok{1}\NormalTok{]), pram.true}\SpecialCharTok{$}\NormalTok{theta, xsim}\SpecialCharTok{$}\NormalTok{time)}
\end{Highlighting}
\end{Shaded}

\begin{verbatim}
## dynamicalModelList model, with derivatives
## Dx and Dtheta appear to be correct
\end{verbatim}

\begin{verbatim}
## $testDx
## [1] TRUE
## 
## $testDtheta
## [1] TRUE
\end{verbatim}

Create inputs for MAGI:

\begin{Shaded}
\begin{Highlighting}[]
\CommentTok{\# Discretization set}
\NormalTok{xsim }\OtherTok{\textless{}{-}} \FunctionTok{setDiscretization}\NormalTok{(xsim.obs, }\AttributeTok{by =}\NormalTok{ config}\SpecialCharTok{$}\NormalTok{fillinterval)}

\CommentTok{\# Rough hyperparameter values based on smoothness and level of each component}
\NormalTok{phiExogenous }\OtherTok{\textless{}{-}} \FunctionTok{cbind}\NormalTok{(}
\FunctionTok{c}\NormalTok{(}\FloatTok{2.5}\NormalTok{, }\DecValTok{600}\NormalTok{),}
\FunctionTok{c}\NormalTok{(}\DecValTok{100}\NormalTok{, }\DecValTok{140}\NormalTok{),}
\FunctionTok{c}\NormalTok{(}\DecValTok{1000}\NormalTok{, }\DecValTok{200}\NormalTok{),}
\FunctionTok{c}\NormalTok{(}\DecValTok{60}\NormalTok{, }\DecValTok{100}\NormalTok{),}
\FunctionTok{c}\NormalTok{(}\FloatTok{0.01}\NormalTok{, }\DecValTok{800}\NormalTok{),}
\FunctionTok{c}\NormalTok{(}\DecValTok{1}\NormalTok{, }\DecValTok{200}\NormalTok{),}
\FunctionTok{c}\NormalTok{(}\DecValTok{500}\NormalTok{, }\DecValTok{300}\NormalTok{),}
\FunctionTok{c}\NormalTok{(}\FloatTok{0.5}\NormalTok{, }\DecValTok{300}\NormalTok{),}
\FunctionTok{c}\NormalTok{(}\DecValTok{1}\NormalTok{, }\DecValTok{400}\NormalTok{),}
\FunctionTok{c}\NormalTok{(}\DecValTok{35}\NormalTok{, }\DecValTok{1000}\NormalTok{)}
\NormalTok{)}
\end{Highlighting}
\end{Shaded}

Run the MAGI method:

\begin{Shaded}
\begin{Highlighting}[]
\NormalTok{gpode }\OtherTok{\textless{}{-}} \FunctionTok{MagiSolver}\NormalTok{(xsim, dynamicalModelList,}
\AttributeTok{control =} \FunctionTok{list}\NormalTok{(}\AttributeTok{niterHmc=}\NormalTok{config}\SpecialCharTok{$}\NormalTok{niterHmc, }\AttributeTok{nstepsHmc =}\NormalTok{ config}\SpecialCharTok{$}\NormalTok{hmcSteps,}
\AttributeTok{phi=}\NormalTok{phiExogenous, }\AttributeTok{sigma=}\NormalTok{config}\SpecialCharTok{$}\NormalTok{noise, }\AttributeTok{useFixedSigma=}\ConstantTok{TRUE}\NormalTok{))}
\end{Highlighting}
\end{Shaded}

Inference for parameters \(k_1,\ldots, k_{16}\):

\begin{Shaded}
\begin{Highlighting}[]
\NormalTok{par.table }\OtherTok{\textless{}{-}} \ControlFlowTok{function}\NormalTok{(res) \{}
\NormalTok{  par.est }\OtherTok{\textless{}{-}} \FunctionTok{apply}\NormalTok{(}\FunctionTok{cbind}\NormalTok{(res}\SpecialCharTok{$}\NormalTok{theta[,}\SpecialCharTok{{-}}\DecValTok{1}\NormalTok{]), }\DecValTok{2}\NormalTok{,}
\ControlFlowTok{function}\NormalTok{(x) }\FunctionTok{c}\NormalTok{(}\FunctionTok{mean}\NormalTok{(x), }\FunctionTok{quantile}\NormalTok{(x, }\FloatTok{0.025}\NormalTok{), }\FunctionTok{quantile}\NormalTok{(x, }\FloatTok{0.975}\NormalTok{)))}
\FunctionTok{colnames}\NormalTok{(par.est) }\OtherTok{\textless{}{-}} \FunctionTok{paste0}\NormalTok{(}\StringTok{\textquotesingle{}k\textquotesingle{}}\NormalTok{, }\DecValTok{1}\SpecialCharTok{:}\DecValTok{16}\NormalTok{)}
\FunctionTok{rownames}\NormalTok{(par.est) }\OtherTok{\textless{}{-}} \FunctionTok{c}\NormalTok{(}\StringTok{"Mean"}\NormalTok{, }\StringTok{"2.5\%"}\NormalTok{, }\StringTok{"97.5\%"}\NormalTok{)}
\FunctionTok{signif}\NormalTok{(par.est, }\DecValTok{3}\NormalTok{)}
\NormalTok{\}}

\FunctionTok{par.table}\NormalTok{(gpode)}
\end{Highlighting}
\end{Shaded}

\begin{verbatim}
##           k1     k2      k3     k4    k5      k6     k7      k8     k9    k10
## Mean  0.0198 0.0966 0.00425 0.0856 0.916 0.00919 0.0982 0.00874 0.0299 0.1000
## 2.5%  0.0193 0.0838 0.00389 0.0786 0.845 0.00842 0.0918 0.00610 0.0291 0.0998
## 97.5% 0.0203 0.1090 0.00465 0.0935 0.989 0.00999 0.1050 0.01140 0.0308 0.1000
##            k11     k12     k13     k14     k15      k16
## Mean  0.001000 0.00990 0.00192 0.00189 0.01000 0.001000
## 2.5%  0.000998 0.00961 0.00161 0.00150 0.00966 0.000992
## 97.5% 0.001000 0.01020 0.00221 0.00227 0.01030 0.001010
\end{verbatim}

Calculate reconstructed trajectories:

\begin{Shaded}
\begin{Highlighting}[]
\NormalTok{tvecsolve }\OtherTok{\textless{}{-}} \FunctionTok{seq}\NormalTok{(config}\SpecialCharTok{$}\NormalTok{t.start,config}\SpecialCharTok{$}\NormalTok{t.end,}\AttributeTok{by =} \FloatTok{0.1}\NormalTok{)}
\NormalTok{calcTraj }\OtherTok{\textless{}{-}} \ControlFlowTok{function}\NormalTok{(res) \{}
\NormalTok{  x0.est }\OtherTok{\textless{}{-}} \FunctionTok{apply}\NormalTok{(res}\SpecialCharTok{$}\NormalTok{xsampled[,}\DecValTok{1}\NormalTok{,],}\DecValTok{2}\NormalTok{,mean)}
\NormalTok{  theta.est }\OtherTok{\textless{}{-}} \FunctionTok{apply}\NormalTok{(res}\SpecialCharTok{$}\NormalTok{theta,}\DecValTok{2}\NormalTok{,mean)}

\NormalTok{  x }\OtherTok{\textless{}{-}}\NormalTok{ deSolve}\SpecialCharTok{::}\FunctionTok{ode}\NormalTok{(}\AttributeTok{y =}\NormalTok{ x0.est, }\AttributeTok{times =}\NormalTok{ tvecsolve,}
\AttributeTok{func =}\NormalTok{ modelODE, }\AttributeTok{parms =}\NormalTok{ theta.est)}
\NormalTok{  x}
\NormalTok{\}}

\NormalTok{recon }\OtherTok{\textless{}{-}} \FunctionTok{calcTraj}\NormalTok{(gpode)}
\NormalTok{recon.obs }\OtherTok{\textless{}{-}} \FunctionTok{subset}\NormalTok{(recon, time }\SpecialCharTok{\%in\%}\NormalTok{ xsim}\SpecialCharTok{$}\NormalTok{time)}
\NormalTok{xtrue.obs }\OtherTok{\textless{}{-}} \FunctionTok{subset}\NormalTok{(xtrue, time }\SpecialCharTok{\%in\%}\NormalTok{ xsim}\SpecialCharTok{$}\NormalTok{time)[,}\SpecialCharTok{{-}}\DecValTok{1}\NormalTok{]}
\end{Highlighting}
\end{Shaded}

\begin{figure}
\centering
\includegraphics{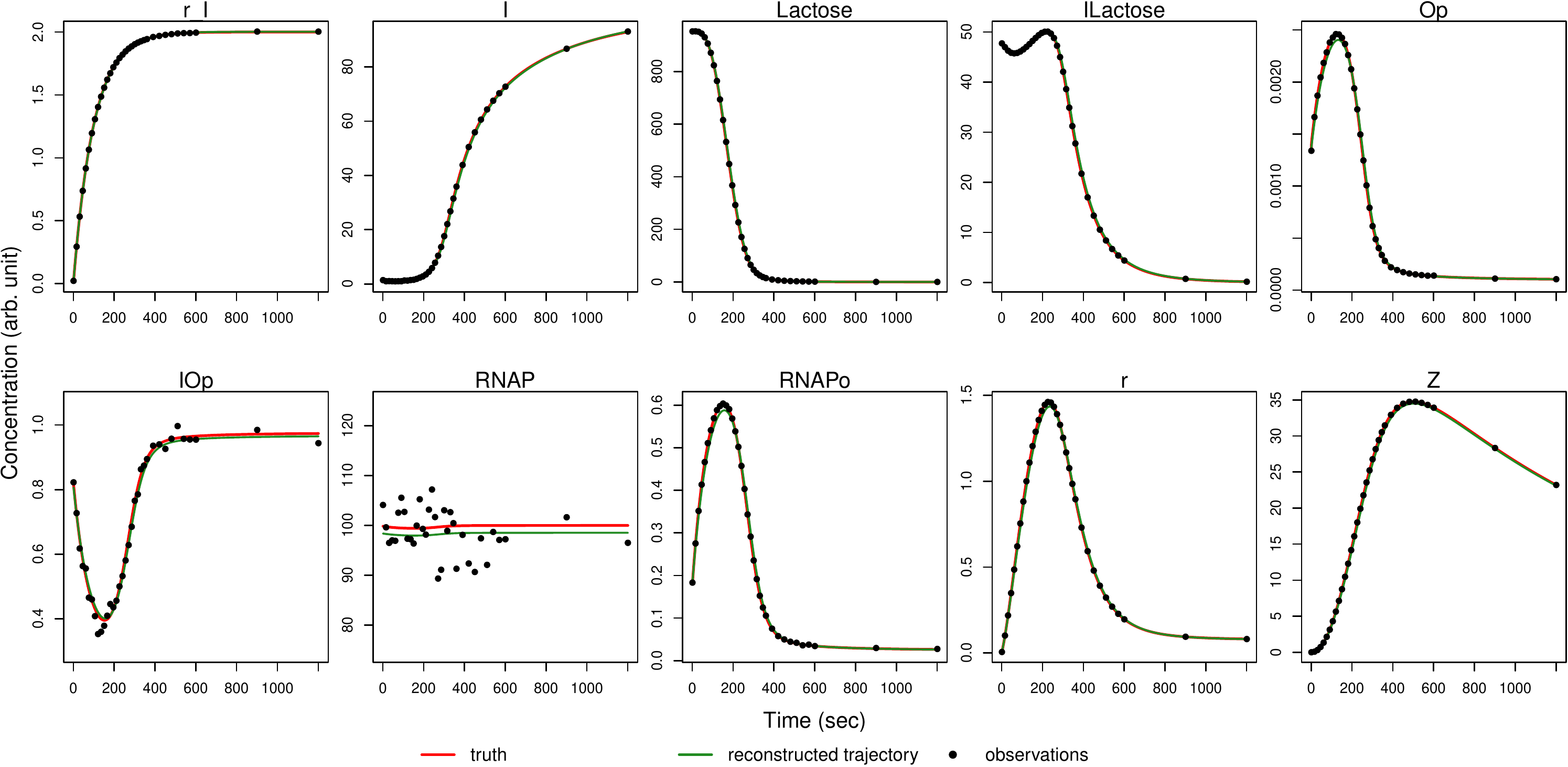}
\caption{\label{fig:lac-recon} \footnotesize  Reconstructed trajectories for a sample dataset simulated from the \(lac\) operon model.}
\end{figure}

Visualize the reconstructed trajectories together with the observations, as shown in Fig \ref{fig:lac-recon}:

\begin{Shaded}
\begin{Highlighting}[]
\NormalTok{compnames }\OtherTok{\textless{}{-}} \FunctionTok{c}\NormalTok{(}\StringTok{"r\_I"}\NormalTok{, }\StringTok{"I"}\NormalTok{, }\StringTok{"Lactose"}\NormalTok{, }\StringTok{"ILactose"}\NormalTok{, }\StringTok{"Op"}\NormalTok{,}
\StringTok{"IOp"}\NormalTok{, }\StringTok{"RNAP"}\NormalTok{, }\StringTok{"RNAPo"}\NormalTok{, }\StringTok{"r"}\NormalTok{, }\StringTok{"Z"}\NormalTok{)}
\FunctionTok{par}\NormalTok{(}\AttributeTok{oma=}\FunctionTok{c}\NormalTok{(}\DecValTok{0}\NormalTok{,}\FloatTok{1.5}\NormalTok{,}\DecValTok{0}\NormalTok{,}\DecValTok{0}\NormalTok{))}
\FunctionTok{layout}\NormalTok{(}\FunctionTok{rbind}\NormalTok{(}\FunctionTok{c}\NormalTok{(}\DecValTok{1}\SpecialCharTok{:}\DecValTok{5}\NormalTok{), }\FunctionTok{c}\NormalTok{(}\DecValTok{6}\SpecialCharTok{:}\DecValTok{10}\NormalTok{), }\FunctionTok{c}\NormalTok{(}\DecValTok{11}\NormalTok{,}\DecValTok{11}\NormalTok{,}\DecValTok{11}\NormalTok{,}\DecValTok{11}\NormalTok{,}\DecValTok{11}\NormalTok{)), }\AttributeTok{heights =} \FunctionTok{c}\NormalTok{(}\DecValTok{8}\NormalTok{,}\DecValTok{8}\NormalTok{,}\DecValTok{1}\NormalTok{))}

\ControlFlowTok{for}\NormalTok{ (i }\ControlFlowTok{in} \DecValTok{1}\SpecialCharTok{:}\DecValTok{10}\NormalTok{) \{}
\FunctionTok{par}\NormalTok{(}\AttributeTok{mar =} \FunctionTok{c}\NormalTok{(}\DecValTok{4}\NormalTok{, }\FloatTok{2.5}\NormalTok{, }\FloatTok{1.75}\NormalTok{, }\FloatTok{0.1}\NormalTok{))}
\FunctionTok{plot}\NormalTok{(}\FunctionTok{c}\NormalTok{(}\FunctionTok{min}\NormalTok{(tvecsolve), }\FunctionTok{max}\NormalTok{(tvecsolve)),}
\FunctionTok{c}\NormalTok{(}\FunctionTok{min}\NormalTok{(}\FunctionTok{c}\NormalTok{(recon[,i}\SpecialCharTok{+}\DecValTok{1}\NormalTok{], xtrue[xtrue}\SpecialCharTok{$}\NormalTok{time }\SpecialCharTok{\textgreater{}=}\DecValTok{1}\NormalTok{,i}\SpecialCharTok{+}\DecValTok{1}\NormalTok{] }\SpecialCharTok{{-}}\NormalTok{ config}\SpecialCharTok{$}\NormalTok{noise[i]}\SpecialCharTok{*}\DecValTok{5}\NormalTok{)),}
\FunctionTok{max}\NormalTok{(}\FunctionTok{c}\NormalTok{(recon[,i}\SpecialCharTok{+}\DecValTok{1}\NormalTok{], xtrue[xtrue}\SpecialCharTok{$}\NormalTok{time}\SpecialCharTok{\textgreater{}=}\DecValTok{1}\NormalTok{,i}\SpecialCharTok{+}\DecValTok{1}\NormalTok{] }\SpecialCharTok{+}\NormalTok{ config}\SpecialCharTok{$}\NormalTok{noise[i]}\SpecialCharTok{*}\DecValTok{5}\NormalTok{))),}
\AttributeTok{type =} \StringTok{\textquotesingle{}n\textquotesingle{}}\NormalTok{, }\AttributeTok{ylab=}\StringTok{\textquotesingle{}\textquotesingle{}}\NormalTok{, }\AttributeTok{xlab=}\StringTok{\textquotesingle{}\textquotesingle{}}\NormalTok{)}
\FunctionTok{mtext}\NormalTok{(compnames[i])}

\FunctionTok{lines}\NormalTok{(xtrue}\SpecialCharTok{$}\NormalTok{time[xtrue}\SpecialCharTok{$}\NormalTok{time }\SpecialCharTok{\textgreater{}=}\DecValTok{1}\NormalTok{], xtrue[xtrue}\SpecialCharTok{$}\NormalTok{time }\SpecialCharTok{\textgreater{}=}\DecValTok{1}\NormalTok{,i}\SpecialCharTok{+}\DecValTok{1}\NormalTok{], }\AttributeTok{col=}\StringTok{"red"}\NormalTok{, }\AttributeTok{lwd=}\DecValTok{2}\NormalTok{)}
\FunctionTok{lines}\NormalTok{(tvecsolve, recon[,i}\SpecialCharTok{+}\DecValTok{1}\NormalTok{], }\AttributeTok{col=}\StringTok{"forestgreen"}\NormalTok{, }\AttributeTok{lwd=}\FloatTok{1.5}\NormalTok{)}

\FunctionTok{points}\NormalTok{(xsim[,}\DecValTok{1}\NormalTok{], xsim[,i}\SpecialCharTok{+}\DecValTok{1}\NormalTok{], }\AttributeTok{pch=}\DecValTok{16}\NormalTok{)}

\ControlFlowTok{if}\NormalTok{ (i }\SpecialCharTok{==} \DecValTok{8}\NormalTok{) }\FunctionTok{title}\NormalTok{(}\AttributeTok{xlab=}\StringTok{\textquotesingle{}Time (sec)\textquotesingle{}}\NormalTok{, }\AttributeTok{cex.lab =} \FloatTok{1.5}\NormalTok{)}
\ControlFlowTok{if}\NormalTok{ (i }\SpecialCharTok{==} \DecValTok{1}\NormalTok{) }\FunctionTok{mtext}\NormalTok{(}\StringTok{"       Concentration (arb. unit)"}\NormalTok{,}\AttributeTok{side=}\DecValTok{2}\NormalTok{,}\AttributeTok{line=}\DecValTok{0}\NormalTok{,}\AttributeTok{outer=}\ConstantTok{TRUE}\NormalTok{)}
\NormalTok{\}}

\FunctionTok{par}\NormalTok{(}\AttributeTok{mar=}\FunctionTok{rep}\NormalTok{(}\DecValTok{0}\NormalTok{,}\DecValTok{4}\NormalTok{))}
\FunctionTok{plot}\NormalTok{(}\DecValTok{1}\NormalTok{,}\AttributeTok{type=}\StringTok{\textquotesingle{}n\textquotesingle{}}\NormalTok{, }\AttributeTok{xaxt=}\StringTok{\textquotesingle{}n\textquotesingle{}}\NormalTok{, }\AttributeTok{yaxt=}\StringTok{\textquotesingle{}n\textquotesingle{}}\NormalTok{, }\AttributeTok{xlab=}\ConstantTok{NA}\NormalTok{, }\AttributeTok{ylab=}\ConstantTok{NA}\NormalTok{, }\AttributeTok{frame.plot =} \ConstantTok{FALSE}\NormalTok{)}

\FunctionTok{legend}\NormalTok{(}\StringTok{"center"}\NormalTok{,  }\FunctionTok{c}\NormalTok{(}\StringTok{"truth"}\NormalTok{, }\StringTok{"reconstructed trajectory"}\NormalTok{, }\StringTok{"observations"}\NormalTok{),}
\AttributeTok{lty =} \FunctionTok{c}\NormalTok{(}\DecValTok{1}\NormalTok{, }\DecValTok{1}\NormalTok{, }\DecValTok{0}\NormalTok{), }\AttributeTok{lwd =} \FunctionTok{c}\NormalTok{(}\DecValTok{2}\NormalTok{, }\DecValTok{2}\NormalTok{, }\DecValTok{0}\NormalTok{), }\AttributeTok{bty =} \StringTok{"n"}\NormalTok{,}
\AttributeTok{col =} \FunctionTok{c}\NormalTok{(}\StringTok{"red"}\NormalTok{, }\StringTok{"forestgreen"}\NormalTok{, }\StringTok{"black"}\NormalTok{), }\AttributeTok{fill =} \FunctionTok{c}\NormalTok{(}\DecValTok{0}\NormalTok{, }\DecValTok{0}\NormalTok{, }\DecValTok{0}\NormalTok{),}
\AttributeTok{border =} \FunctionTok{c}\NormalTok{(}\DecValTok{0}\NormalTok{, }\DecValTok{0}\NormalTok{, }\DecValTok{0}\NormalTok{), }\AttributeTok{pch =} \FunctionTok{c}\NormalTok{(}\ConstantTok{NA}\NormalTok{, }\ConstantTok{NA}\NormalTok{, }\DecValTok{16}\NormalTok{), }\AttributeTok{horiz =} \ConstantTok{TRUE}\NormalTok{, }\AttributeTok{cex =} \FloatTok{1.25}\NormalTok{)}
\end{Highlighting}
\end{Shaded}

\hypertarget{model-comparison}{%
\subsection{Model comparison}\label{model-comparison}}

We begin by defining a function that codes the ODEs for the inhibitor model:

\begin{Shaded}
\begin{Highlighting}[]
\NormalTok{RMichaelisMentenInhibitor6ODE }\OtherTok{\textless{}{-}} \ControlFlowTok{function}\NormalTok{(theta, x, tvec) \{}
\NormalTok{  resultdt }\OtherTok{\textless{}{-}}  \FunctionTok{array}\NormalTok{(}\DecValTok{0}\NormalTok{, }\FunctionTok{c}\NormalTok{(}\FunctionTok{nrow}\NormalTok{(x),}\FunctionTok{ncol}\NormalTok{(x)))}

\NormalTok{  e0 }\OtherTok{=} \FloatTok{0.1}
\NormalTok{  e }\OtherTok{=}\NormalTok{ x[,}\DecValTok{1}\NormalTok{]}
\NormalTok{  s }\OtherTok{=}\NormalTok{ x[,}\DecValTok{2}\NormalTok{]}
\NormalTok{  p }\OtherTok{=}\NormalTok{ x[,}\DecValTok{3}\NormalTok{]}
\NormalTok{  i }\OtherTok{=}\NormalTok{ x[,}\DecValTok{4}\NormalTok{]}
\NormalTok{  ei }\OtherTok{=}\NormalTok{ x[,}\DecValTok{5}\NormalTok{]}
\NormalTok{  es }\OtherTok{=}\NormalTok{ x[,}\DecValTok{6}\NormalTok{]}

\NormalTok{  resultdt[,}\DecValTok{1}\NormalTok{] }\OtherTok{=} \SpecialCharTok{{-}}\NormalTok{theta[}\DecValTok{1}\NormalTok{] }\SpecialCharTok{*}\NormalTok{ e }\SpecialCharTok{*}\NormalTok{ s }\SpecialCharTok{+}\NormalTok{ (theta[}\DecValTok{2}\NormalTok{]}\SpecialCharTok{+}\NormalTok{theta[}\DecValTok{3}\NormalTok{]) }\SpecialCharTok{*}\NormalTok{ es }\SpecialCharTok{{-}}
\NormalTok{    theta[}\DecValTok{4}\NormalTok{] }\SpecialCharTok{*}\NormalTok{ i }\SpecialCharTok{*}\NormalTok{ e }\SpecialCharTok{+}\NormalTok{ theta[}\DecValTok{5}\NormalTok{] }\SpecialCharTok{*}\NormalTok{ ei}
\NormalTok{  resultdt[,}\DecValTok{2}\NormalTok{] }\OtherTok{=} \SpecialCharTok{{-}}\NormalTok{theta[}\DecValTok{1}\NormalTok{] }\SpecialCharTok{*}\NormalTok{ e }\SpecialCharTok{*}\NormalTok{ s }\SpecialCharTok{+}\NormalTok{ (theta[}\DecValTok{2}\NormalTok{]) }\SpecialCharTok{*}\NormalTok{ es}
\NormalTok{  resultdt[,}\DecValTok{3}\NormalTok{] }\OtherTok{=}\NormalTok{ theta[}\DecValTok{3}\NormalTok{] }\SpecialCharTok{*}\NormalTok{ es}
\NormalTok{  resultdt[,}\DecValTok{4}\NormalTok{] }\OtherTok{=} \SpecialCharTok{{-}}\NormalTok{theta[}\DecValTok{4}\NormalTok{] }\SpecialCharTok{*}\NormalTok{ i }\SpecialCharTok{*}\NormalTok{ e }\SpecialCharTok{+}\NormalTok{ theta[}\DecValTok{5}\NormalTok{] }\SpecialCharTok{*}\NormalTok{ ei}
\NormalTok{  resultdt[,}\DecValTok{5}\NormalTok{] }\OtherTok{=}\NormalTok{ theta[}\DecValTok{4}\NormalTok{] }\SpecialCharTok{*}\NormalTok{ i }\SpecialCharTok{*}\NormalTok{ e }\SpecialCharTok{{-}}\NormalTok{ theta[}\DecValTok{5}\NormalTok{] }\SpecialCharTok{*}\NormalTok{ ei}
\NormalTok{  resultdt[,}\DecValTok{6}\NormalTok{] }\OtherTok{=}\NormalTok{ theta[}\DecValTok{1}\NormalTok{] }\SpecialCharTok{*}\NormalTok{ e }\SpecialCharTok{*}\NormalTok{ s  }\SpecialCharTok{{-}}\NormalTok{ (theta[}\DecValTok{2}\NormalTok{]}\SpecialCharTok{+}\NormalTok{theta[}\DecValTok{3}\NormalTok{]) }\SpecialCharTok{*}\NormalTok{ es}

\NormalTok{  resultdt}
\NormalTok{\}}
\end{Highlighting}
\end{Shaded}

Next, we provide the gradients of the ODEs with respect to the system components \(X\) and the parameters \(\theta\).

\begin{Shaded}
\begin{Highlighting}[]
\NormalTok{RMichaelisMentenInhibitor6Dx }\OtherTok{\textless{}{-}} \ControlFlowTok{function}\NormalTok{(theta, x, tvec) \{}
\NormalTok{  resultDx }\OtherTok{\textless{}{-}} \FunctionTok{array}\NormalTok{(}\DecValTok{0}\NormalTok{, }\FunctionTok{c}\NormalTok{(}\FunctionTok{nrow}\NormalTok{(x), }\FunctionTok{ncol}\NormalTok{(x), }\FunctionTok{ncol}\NormalTok{(x)))}

\NormalTok{  e0 }\OtherTok{=} \FloatTok{0.1}
\NormalTok{  e }\OtherTok{=}\NormalTok{ x[,}\DecValTok{1}\NormalTok{]}
\NormalTok{  s }\OtherTok{=}\NormalTok{ x[,}\DecValTok{2}\NormalTok{]}
\NormalTok{  p }\OtherTok{=}\NormalTok{ x[,}\DecValTok{3}\NormalTok{]}
\NormalTok{  i }\OtherTok{=}\NormalTok{ x[,}\DecValTok{4}\NormalTok{]}
\NormalTok{  ei }\OtherTok{=}\NormalTok{ x[,}\DecValTok{5}\NormalTok{]}
\NormalTok{  es }\OtherTok{=}\NormalTok{ x[,}\DecValTok{6}\NormalTok{]}

\NormalTok{  resultDx[,}\DecValTok{1}\NormalTok{,}\DecValTok{1}\NormalTok{] }\OtherTok{=} \SpecialCharTok{{-}}\NormalTok{theta[}\DecValTok{1}\NormalTok{] }\SpecialCharTok{*}\NormalTok{ s }\SpecialCharTok{{-}}\NormalTok{ theta[}\DecValTok{4}\NormalTok{] }\SpecialCharTok{*}\NormalTok{ i}
\NormalTok{  resultDx[,}\DecValTok{2}\NormalTok{,}\DecValTok{1}\NormalTok{] }\OtherTok{=} \SpecialCharTok{{-}}\NormalTok{theta[}\DecValTok{1}\NormalTok{] }\SpecialCharTok{*}\NormalTok{ e}
\NormalTok{  resultDx[,}\DecValTok{4}\NormalTok{,}\DecValTok{1}\NormalTok{] }\OtherTok{=} \SpecialCharTok{{-}}\NormalTok{theta[}\DecValTok{4}\NormalTok{] }\SpecialCharTok{*}\NormalTok{ e}
\NormalTok{  resultDx[,}\DecValTok{5}\NormalTok{,}\DecValTok{1}\NormalTok{] }\OtherTok{=}\NormalTok{ (theta[}\DecValTok{5}\NormalTok{])}
\NormalTok{  resultDx[,}\DecValTok{6}\NormalTok{,}\DecValTok{1}\NormalTok{] }\OtherTok{=}\NormalTok{ (theta[}\DecValTok{2}\NormalTok{]}\SpecialCharTok{+}\NormalTok{theta[}\DecValTok{3}\NormalTok{])}

\NormalTok{  resultDx[,}\DecValTok{1}\NormalTok{,}\DecValTok{2}\NormalTok{] }\OtherTok{=} \SpecialCharTok{{-}}\NormalTok{theta[}\DecValTok{1}\NormalTok{] }\SpecialCharTok{*}\NormalTok{ s}
\NormalTok{  resultDx[,}\DecValTok{2}\NormalTok{,}\DecValTok{2}\NormalTok{] }\OtherTok{=} \SpecialCharTok{{-}}\NormalTok{theta[}\DecValTok{1}\NormalTok{] }\SpecialCharTok{*}\NormalTok{ e}
\NormalTok{  resultDx[,}\DecValTok{6}\NormalTok{,}\DecValTok{2}\NormalTok{] }\OtherTok{=}\NormalTok{ (theta[}\DecValTok{2}\NormalTok{])}

\NormalTok{  resultDx[,}\DecValTok{6}\NormalTok{,}\DecValTok{3}\NormalTok{] }\OtherTok{=}\NormalTok{ (theta[}\DecValTok{3}\NormalTok{])}

\NormalTok{  resultDx[,}\DecValTok{1}\NormalTok{,}\DecValTok{4}\NormalTok{] }\OtherTok{=} \SpecialCharTok{{-}}\NormalTok{theta[}\DecValTok{4}\NormalTok{] }\SpecialCharTok{*}\NormalTok{ i}
\NormalTok{  resultDx[,}\DecValTok{4}\NormalTok{,}\DecValTok{4}\NormalTok{] }\OtherTok{=} \SpecialCharTok{{-}}\NormalTok{theta[}\DecValTok{4}\NormalTok{] }\SpecialCharTok{*}\NormalTok{ e}
\NormalTok{  resultDx[,}\DecValTok{5}\NormalTok{,}\DecValTok{4}\NormalTok{] }\OtherTok{=}\NormalTok{ (theta[}\DecValTok{5}\NormalTok{])}

\NormalTok{  resultDx[,}\DecValTok{1}\NormalTok{,}\DecValTok{5}\NormalTok{] }\OtherTok{=}\NormalTok{ theta[}\DecValTok{4}\NormalTok{] }\SpecialCharTok{*}\NormalTok{ i}
\NormalTok{  resultDx[,}\DecValTok{4}\NormalTok{,}\DecValTok{5}\NormalTok{] }\OtherTok{=}\NormalTok{ theta[}\DecValTok{4}\NormalTok{] }\SpecialCharTok{*}\NormalTok{ e}
\NormalTok{  resultDx[,}\DecValTok{5}\NormalTok{,}\DecValTok{5}\NormalTok{] }\OtherTok{=}\NormalTok{ (}\SpecialCharTok{{-}}\NormalTok{theta[}\DecValTok{5}\NormalTok{])}

\NormalTok{  resultDx[,}\DecValTok{1}\NormalTok{,}\DecValTok{6}\NormalTok{] }\OtherTok{=}\NormalTok{ theta[}\DecValTok{1}\NormalTok{] }\SpecialCharTok{*}\NormalTok{ s}
\NormalTok{  resultDx[,}\DecValTok{2}\NormalTok{,}\DecValTok{6}\NormalTok{] }\OtherTok{=}\NormalTok{ theta[}\DecValTok{1}\NormalTok{] }\SpecialCharTok{*}\NormalTok{ e}
\NormalTok{  resultDx[,}\DecValTok{6}\NormalTok{,}\DecValTok{6}\NormalTok{] }\OtherTok{=}\NormalTok{ (}\SpecialCharTok{{-}}\NormalTok{(theta[}\DecValTok{2}\NormalTok{]}\SpecialCharTok{+}\NormalTok{theta[}\DecValTok{3}\NormalTok{]))}

\NormalTok{  resultDx}
\NormalTok{\}}

\NormalTok{RMichaelisMentenInhibitor6Dtheta }\OtherTok{\textless{}{-}} \ControlFlowTok{function}\NormalTok{(theta, x, tvec) \{}
\NormalTok{  resultDtheta }\OtherTok{\textless{}{-}} \FunctionTok{array}\NormalTok{(}\DecValTok{0}\NormalTok{, }\FunctionTok{c}\NormalTok{(}\FunctionTok{nrow}\NormalTok{(x), }\FunctionTok{length}\NormalTok{(theta), }\FunctionTok{ncol}\NormalTok{(x)))}

\NormalTok{  e0 }\OtherTok{=} \FloatTok{0.1}
\NormalTok{  e }\OtherTok{=}\NormalTok{ x[,}\DecValTok{1}\NormalTok{]}
\NormalTok{  s }\OtherTok{=}\NormalTok{ x[,}\DecValTok{2}\NormalTok{]}
\NormalTok{  p }\OtherTok{=}\NormalTok{ x[,}\DecValTok{3}\NormalTok{]}
\NormalTok{  i }\OtherTok{=}\NormalTok{ x[,}\DecValTok{4}\NormalTok{]}
\NormalTok{  ei }\OtherTok{=}\NormalTok{ x[,}\DecValTok{5}\NormalTok{]}
\NormalTok{  es }\OtherTok{=}\NormalTok{ x[,}\DecValTok{6}\NormalTok{]}

\NormalTok{  resultDtheta[,}\DecValTok{1}\NormalTok{,}\DecValTok{1}\NormalTok{] }\OtherTok{=} \SpecialCharTok{{-}}\NormalTok{e }\SpecialCharTok{*}\NormalTok{ s}
\NormalTok{  resultDtheta[,}\DecValTok{2}\NormalTok{,}\DecValTok{1}\NormalTok{] }\OtherTok{=}\NormalTok{ es}
\NormalTok{  resultDtheta[,}\DecValTok{3}\NormalTok{,}\DecValTok{1}\NormalTok{] }\OtherTok{=}\NormalTok{ es}
\NormalTok{  resultDtheta[,}\DecValTok{4}\NormalTok{,}\DecValTok{1}\NormalTok{] }\OtherTok{=} \SpecialCharTok{{-}}\NormalTok{i }\SpecialCharTok{*}\NormalTok{ e}
\NormalTok{  resultDtheta[,}\DecValTok{5}\NormalTok{,}\DecValTok{1}\NormalTok{] }\OtherTok{=}\NormalTok{ ei}

\NormalTok{  resultDtheta[,}\DecValTok{1}\NormalTok{,}\DecValTok{2}\NormalTok{] }\OtherTok{=} \SpecialCharTok{{-}}\NormalTok{e }\SpecialCharTok{*}\NormalTok{ s}
\NormalTok{  resultDtheta[,}\DecValTok{2}\NormalTok{,}\DecValTok{2}\NormalTok{] }\OtherTok{=}\NormalTok{ es}

\NormalTok{  resultDtheta[,}\DecValTok{3}\NormalTok{,}\DecValTok{3}\NormalTok{] }\OtherTok{=}\NormalTok{ es}

\NormalTok{  resultDtheta[,}\DecValTok{4}\NormalTok{,}\DecValTok{4}\NormalTok{] }\OtherTok{=} \SpecialCharTok{{-}}\NormalTok{i }\SpecialCharTok{*}\NormalTok{ e}
\NormalTok{  resultDtheta[,}\DecValTok{5}\NormalTok{,}\DecValTok{4}\NormalTok{] }\OtherTok{=}\NormalTok{ ei}

\NormalTok{  resultDtheta[,}\DecValTok{4}\NormalTok{,}\DecValTok{5}\NormalTok{] }\OtherTok{=}\NormalTok{ i }\SpecialCharTok{*}\NormalTok{ e}
\NormalTok{  resultDtheta[,}\DecValTok{5}\NormalTok{,}\DecValTok{5}\NormalTok{] }\OtherTok{=} \SpecialCharTok{{-}}\NormalTok{ei}

\NormalTok{  resultDtheta[,}\DecValTok{1}\NormalTok{,}\DecValTok{6}\NormalTok{] }\OtherTok{=}\NormalTok{ e }\SpecialCharTok{*}\NormalTok{ s}
\NormalTok{  resultDtheta[,}\DecValTok{2}\NormalTok{,}\DecValTok{6}\NormalTok{] }\OtherTok{=} \SpecialCharTok{{-}}\NormalTok{es}
\NormalTok{  resultDtheta[,}\DecValTok{3}\NormalTok{,}\DecValTok{6}\NormalTok{] }\OtherTok{=} \SpecialCharTok{{-}}\NormalTok{es}

\NormalTok{  resultDtheta}
\NormalTok{\}}
\end{Highlighting}
\end{Shaded}

Define parameters and settings for the experiment:

\begin{Shaded}
\begin{Highlighting}[]
\CommentTok{\# Train time points (used for fitting)}
\NormalTok{obs.times }\OtherTok{\textless{}{-}} \FunctionTok{c}\NormalTok{(}\FloatTok{2.5}\NormalTok{, }\FloatTok{4.5}\NormalTok{, }\DecValTok{7}\NormalTok{, }\FloatTok{9.5}\NormalTok{, }\DecValTok{11}\NormalTok{, }\FloatTok{13.5}\NormalTok{, }\DecValTok{15}\NormalTok{, }\DecValTok{16}\NormalTok{, }\DecValTok{18}\NormalTok{, }\DecValTok{20}\NormalTok{)}
\CommentTok{\# Test time points (not used for fiting)}
\NormalTok{test.times }\OtherTok{\textless{}{-}} \FunctionTok{c}\NormalTok{(}\FloatTok{21.5}\NormalTok{, }\DecValTok{24}\NormalTok{, }\DecValTok{27}\NormalTok{, }\FloatTok{29.5}\NormalTok{, }\FloatTok{32.5}\NormalTok{, }\FloatTok{35.5}\NormalTok{, }\FloatTok{39.5}\NormalTok{, }\DecValTok{45}\NormalTok{, }\DecValTok{55}\NormalTok{, }\DecValTok{69}\NormalTok{)}

\NormalTok{config }\OtherTok{\textless{}{-}} \FunctionTok{list}\NormalTok{(}
\AttributeTok{nobs =} \FunctionTok{length}\NormalTok{(obs.times),}
\AttributeTok{noise =} \FunctionTok{c}\NormalTok{(}\ConstantTok{NA}\NormalTok{, }\FloatTok{0.02}\NormalTok{, }\FloatTok{0.02}\NormalTok{, }\ConstantTok{NA}\NormalTok{, }\ConstantTok{NA}\NormalTok{, }\ConstantTok{NA}\NormalTok{),}
\AttributeTok{kernel =} \StringTok{"generalMatern"}\NormalTok{,}
\AttributeTok{seed =} \DecValTok{669097609}\NormalTok{,}
\AttributeTok{n.iter =} \DecValTok{20001}\NormalTok{,}
\AttributeTok{linfillspace =} \FloatTok{0.5}\NormalTok{, }
\AttributeTok{t.end =} \DecValTok{70}\NormalTok{,}
\AttributeTok{modelName =} \StringTok{"MM{-}Inhibitor"}
\NormalTok{)}

\NormalTok{pram.true }\OtherTok{\textless{}{-}} \FunctionTok{list}\NormalTok{( }
\AttributeTok{theta=}\FunctionTok{c}\NormalTok{(}\FloatTok{0.9}\NormalTok{, }\FloatTok{0.75}\NormalTok{, }\FloatTok{2.54}\NormalTok{, }\DecValTok{1}\NormalTok{, }\FloatTok{0.5}\NormalTok{),}
\AttributeTok{x0 =} \FunctionTok{c}\NormalTok{(}\FloatTok{0.1}\NormalTok{, }\DecValTok{1}\NormalTok{, }\DecValTok{0}\NormalTok{, }\FloatTok{0.2}\NormalTok{, }\DecValTok{0}\NormalTok{, }\DecValTok{0}\NormalTok{),}
\AttributeTok{phi =} \FunctionTok{cbind}\NormalTok{(}\FunctionTok{c}\NormalTok{(}\FloatTok{0.1}\NormalTok{, }\DecValTok{70}\NormalTok{), }\FunctionTok{c}\NormalTok{(}\DecValTok{1}\NormalTok{, }\DecValTok{30}\NormalTok{), }\FunctionTok{c}\NormalTok{(}\DecValTok{1}\NormalTok{, }\DecValTok{30}\NormalTok{), }\FunctionTok{c}\NormalTok{(}\DecValTok{1}\NormalTok{, }\DecValTok{70}\NormalTok{), }\FunctionTok{c}\NormalTok{(}\DecValTok{1}\NormalTok{, }\DecValTok{70}\NormalTok{), }\FunctionTok{c}\NormalTok{(}\DecValTok{1}\NormalTok{, }\DecValTok{70}\NormalTok{)),}
\AttributeTok{sigma=}\NormalTok{config}\SpecialCharTok{$}\NormalTok{noise}
\NormalTok{)}
\end{Highlighting}
\end{Shaded}

Use a numerical solver to generate the true trajectories under the inhibitor model:

\begin{Shaded}
\begin{Highlighting}[]
\NormalTok{times }\OtherTok{\textless{}{-}} \FunctionTok{seq}\NormalTok{(}\DecValTok{0}\NormalTok{,config}\SpecialCharTok{$}\NormalTok{t.end,}\AttributeTok{length=}\DecValTok{1001}\NormalTok{)}

\NormalTok{modelODE }\OtherTok{\textless{}{-}} \ControlFlowTok{function}\NormalTok{(t, state, parameters) \{}
\FunctionTok{list}\NormalTok{(}\FunctionTok{as.vector}\NormalTok{(}\FunctionTok{RMichaelisMentenInhibitor6ODE}\NormalTok{(parameters, }\FunctionTok{t}\NormalTok{(state), t)))}
\NormalTok{\}}

\NormalTok{xtrue }\OtherTok{\textless{}{-}}\NormalTok{ deSolve}\SpecialCharTok{::}\FunctionTok{ode}\NormalTok{(}\AttributeTok{y =}\NormalTok{ pram.true}\SpecialCharTok{$}\NormalTok{x0, }\AttributeTok{times =}\NormalTok{ times,}
\AttributeTok{func =}\NormalTok{ modelODE, }\AttributeTok{parms =}\NormalTok{ pram.true}\SpecialCharTok{$}\NormalTok{theta)}
\NormalTok{xtrue }\OtherTok{\textless{}{-}} \FunctionTok{data.frame}\NormalTok{(xtrue)}

\NormalTok{xtrueFunc }\OtherTok{\textless{}{-}} \FunctionTok{lapply}\NormalTok{(}\DecValTok{2}\SpecialCharTok{:}\FunctionTok{ncol}\NormalTok{(xtrue), }\ControlFlowTok{function}\NormalTok{(j)}
\FunctionTok{approxfun}\NormalTok{(xtrue[, }\StringTok{"time"}\NormalTok{], xtrue[, j]))}

\NormalTok{xsim }\OtherTok{\textless{}{-}} \FunctionTok{data.frame}\NormalTok{(}\AttributeTok{time =} \FunctionTok{round}\NormalTok{(}\FunctionTok{c}\NormalTok{(obs.times,test.times) }\SpecialCharTok{/}\NormalTok{ config}\SpecialCharTok{$}\NormalTok{linfillspace) }\SpecialCharTok{*}
\NormalTok{                     config}\SpecialCharTok{$}\NormalTok{linfillspace)}
\NormalTok{xsim }\OtherTok{\textless{}{-}} \FunctionTok{cbind}\NormalTok{(xsim, }\FunctionTok{sapply}\NormalTok{(xtrueFunc, }\ControlFlowTok{function}\NormalTok{(f) }\FunctionTok{f}\NormalTok{(xsim}\SpecialCharTok{$}\NormalTok{time)))}
\NormalTok{xtestDS }\OtherTok{\textless{}{-}}\NormalTok{ xsim}
\end{Highlighting}
\end{Shaded}

Create simulated noisy data, and then divide into train/test parts:

\begin{Shaded}
\begin{Highlighting}[]
\FunctionTok{set.seed}\NormalTok{(config}\SpecialCharTok{$}\NormalTok{seed)}
\ControlFlowTok{for}\NormalTok{(j }\ControlFlowTok{in} \DecValTok{1}\SpecialCharTok{:}\NormalTok{(}\FunctionTok{ncol}\NormalTok{(xsim)}\SpecialCharTok{{-}}\DecValTok{1}\NormalTok{))\{}
\NormalTok{  xsim[,}\DecValTok{1}\SpecialCharTok{+}\NormalTok{j] }\OtherTok{\textless{}{-}}\NormalTok{ xsim[,}\DecValTok{1}\SpecialCharTok{+}\NormalTok{j]}\SpecialCharTok{+}\FunctionTok{rnorm}\NormalTok{(}\FunctionTok{nrow}\NormalTok{(xsim), }\AttributeTok{sd=}\NormalTok{config}\SpecialCharTok{$}\NormalTok{noise[j])}
\NormalTok{\}}

\CommentTok{\# Divide into train/test}
\NormalTok{xtest }\OtherTok{\textless{}{-}}\NormalTok{ xsim[xsim}\SpecialCharTok{$}\NormalTok{time }\SpecialCharTok{\%in\%}\NormalTok{ test.times,]}
\NormalTok{xsim }\OtherTok{\textless{}{-}}\NormalTok{ xsim[xsim}\SpecialCharTok{$}\NormalTok{time }\SpecialCharTok{\%in\%}\NormalTok{ obs.times,]}

\NormalTok{xsim.obs }\OtherTok{\textless{}{-}} \FunctionTok{rbind}\NormalTok{(}\FunctionTok{c}\NormalTok{(}\DecValTok{0}\NormalTok{, pram.true}\SpecialCharTok{$}\NormalTok{x0), xsim) }\CommentTok{\# tack on initial conditions}
\end{Highlighting}
\end{Shaded}

Create the \texttt{odeModel} list, then confirm ODEs and derivatives are correct:

\begin{Shaded}
\begin{Highlighting}[]
\NormalTok{dynamicalModelList }\OtherTok{\textless{}{-}} \FunctionTok{list}\NormalTok{(}
\AttributeTok{fOde=}\NormalTok{RMichaelisMentenInhibitor6ODE,}
\AttributeTok{fOdeDx=}\NormalTok{RMichaelisMentenInhibitor6Dx,}
\AttributeTok{fOdeDtheta=}\NormalTok{RMichaelisMentenInhibitor6Dtheta,}
\AttributeTok{thetaLowerBound=}\FunctionTok{c}\NormalTok{(}\DecValTok{0}\NormalTok{,}\SpecialCharTok{{-}}\DecValTok{100}\NormalTok{,}\DecValTok{0}\NormalTok{,}\DecValTok{0}\NormalTok{,}\SpecialCharTok{{-}}\DecValTok{100}\NormalTok{),}
\AttributeTok{thetaUpperBound=}\FunctionTok{c}\NormalTok{(}\ConstantTok{Inf}\NormalTok{,}\ConstantTok{Inf}\NormalTok{,}\ConstantTok{Inf}\NormalTok{,}\ConstantTok{Inf}\NormalTok{,}\ConstantTok{Inf}\NormalTok{)}
\NormalTok{)}

\FunctionTok{testDynamicalModel}\NormalTok{(dynamicalModelList}\SpecialCharTok{$}\NormalTok{fOde, dynamicalModelList}\SpecialCharTok{$}\NormalTok{fOdeDx,}
\NormalTok{                   dynamicalModelList}\SpecialCharTok{$}\NormalTok{fOdeDtheta, }\StringTok{"dynamicalModelList"}\NormalTok{,}
\FunctionTok{data.matrix}\NormalTok{(xtestDS[,}\SpecialCharTok{{-}}\DecValTok{1}\NormalTok{]), pram.true}\SpecialCharTok{$}\NormalTok{theta, xtestDS}\SpecialCharTok{$}\NormalTok{time)}
\end{Highlighting}
\end{Shaded}

\begin{verbatim}
## dynamicalModelList model, with derivatives
## Dx and Dtheta appear to be correct
\end{verbatim}

\begin{verbatim}
## $testDx
## [1] TRUE
## 
## $testDtheta
## [1] TRUE
\end{verbatim}

Create inputs to MAGI for inhibitor model:

\begin{Shaded}
\begin{Highlighting}[]
\CommentTok{\# Discretization set}
\NormalTok{xsim }\OtherTok{\textless{}{-}} \FunctionTok{setDiscretization}\NormalTok{(}\FunctionTok{rbind}\NormalTok{(xsim.obs, }\FunctionTok{c}\NormalTok{(config}\SpecialCharTok{$}\NormalTok{t.end, }\FunctionTok{rep}\NormalTok{(}\ConstantTok{NaN}\NormalTok{,}\FunctionTok{ncol}\NormalTok{(xsim)}\SpecialCharTok{{-}}\DecValTok{1}\NormalTok{))),}
\AttributeTok{by=}\NormalTok{config}\SpecialCharTok{$}\NormalTok{linfillspace)}
\NormalTok{xsim[}\DecValTok{1}\NormalTok{,}\DecValTok{5}\NormalTok{] }\OtherTok{\textless{}{-}} \ConstantTok{NaN}  \CommentTok{\# do not observe initial I value, other initial conditions known}

\CommentTok{\# Use setSizeFactor=0 to fix initial conditions}
\CommentTok{\# [E]=0.1, [S]=1, [P]=0, [ES]=0, except for unknown I}
\NormalTok{stepSizeFactor }\OtherTok{\textless{}{-}} \FunctionTok{rep}\NormalTok{(}\FloatTok{0.01}\NormalTok{, }\FunctionTok{nrow}\NormalTok{(xsim)}\SpecialCharTok{*}\FunctionTok{length}\NormalTok{(pram.true}\SpecialCharTok{$}\NormalTok{x0) }\SpecialCharTok{+}
\FunctionTok{length}\NormalTok{(dynamicalModelList}\SpecialCharTok{$}\NormalTok{thetaLowerBound) }\SpecialCharTok{+} \FunctionTok{length}\NormalTok{(pram.true}\SpecialCharTok{$}\NormalTok{x0))}
\ControlFlowTok{for}\NormalTok{(j }\ControlFlowTok{in} \FunctionTok{c}\NormalTok{(}\DecValTok{1}\NormalTok{,}\DecValTok{2}\NormalTok{,}\DecValTok{3}\NormalTok{,}\DecValTok{5}\NormalTok{,}\DecValTok{6}\NormalTok{))\{}
\ControlFlowTok{for}\NormalTok{(incre }\ControlFlowTok{in} \DecValTok{1}\SpecialCharTok{:}\DecValTok{1}\NormalTok{)\{}
\NormalTok{    stepSizeFactor[(j}\DecValTok{{-}1}\NormalTok{)}\SpecialCharTok{*}\FunctionTok{nrow}\NormalTok{(xsim) }\SpecialCharTok{+}\NormalTok{ incre] }\OtherTok{\textless{}{-}} \DecValTok{0}
\NormalTok{  \}}
\NormalTok{\}}

\CommentTok{\# Initialize X matrix for HMC sampling with some naive values}
\NormalTok{xInitExogenous }\OtherTok{\textless{}{-}} \FunctionTok{matrix}\NormalTok{(}\ConstantTok{NA}\NormalTok{, }\AttributeTok{nrow=}\FunctionTok{nrow}\NormalTok{(xsim[,}\SpecialCharTok{{-}}\DecValTok{1}\NormalTok{]), }\AttributeTok{ncol=}\FunctionTok{ncol}\NormalTok{(xsim[,}\SpecialCharTok{{-}}\DecValTok{1}\NormalTok{]))}
\NormalTok{xInitExogenous[,}\DecValTok{1}\NormalTok{] }\OtherTok{\textless{}{-}} \FloatTok{0.1}
\NormalTok{xInitExogenous[,}\DecValTok{2}\NormalTok{] }\OtherTok{\textless{}{-}} \DecValTok{1}
\NormalTok{xInitExogenous[,}\DecValTok{3}\NormalTok{] }\OtherTok{\textless{}{-}} \DecValTok{0}
\NormalTok{xInitExogenous[,}\DecValTok{4}\NormalTok{] }\OtherTok{\textless{}{-}} \FloatTok{0.1}
\NormalTok{xInitExogenous[,}\DecValTok{5}\NormalTok{] }\OtherTok{\textless{}{-}} \DecValTok{0}
\NormalTok{xInitExogenous[,}\DecValTok{6}\NormalTok{] }\OtherTok{\textless{}{-}} \DecValTok{0}
\end{Highlighting}
\end{Shaded}

Run the MAGI method under the inhibitor model:

\begin{Shaded}
\begin{Highlighting}[]
\CommentTok{\# Use the option positiveSystem = TRUE since all components are non{-}negative}
\NormalTok{gpode }\OtherTok{\textless{}{-}} \FunctionTok{MagiSolver}\NormalTok{(xsim, dynamicalModelList, }
\AttributeTok{control =} \FunctionTok{list}\NormalTok{(}\AttributeTok{xInit =}\NormalTok{ xInitExogenous, }\AttributeTok{niterHmc=}\NormalTok{config}\SpecialCharTok{$}\NormalTok{n.iter,}
\AttributeTok{stepSizeFactor =}\NormalTok{ stepSizeFactor, }\AttributeTok{positiveSystem =} \ConstantTok{TRUE}\NormalTok{,}
\AttributeTok{skipMissingComponentOptimization =} \ConstantTok{TRUE}\NormalTok{, }
\AttributeTok{phi =}\NormalTok{ pram.true}\SpecialCharTok{$}\NormalTok{phi, }\AttributeTok{sigma=}\NormalTok{config}\SpecialCharTok{$}\NormalTok{noise, }\AttributeTok{useFixedSigma=}\ConstantTok{TRUE}\NormalTok{))}
\end{Highlighting}
\end{Shaded}

Calculate sum of square errors for fitting and prediction for inhibitor model:

\begin{Shaded}
\begin{Highlighting}[]
\NormalTok{xMean }\OtherTok{\textless{}{-}} \FunctionTok{apply}\NormalTok{(gpode}\SpecialCharTok{$}\NormalTok{xsampled, }\FunctionTok{c}\NormalTok{(}\DecValTok{2}\NormalTok{, }\DecValTok{3}\NormalTok{), mean)}
\NormalTok{fit\_inhib }\OtherTok{\textless{}{-}}\NormalTok{ xMean[gpode}\SpecialCharTok{$}\NormalTok{tvec }\SpecialCharTok{\%in\%}\NormalTok{ xsim.obs[,}\StringTok{"time"}\NormalTok{],}\DecValTok{2}\SpecialCharTok{:}\DecValTok{3}\NormalTok{]}
\NormalTok{pred\_inhib }\OtherTok{\textless{}{-}}\NormalTok{ xMean[gpode}\SpecialCharTok{$}\NormalTok{tvec }\SpecialCharTok{\%in\%}\NormalTok{ xtest[,}\StringTok{"time"}\NormalTok{],}\DecValTok{2}\SpecialCharTok{:}\DecValTok{3}\NormalTok{]}
\NormalTok{sse\_train\_inhib }\OtherTok{\textless{}{-}} \FunctionTok{sum}\NormalTok{((fit\_inhib }\SpecialCharTok{{-}}\NormalTok{ xsim.obs[,}\DecValTok{3}\SpecialCharTok{:}\DecValTok{4}\NormalTok{])}\SpecialCharTok{\^{}}\DecValTok{2}\NormalTok{)}
\NormalTok{sse\_test\_inhib }\OtherTok{\textless{}{-}}  \FunctionTok{sum}\NormalTok{((pred\_inhib }\SpecialCharTok{{-}}\NormalTok{ xtest[,}\DecValTok{3}\SpecialCharTok{:}\DecValTok{4}\NormalTok{])}\SpecialCharTok{\^{}}\DecValTok{2}\NormalTok{)}

\NormalTok{ourEst\_inhib }\OtherTok{\textless{}{-}} \FunctionTok{apply}\NormalTok{(gpode}\SpecialCharTok{$}\NormalTok{xsampled, }\FunctionTok{c}\NormalTok{(}\DecValTok{2}\NormalTok{, }\DecValTok{3}\NormalTok{), mean)}
\NormalTok{ourLB\_inhib }\OtherTok{\textless{}{-}} \FunctionTok{apply}\NormalTok{(gpode}\SpecialCharTok{$}\NormalTok{xsampled, }\FunctionTok{c}\NormalTok{(}\DecValTok{2}\NormalTok{, }\DecValTok{3}\NormalTok{), }\ControlFlowTok{function}\NormalTok{(x) }\FunctionTok{quantile}\NormalTok{(x, }\FloatTok{0.025}\NormalTok{))}
\NormalTok{ourUB\_inhib }\OtherTok{\textless{}{-}} \FunctionTok{apply}\NormalTok{(gpode}\SpecialCharTok{$}\NormalTok{xsampled, }\FunctionTok{c}\NormalTok{(}\DecValTok{2}\NormalTok{, }\DecValTok{3}\NormalTok{), }\ControlFlowTok{function}\NormalTok{(x) }\FunctionTok{quantile}\NormalTok{(x, }\FloatTok{0.975}\NormalTok{))}
\end{Highlighting}
\end{Shaded}

Create inputs to MAGI for fitting with Michaelis-Menten model:

\begin{Shaded}
\begin{Highlighting}[]
\NormalTok{dynamicalModelListReduced }\OtherTok{\textless{}{-}} \FunctionTok{list}\NormalTok{(}
\AttributeTok{fOde=}\NormalTok{RMichaelisMentenReducedODE,}
\AttributeTok{fOdeDx=}\NormalTok{RMichaelisMentenReducedDx,}
\AttributeTok{fOdeDtheta=}\NormalTok{RMichaelisMentenReducedDtheta,}
\AttributeTok{thetaLowerBound=}\FunctionTok{c}\NormalTok{(}\DecValTok{0}\NormalTok{,}\SpecialCharTok{{-}}\DecValTok{100}\NormalTok{,}\DecValTok{0}\NormalTok{),}
\AttributeTok{thetaUpperBound=}\FunctionTok{c}\NormalTok{(}\ConstantTok{Inf}\NormalTok{,}\ConstantTok{Inf}\NormalTok{,}\ConstantTok{Inf}\NormalTok{)}
\NormalTok{)}

\CommentTok{\# Use setSizeFactor=0 to fix initial conditions [E]=0.1, [S]=1, [P]=0}
\NormalTok{stepSizeFactor }\OtherTok{\textless{}{-}} \FunctionTok{rep}\NormalTok{(}\FloatTok{0.01}\NormalTok{, (}\FunctionTok{nrow}\NormalTok{(xsim))}\SpecialCharTok{*}\NormalTok{(}\FunctionTok{length}\NormalTok{(pram.true}\SpecialCharTok{$}\NormalTok{x0)}\SpecialCharTok{{-}}\DecValTok{3}\NormalTok{) }\SpecialCharTok{+} 
\FunctionTok{length}\NormalTok{(dynamicalModelListReduced}\SpecialCharTok{$}\NormalTok{thetaLowerBound) }\SpecialCharTok{+} \FunctionTok{length}\NormalTok{(pram.true}\SpecialCharTok{$}\NormalTok{x0) }\SpecialCharTok{{-}} \DecValTok{3}\NormalTok{)}
\ControlFlowTok{for}\NormalTok{(j }\ControlFlowTok{in} \DecValTok{1}\SpecialCharTok{:}\DecValTok{3}\NormalTok{)\{}
\ControlFlowTok{for}\NormalTok{(incre }\ControlFlowTok{in} \DecValTok{1}\SpecialCharTok{:}\DecValTok{1}\NormalTok{)\{}
\NormalTok{    stepSizeFactor[(j}\DecValTok{{-}1}\NormalTok{)}\SpecialCharTok{*}\FunctionTok{nrow}\NormalTok{(xsim) }\SpecialCharTok{+}\NormalTok{ incre] }\OtherTok{\textless{}{-}} \DecValTok{0}
\NormalTok{  \}}
\NormalTok{\}}
\end{Highlighting}
\end{Shaded}

Run the MAGI method under the Michaelis-Menten model:

\begin{Shaded}
\begin{Highlighting}[]
\NormalTok{gpode }\OtherTok{\textless{}{-}} \FunctionTok{MagiSolver}\NormalTok{(xsim[,}\FunctionTok{c}\NormalTok{(}\DecValTok{2}\NormalTok{,}\DecValTok{3}\NormalTok{,}\DecValTok{4}\NormalTok{)], dynamicalModelListReduced, xsim}\SpecialCharTok{$}\NormalTok{time,}
\AttributeTok{control =} \FunctionTok{list}\NormalTok{(}\AttributeTok{xInit =}\NormalTok{ xInitExogenous[,}\DecValTok{1}\SpecialCharTok{:}\DecValTok{3}\NormalTok{], }\AttributeTok{niterHmc=}\NormalTok{config}\SpecialCharTok{$}\NormalTok{n.iter,}
\AttributeTok{stepSizeFactor =}\NormalTok{ stepSizeFactor, }\AttributeTok{positiveSystem =} \ConstantTok{TRUE}\NormalTok{,}
\AttributeTok{skipMissingComponentOptimization =} \ConstantTok{TRUE}\NormalTok{,}
\AttributeTok{phi =}\NormalTok{ pram.true}\SpecialCharTok{$}\NormalTok{phi[,}\DecValTok{1}\SpecialCharTok{:}\DecValTok{3}\NormalTok{], }\AttributeTok{sigma=}\NormalTok{config}\SpecialCharTok{$}\NormalTok{noise[}\DecValTok{1}\SpecialCharTok{:}\DecValTok{3}\NormalTok{],}
\AttributeTok{useFixedSigma=}\ConstantTok{TRUE}\NormalTok{))}
\end{Highlighting}
\end{Shaded}

Calculate sum of square errors for fitting and prediction for Michaelis-Menten model:

\begin{Shaded}
\begin{Highlighting}[]
\NormalTok{xMean }\OtherTok{\textless{}{-}} \FunctionTok{apply}\NormalTok{(gpode}\SpecialCharTok{$}\NormalTok{xsampled, }\FunctionTok{c}\NormalTok{(}\DecValTok{2}\NormalTok{, }\DecValTok{3}\NormalTok{), mean)}
\NormalTok{fit\_vanil }\OtherTok{\textless{}{-}}\NormalTok{ xMean[gpode}\SpecialCharTok{$}\NormalTok{tvec }\SpecialCharTok{\%in\%}\NormalTok{ xsim.obs[,}\StringTok{"time"}\NormalTok{],}\DecValTok{2}\SpecialCharTok{:}\DecValTok{3}\NormalTok{]}
\NormalTok{pred\_vanil }\OtherTok{\textless{}{-}}\NormalTok{ xMean[gpode}\SpecialCharTok{$}\NormalTok{tvec }\SpecialCharTok{\%in\%}\NormalTok{ xtest[,}\StringTok{"time"}\NormalTok{],}\DecValTok{2}\SpecialCharTok{:}\DecValTok{3}\NormalTok{]}
\NormalTok{sse\_train\_vanil }\OtherTok{\textless{}{-}} \FunctionTok{sum}\NormalTok{((fit\_vanil }\SpecialCharTok{{-}}\NormalTok{ xsim.obs[,}\DecValTok{3}\SpecialCharTok{:}\DecValTok{4}\NormalTok{])}\SpecialCharTok{\^{}}\DecValTok{2}\NormalTok{)}
\NormalTok{sse\_test\_vanil }\OtherTok{\textless{}{-}}  \FunctionTok{sum}\NormalTok{((pred\_vanil }\SpecialCharTok{{-}}\NormalTok{ xtest[,}\DecValTok{3}\SpecialCharTok{:}\DecValTok{4}\NormalTok{])}\SpecialCharTok{\^{}}\DecValTok{2}\NormalTok{)}

\NormalTok{ourEst\_vanil }\OtherTok{\textless{}{-}} \FunctionTok{apply}\NormalTok{(gpode}\SpecialCharTok{$}\NormalTok{xsampled, }\FunctionTok{c}\NormalTok{(}\DecValTok{2}\NormalTok{, }\DecValTok{3}\NormalTok{), mean)}
\NormalTok{ourLB\_vanil }\OtherTok{\textless{}{-}} \FunctionTok{apply}\NormalTok{(gpode}\SpecialCharTok{$}\NormalTok{xsampled, }\FunctionTok{c}\NormalTok{(}\DecValTok{2}\NormalTok{, }\DecValTok{3}\NormalTok{), }\ControlFlowTok{function}\NormalTok{(x) }\FunctionTok{quantile}\NormalTok{(x, }\FloatTok{0.025}\NormalTok{))}
\NormalTok{ourUB\_vanil }\OtherTok{\textless{}{-}} \FunctionTok{apply}\NormalTok{(gpode}\SpecialCharTok{$}\NormalTok{xsampled, }\FunctionTok{c}\NormalTok{(}\DecValTok{2}\NormalTok{, }\DecValTok{3}\NormalTok{), }\ControlFlowTok{function}\NormalTok{(x) }\FunctionTok{quantile}\NormalTok{(x, }\FloatTok{0.975}\NormalTok{))}
\end{Highlighting}
\end{Shaded}

Summary of fitting and prediction errors for each model:

\begin{Shaded}
\begin{Highlighting}[]
\FunctionTok{print}\NormalTok{(}\FunctionTok{paste0}\NormalTok{(}\StringTok{"Inhibitor: SSE(train) = "}\NormalTok{, }\FunctionTok{round}\NormalTok{(sse\_train\_inhib,}\DecValTok{3}\NormalTok{),}
\StringTok{", SSE(test) = "}\NormalTok{, }\FunctionTok{round}\NormalTok{(sse\_test\_inhib,}\DecValTok{3}\NormalTok{)))}
\end{Highlighting}
\end{Shaded}

\begin{verbatim}
## [1] "Inhibitor: SSE(train) = 0.009, SSE(test) = 0.009"
\end{verbatim}

\begin{Shaded}
\begin{Highlighting}[]
\FunctionTok{print}\NormalTok{(}\FunctionTok{paste0}\NormalTok{(}\StringTok{"M{-}M: SSE(train) = "}\NormalTok{, }\FunctionTok{round}\NormalTok{(sse\_train\_vanil,}\DecValTok{3}\NormalTok{),}
\StringTok{", SSE(test) = "}\NormalTok{, }\FunctionTok{round}\NormalTok{(sse\_test\_vanil,}\DecValTok{3}\NormalTok{)))}
\end{Highlighting}
\end{Shaded}

\begin{verbatim}
## [1] "M-M: SSE(train) = 0.016, SSE(test) = 0.035"
\end{verbatim}

\begin{figure}
\centering
\includegraphics{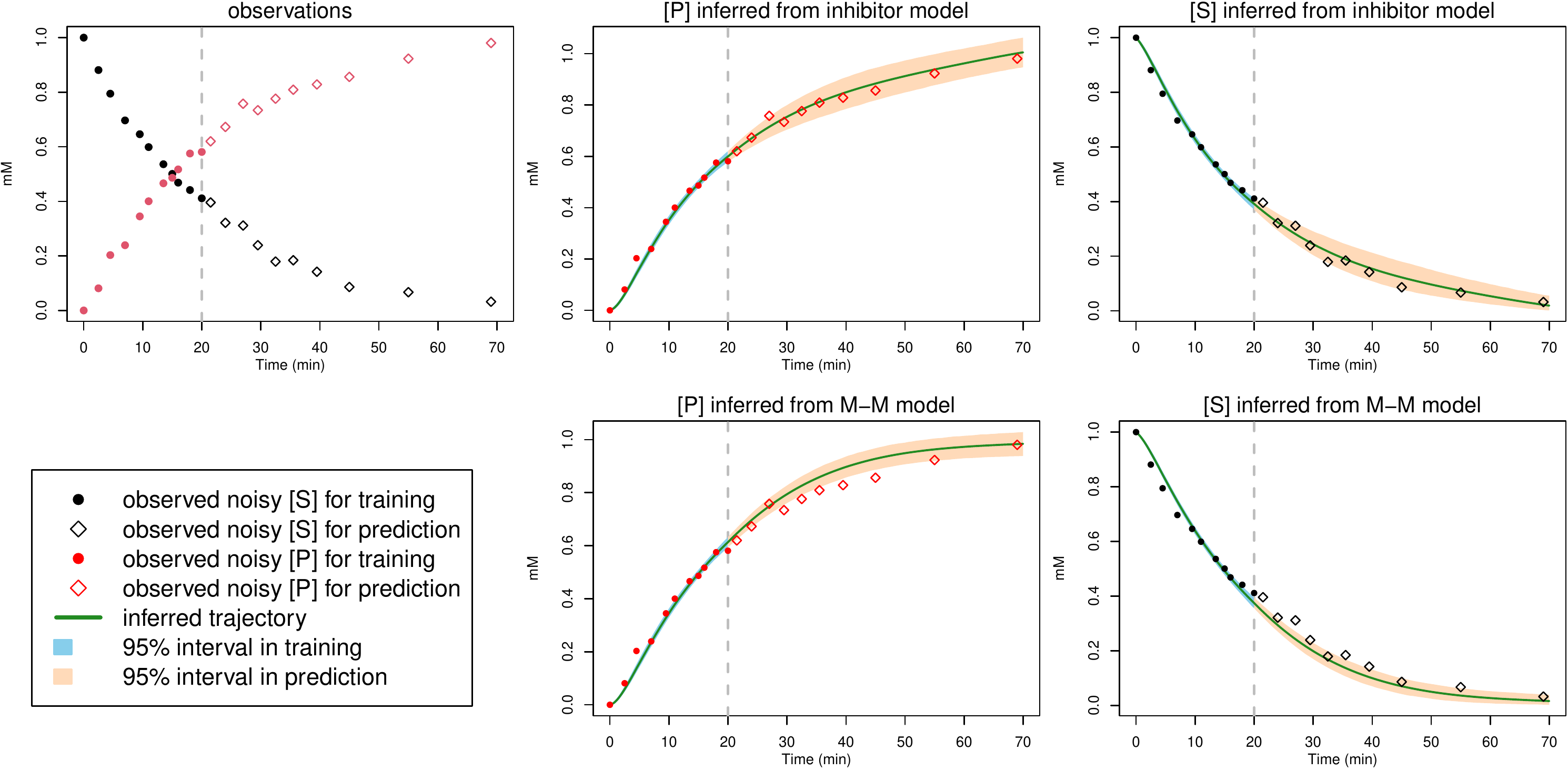}
\caption{\label{fig:mm-compare} \footnotesize Michaelis-Menten model comparison for a sample dataset simulated with inhibitor.}
\end{figure}

Visualizations of the observations, model fits and predictions, as shown in Fig \ref{fig:mm-compare}:

\begin{Shaded}
\begin{Highlighting}[]
\NormalTok{compnames }\OtherTok{\textless{}{-}} \FunctionTok{c}\NormalTok{(}\StringTok{""}\NormalTok{, }\StringTok{"[S]"}\NormalTok{, }\StringTok{"[P]"}\NormalTok{)}

\FunctionTok{layout}\NormalTok{(}\FunctionTok{cbind}\NormalTok{(}\FunctionTok{c}\NormalTok{(}\DecValTok{1}\NormalTok{,}\DecValTok{1}\NormalTok{,}\DecValTok{6}\NormalTok{,}\DecValTok{6}\NormalTok{),}\FunctionTok{c}\NormalTok{(}\DecValTok{2}\NormalTok{,}\DecValTok{2}\NormalTok{,}\DecValTok{4}\NormalTok{,}\DecValTok{4}\NormalTok{),}\FunctionTok{c}\NormalTok{(}\DecValTok{3}\NormalTok{,}\DecValTok{3}\NormalTok{,}\DecValTok{5}\NormalTok{,}\DecValTok{5}\NormalTok{)))}
\FunctionTok{par}\NormalTok{(}\AttributeTok{mar =} \FunctionTok{c}\NormalTok{(}\DecValTok{4}\NormalTok{, }\FloatTok{4.5}\NormalTok{, }\FloatTok{1.75}\NormalTok{, }\FloatTok{0.1}\NormalTok{))}

\FunctionTok{matplot}\NormalTok{(xtrue[, }\StringTok{"time"}\NormalTok{], (xtrue[, }\SpecialCharTok{{-}}\DecValTok{1}\NormalTok{]), }\AttributeTok{type=}\StringTok{"n"}\NormalTok{, }\AttributeTok{lty=}\DecValTok{1}\NormalTok{, }\AttributeTok{col=}\DecValTok{0}\NormalTok{, }\AttributeTok{xlab=}\StringTok{\textquotesingle{}\textquotesingle{}}\NormalTok{, }\AttributeTok{ylab=}\StringTok{\textquotesingle{}mM\textquotesingle{}}\NormalTok{)}
\FunctionTok{title}\NormalTok{(}\AttributeTok{xlab=}\StringTok{"Time (min)"}\NormalTok{, }\AttributeTok{line=}\DecValTok{2}\NormalTok{, }\AttributeTok{cex.lab=}\DecValTok{1}\NormalTok{)}
\FunctionTok{abline}\NormalTok{(}\AttributeTok{v =} \FunctionTok{max}\NormalTok{(obs.times), }\AttributeTok{col=}\StringTok{"grey"}\NormalTok{, }\AttributeTok{lty=}\DecValTok{2}\NormalTok{, }\AttributeTok{lwd=}\DecValTok{2}\NormalTok{)}
\FunctionTok{matplot}\NormalTok{(xsim.obs}\SpecialCharTok{$}\NormalTok{time, (xsim.obs[,}\DecValTok{3}\SpecialCharTok{:}\DecValTok{4}\NormalTok{]), }\AttributeTok{type=}\StringTok{"p"}\NormalTok{, }\AttributeTok{col=}\FunctionTok{c}\NormalTok{(}\DecValTok{1}\NormalTok{,}\DecValTok{2}\NormalTok{), }\AttributeTok{pch=}\DecValTok{19}\NormalTok{, }\AttributeTok{add =} \ConstantTok{TRUE}\NormalTok{)}
\FunctionTok{matplot}\NormalTok{(xtest}\SpecialCharTok{$}\NormalTok{time, xtest[,}\DecValTok{3}\SpecialCharTok{:}\DecValTok{4}\NormalTok{], }\AttributeTok{type=}\StringTok{"p"}\NormalTok{, }\AttributeTok{col=}\FunctionTok{c}\NormalTok{(}\DecValTok{1}\NormalTok{,}\DecValTok{2}\NormalTok{), }\AttributeTok{pch=}\DecValTok{5}\NormalTok{, }\AttributeTok{add =} \ConstantTok{TRUE}\NormalTok{)}

\FunctionTok{mtext}\NormalTok{(}\StringTok{\textquotesingle{}observations\textquotesingle{}}\NormalTok{, }\AttributeTok{line =} \FloatTok{0.3}\NormalTok{)}

\ControlFlowTok{for}\NormalTok{ (ii }\ControlFlowTok{in} \DecValTok{3}\SpecialCharTok{:}\DecValTok{2}\NormalTok{) \{}

\FunctionTok{par}\NormalTok{(}\AttributeTok{mar =} \FunctionTok{c}\NormalTok{(}\DecValTok{4}\NormalTok{, }\FloatTok{4.5}\NormalTok{, }\FloatTok{1.75}\NormalTok{, }\FloatTok{0.1}\NormalTok{))}
\NormalTok{  ourEstp }\OtherTok{\textless{}{-}}\NormalTok{ magi}\SpecialCharTok{:::}\FunctionTok{getMeanCurve}\NormalTok{(xsim}\SpecialCharTok{$}\NormalTok{time, ourEst\_inhib[,ii], xtrue[,}\DecValTok{1}\NormalTok{],}
\FunctionTok{t}\NormalTok{(pram.true}\SpecialCharTok{$}\NormalTok{phi[,ii]), }\DecValTok{0}\NormalTok{,}
\AttributeTok{kerneltype=}\NormalTok{config}\SpecialCharTok{$}\NormalTok{kernel, }\AttributeTok{deriv =} \ConstantTok{FALSE}\NormalTok{)}

\NormalTok{  ourUBp }\OtherTok{\textless{}{-}}\NormalTok{ magi}\SpecialCharTok{:::}\FunctionTok{getMeanCurve}\NormalTok{(xsim}\SpecialCharTok{$}\NormalTok{time, ourUB\_inhib[,ii], xtrue[,}\DecValTok{1}\NormalTok{],}
\FunctionTok{t}\NormalTok{(pram.true}\SpecialCharTok{$}\NormalTok{phi[,ii]), }\DecValTok{0}\NormalTok{,}
\AttributeTok{kerneltype=}\NormalTok{config}\SpecialCharTok{$}\NormalTok{kernel, }\AttributeTok{deriv =} \ConstantTok{FALSE}\NormalTok{)}

\NormalTok{  ourLBp }\OtherTok{\textless{}{-}}\NormalTok{ magi}\SpecialCharTok{:::}\FunctionTok{getMeanCurve}\NormalTok{(xsim}\SpecialCharTok{$}\NormalTok{time, ourLB\_inhib[,ii], xtrue[,}\DecValTok{1}\NormalTok{],}
\FunctionTok{t}\NormalTok{(pram.true}\SpecialCharTok{$}\NormalTok{phi[,ii]), }\DecValTok{0}\NormalTok{,}
\AttributeTok{kerneltype=}\NormalTok{config}\SpecialCharTok{$}\NormalTok{kernel, }\AttributeTok{deriv =} \ConstantTok{FALSE}\NormalTok{)}
\FunctionTok{plot}\NormalTok{( }\FunctionTok{c}\NormalTok{(}\FunctionTok{min}\NormalTok{(xtrue}\SpecialCharTok{$}\NormalTok{time),}\FunctionTok{max}\NormalTok{(xtrue}\SpecialCharTok{$}\NormalTok{time)), }\FunctionTok{c}\NormalTok{(}\FunctionTok{min}\NormalTok{(ourLBp), }\FunctionTok{min}\NormalTok{(}\FunctionTok{max}\NormalTok{(ourUBp),}\DecValTok{175}\NormalTok{)),}
\AttributeTok{type=}\StringTok{\textquotesingle{}n\textquotesingle{}}\NormalTok{, }\AttributeTok{xlab=}\StringTok{\textquotesingle{}\textquotesingle{}}\NormalTok{, }\AttributeTok{ylab=}\StringTok{\textquotesingle{}mM\textquotesingle{}}\NormalTok{)}
\FunctionTok{title}\NormalTok{(}\AttributeTok{xlab=}\StringTok{"Time (min)"}\NormalTok{, }\AttributeTok{line=}\DecValTok{2}\NormalTok{, }\AttributeTok{cex.lab=}\DecValTok{1}\NormalTok{)}
\FunctionTok{abline}\NormalTok{(}\AttributeTok{v =} \FunctionTok{max}\NormalTok{(obs.times), }\AttributeTok{col=}\StringTok{"grey"}\NormalTok{, }\AttributeTok{lty=}\DecValTok{2}\NormalTok{, }\AttributeTok{lwd=}\DecValTok{2}\NormalTok{)}

\FunctionTok{polygon}\NormalTok{(}\FunctionTok{c}\NormalTok{(xtrue[xtrue[,}\DecValTok{1}\NormalTok{] }\SpecialCharTok{\textless{}=} \FunctionTok{max}\NormalTok{(obs.times),}\DecValTok{1}\NormalTok{],}
\FunctionTok{rev}\NormalTok{(xtrue[xtrue[,}\DecValTok{1}\NormalTok{] }\SpecialCharTok{\textless{}=} \FunctionTok{max}\NormalTok{(obs.times),}\DecValTok{1}\NormalTok{])),}
\FunctionTok{c}\NormalTok{(ourUBp[xtrue[,}\DecValTok{1}\NormalTok{] }\SpecialCharTok{\textless{}=} \FunctionTok{max}\NormalTok{(obs.times)],}
\FunctionTok{rev}\NormalTok{(ourLBp[xtrue[,}\DecValTok{1}\NormalTok{] }\SpecialCharTok{\textless{}=} \FunctionTok{max}\NormalTok{(obs.times)])), }\AttributeTok{col =} \StringTok{"skyblue"}\NormalTok{, }\AttributeTok{border =} \ConstantTok{NA}\NormalTok{)    }
\FunctionTok{polygon}\NormalTok{(}\FunctionTok{c}\NormalTok{(xtrue[xtrue[,}\DecValTok{1}\NormalTok{] }\SpecialCharTok{\textgreater{}} \FunctionTok{max}\NormalTok{(obs.times),}\DecValTok{1}\NormalTok{],}
\FunctionTok{rev}\NormalTok{(xtrue[xtrue[,}\DecValTok{1}\NormalTok{] }\SpecialCharTok{\textgreater{}} \FunctionTok{max}\NormalTok{(obs.times),}\DecValTok{1}\NormalTok{])),}
\FunctionTok{c}\NormalTok{(ourUBp[xtrue[,}\DecValTok{1}\NormalTok{] }\SpecialCharTok{\textgreater{}} \FunctionTok{max}\NormalTok{(obs.times)],}
\FunctionTok{rev}\NormalTok{(ourLBp[xtrue[,}\DecValTok{1}\NormalTok{] }\SpecialCharTok{\textgreater{}} \FunctionTok{max}\NormalTok{(obs.times)])), }\AttributeTok{col =} \StringTok{"peachpuff"}\NormalTok{, }\AttributeTok{border =} \ConstantTok{NA}\NormalTok{)      }

\FunctionTok{lines}\NormalTok{(xtrue[,}\DecValTok{1}\NormalTok{], ourEstp, }\AttributeTok{col=}\StringTok{\textquotesingle{}forestgreen\textquotesingle{}}\NormalTok{, }\AttributeTok{lwd=}\FloatTok{1.5}\NormalTok{)}
\FunctionTok{mtext}\NormalTok{(}\FunctionTok{paste}\NormalTok{(compnames[ii], }\StringTok{"inferred from inhibitor model"}\NormalTok{), }\AttributeTok{line =} \FloatTok{0.3}\NormalTok{)}

\ControlFlowTok{if}\NormalTok{(compnames[ii] }\SpecialCharTok{==} \StringTok{"[P]"}\NormalTok{)\{}
\NormalTok{    point\_col }\OtherTok{=} \StringTok{"red"}
\NormalTok{  \}}\ControlFlowTok{else}\NormalTok{\{}
\NormalTok{    point\_col }\OtherTok{=} \StringTok{"black"}
\NormalTok{  \}}
\FunctionTok{points}\NormalTok{(xsim}\SpecialCharTok{$}\NormalTok{time, xsim[,ii}\SpecialCharTok{+}\DecValTok{1}\NormalTok{], }\AttributeTok{col=}\NormalTok{point\_col, }\AttributeTok{pch=}\DecValTok{16}\NormalTok{)}
\FunctionTok{points}\NormalTok{(xtest}\SpecialCharTok{$}\NormalTok{time, xtest[,ii}\SpecialCharTok{+}\DecValTok{1}\NormalTok{], }\AttributeTok{col=}\NormalTok{point\_col, }\AttributeTok{pch=}\DecValTok{5}\NormalTok{)}
\NormalTok{\}}

\ControlFlowTok{for}\NormalTok{ (ii }\ControlFlowTok{in} \DecValTok{3}\SpecialCharTok{:}\DecValTok{2}\NormalTok{) \{}

\FunctionTok{par}\NormalTok{(}\AttributeTok{mar =} \FunctionTok{c}\NormalTok{(}\DecValTok{4}\NormalTok{, }\FloatTok{4.5}\NormalTok{, }\FloatTok{1.75}\NormalTok{, }\FloatTok{0.1}\NormalTok{))}
\NormalTok{  ourEstp }\OtherTok{\textless{}{-}}\NormalTok{ magi}\SpecialCharTok{:::}\FunctionTok{getMeanCurve}\NormalTok{(xsim}\SpecialCharTok{$}\NormalTok{time, ourEst\_vanil[,ii], xtrue[,}\DecValTok{1}\NormalTok{],}
\FunctionTok{t}\NormalTok{(pram.true}\SpecialCharTok{$}\NormalTok{phi[,ii]), }\DecValTok{0}\NormalTok{,}
\AttributeTok{kerneltype=}\NormalTok{config}\SpecialCharTok{$}\NormalTok{kernel, }\AttributeTok{deriv =} \ConstantTok{FALSE}\NormalTok{)}

\NormalTok{  ourUBp }\OtherTok{\textless{}{-}}\NormalTok{ magi}\SpecialCharTok{:::}\FunctionTok{getMeanCurve}\NormalTok{(xsim}\SpecialCharTok{$}\NormalTok{time, ourUB\_vanil[,ii], xtrue[,}\DecValTok{1}\NormalTok{],}
\FunctionTok{t}\NormalTok{(pram.true}\SpecialCharTok{$}\NormalTok{phi[,ii]), }\DecValTok{0}\NormalTok{,}
\AttributeTok{kerneltype=}\NormalTok{config}\SpecialCharTok{$}\NormalTok{kernel, }\AttributeTok{deriv =} \ConstantTok{FALSE}\NormalTok{)}

\NormalTok{  ourLBp }\OtherTok{\textless{}{-}}\NormalTok{ magi}\SpecialCharTok{:::}\FunctionTok{getMeanCurve}\NormalTok{(xsim}\SpecialCharTok{$}\NormalTok{time, ourLB\_vanil[,ii], xtrue[,}\DecValTok{1}\NormalTok{],}
\FunctionTok{t}\NormalTok{(pram.true}\SpecialCharTok{$}\NormalTok{phi[,ii]), }\DecValTok{0}\NormalTok{,}
\AttributeTok{kerneltype=}\NormalTok{config}\SpecialCharTok{$}\NormalTok{kernel, }\AttributeTok{deriv =} \ConstantTok{FALSE}\NormalTok{)}
\FunctionTok{plot}\NormalTok{( }\FunctionTok{c}\NormalTok{(}\FunctionTok{min}\NormalTok{(xtrue}\SpecialCharTok{$}\NormalTok{time),}\FunctionTok{max}\NormalTok{(xtrue}\SpecialCharTok{$}\NormalTok{time)), }\FunctionTok{c}\NormalTok{(}\FunctionTok{min}\NormalTok{(ourLBp), }\FunctionTok{min}\NormalTok{(}\FunctionTok{max}\NormalTok{(ourUBp),}\DecValTok{175}\NormalTok{)),}
\AttributeTok{type=}\StringTok{\textquotesingle{}n\textquotesingle{}}\NormalTok{, }\AttributeTok{xlab=}\StringTok{\textquotesingle{}\textquotesingle{}}\NormalTok{, }\AttributeTok{ylab=}\StringTok{\textquotesingle{}mM\textquotesingle{}}\NormalTok{)}
\FunctionTok{title}\NormalTok{(}\AttributeTok{xlab=}\StringTok{"Time (min)"}\NormalTok{, }\AttributeTok{line=}\DecValTok{2}\NormalTok{, }\AttributeTok{cex.lab=}\DecValTok{1}\NormalTok{)}
\FunctionTok{abline}\NormalTok{(}\AttributeTok{v =} \FunctionTok{max}\NormalTok{(obs.times), }\AttributeTok{col=}\StringTok{"grey"}\NormalTok{, }\AttributeTok{lty=}\DecValTok{2}\NormalTok{, }\AttributeTok{lwd=}\DecValTok{2}\NormalTok{)  }

\FunctionTok{polygon}\NormalTok{(}\FunctionTok{c}\NormalTok{(xtrue[xtrue[,}\DecValTok{1}\NormalTok{] }\SpecialCharTok{\textless{}=} \FunctionTok{max}\NormalTok{(obs.times),}\DecValTok{1}\NormalTok{],}
\FunctionTok{rev}\NormalTok{(xtrue[xtrue[,}\DecValTok{1}\NormalTok{] }\SpecialCharTok{\textless{}=} \FunctionTok{max}\NormalTok{(obs.times),}\DecValTok{1}\NormalTok{])),}
\FunctionTok{c}\NormalTok{(ourUBp[xtrue[,}\DecValTok{1}\NormalTok{] }\SpecialCharTok{\textless{}=} \FunctionTok{max}\NormalTok{(obs.times)],}
\FunctionTok{rev}\NormalTok{(ourLBp[xtrue[,}\DecValTok{1}\NormalTok{] }\SpecialCharTok{\textless{}=} \FunctionTok{max}\NormalTok{(obs.times)])), }\AttributeTok{col =} \StringTok{"skyblue"}\NormalTok{, }\AttributeTok{border =} \ConstantTok{NA}\NormalTok{)    }
\FunctionTok{polygon}\NormalTok{(}\FunctionTok{c}\NormalTok{(xtrue[xtrue[,}\DecValTok{1}\NormalTok{] }\SpecialCharTok{\textgreater{}} \FunctionTok{max}\NormalTok{(obs.times),}\DecValTok{1}\NormalTok{],}
\FunctionTok{rev}\NormalTok{(xtrue[xtrue[,}\DecValTok{1}\NormalTok{] }\SpecialCharTok{\textgreater{}} \FunctionTok{max}\NormalTok{(obs.times),}\DecValTok{1}\NormalTok{])),}
\FunctionTok{c}\NormalTok{(ourUBp[xtrue[,}\DecValTok{1}\NormalTok{] }\SpecialCharTok{\textgreater{}} \FunctionTok{max}\NormalTok{(obs.times)],}
\FunctionTok{rev}\NormalTok{(ourLBp[xtrue[,}\DecValTok{1}\NormalTok{] }\SpecialCharTok{\textgreater{}} \FunctionTok{max}\NormalTok{(obs.times)])), }\AttributeTok{col =} \StringTok{"peachpuff"}\NormalTok{, }\AttributeTok{border =} \ConstantTok{NA}\NormalTok{)    }

\FunctionTok{lines}\NormalTok{(xtrue[,}\DecValTok{1}\NormalTok{], ourEstp, }\AttributeTok{col=}\StringTok{\textquotesingle{}forestgreen\textquotesingle{}}\NormalTok{, }\AttributeTok{lwd=}\FloatTok{1.5}\NormalTok{)}
\FunctionTok{mtext}\NormalTok{(}\FunctionTok{paste}\NormalTok{(compnames[ii], }\StringTok{"inferred from M{-}M model"}\NormalTok{), }\AttributeTok{line =} \FloatTok{0.3}\NormalTok{)}

\ControlFlowTok{if}\NormalTok{(compnames[ii] }\SpecialCharTok{==} \StringTok{"[P]"}\NormalTok{)\{}
\NormalTok{    point\_col }\OtherTok{=} \StringTok{"red"}
\NormalTok{  \}}\ControlFlowTok{else}\NormalTok{\{}
\NormalTok{    point\_col }\OtherTok{=} \StringTok{"black"}
\NormalTok{  \}}
\FunctionTok{points}\NormalTok{(xsim}\SpecialCharTok{$}\NormalTok{time, xsim[,ii}\SpecialCharTok{+}\DecValTok{1}\NormalTok{], }\AttributeTok{col=}\NormalTok{point\_col, }\AttributeTok{pch=}\DecValTok{16}\NormalTok{)}
\FunctionTok{points}\NormalTok{(xtest}\SpecialCharTok{$}\NormalTok{time, xtest[,ii}\SpecialCharTok{+}\DecValTok{1}\NormalTok{], }\AttributeTok{col=}\NormalTok{point\_col, }\AttributeTok{pch=}\DecValTok{5}\NormalTok{)}

\NormalTok{\}}

\FunctionTok{par}\NormalTok{(}\AttributeTok{mar=}\FunctionTok{rep}\NormalTok{(}\DecValTok{0}\NormalTok{,}\DecValTok{4}\NormalTok{))}
\FunctionTok{plot}\NormalTok{(}\DecValTok{1}\NormalTok{,}\AttributeTok{type=}\StringTok{\textquotesingle{}n\textquotesingle{}}\NormalTok{, }\AttributeTok{xaxt=}\StringTok{\textquotesingle{}n\textquotesingle{}}\NormalTok{, }\AttributeTok{yaxt=}\StringTok{\textquotesingle{}n\textquotesingle{}}\NormalTok{, }\AttributeTok{xlab=}\ConstantTok{NA}\NormalTok{, }\AttributeTok{ylab=}\ConstantTok{NA}\NormalTok{, }\AttributeTok{frame.plot =} \ConstantTok{FALSE}\NormalTok{)}

\NormalTok{oos\_bg\_col }\OtherTok{=} \StringTok{"peachpuff"}

\FunctionTok{legend}\NormalTok{(}\StringTok{"center"}\NormalTok{, }\FunctionTok{c}\NormalTok{(}\StringTok{"observed noisy [S] for training"}\NormalTok{, }\StringTok{"observed noisy [S] for prediction"}\NormalTok{,}
\StringTok{"observed noisy [P] for training"}\NormalTok{, }\StringTok{"observed noisy [P] for prediction"}\NormalTok{,}
\StringTok{"inferred trajectory"}\NormalTok{, }\StringTok{"95\% interval in training"}\NormalTok{,}
\StringTok{"95\% interval in prediction"}\NormalTok{), }
\AttributeTok{lty=}\FunctionTok{c}\NormalTok{(}\DecValTok{0}\NormalTok{,}\DecValTok{0}\NormalTok{,}\DecValTok{0}\NormalTok{,}\DecValTok{0}\NormalTok{,}\DecValTok{1}\NormalTok{,}\DecValTok{0}\NormalTok{,}\DecValTok{0}\NormalTok{), }\AttributeTok{lwd=}\FunctionTok{c}\NormalTok{(}\DecValTok{0}\NormalTok{,}\DecValTok{1}\NormalTok{,}\DecValTok{0}\NormalTok{,}\DecValTok{1}\NormalTok{,}\DecValTok{3}\NormalTok{,}\DecValTok{0}\NormalTok{,}\DecValTok{0}\NormalTok{),}
\AttributeTok{col =} \FunctionTok{c}\NormalTok{(}\DecValTok{1}\NormalTok{,}\DecValTok{1}\NormalTok{,}\StringTok{"red"}\NormalTok{,}\StringTok{"red"}\NormalTok{, }\StringTok{"forestgreen"}\NormalTok{, }\ConstantTok{NA}\NormalTok{, }\ConstantTok{NA}\NormalTok{),}
\AttributeTok{fill=}\FunctionTok{c}\NormalTok{(}\DecValTok{0}\NormalTok{,}\DecValTok{0}\NormalTok{,}\DecValTok{0}\NormalTok{,}\DecValTok{0}\NormalTok{, }\DecValTok{0}\NormalTok{,}\StringTok{"skyblue"}\NormalTok{,oos\_bg\_col),}
\AttributeTok{border=}\FunctionTok{c}\NormalTok{(}\DecValTok{0}\NormalTok{,}\DecValTok{0}\NormalTok{,}\DecValTok{0}\NormalTok{,}\DecValTok{0}\NormalTok{, }\DecValTok{0}\NormalTok{, }\StringTok{"skyblue"}\NormalTok{,oos\_bg\_col), }\AttributeTok{pch=}\FunctionTok{c}\NormalTok{(}\DecValTok{19}\NormalTok{,}\DecValTok{5}\NormalTok{,}\DecValTok{19}\NormalTok{,}\DecValTok{5}\NormalTok{, }\ConstantTok{NA}\NormalTok{, }\DecValTok{15}\NormalTok{, }\DecValTok{15}\NormalTok{), }\AttributeTok{cex=}\FloatTok{1.7}\NormalTok{)}
\end{Highlighting}
\end{Shaded}

